\documentclass[final,5p,times,twocolumn,authoryear]{elsarticle}

\usepackage[T1]{fontenc}

\pdfoutput=1 %

\usepackage{lineno}
\usepackage{siunitx}
\DeclareSIUnit\pe{p.e.}
\DeclareSIUnit\erg{erg}
\DeclareSIUnit\diffflux{\erg \per \centi \metre \squared \per \second \per \TeV}

\usepackage{xfrac}
\usepackage{xcolor}
\usepackage{multirow}

\usepackage{aas_macros}
\usepackage{eurosym}

\usepackage[colorlinks=true]{hyperref}

\usepackage{amsmath}
\usepackage{amssymb}
\usepackage{dcolumn}

\newcommand*{\rdash}{--}

\providecommand*{\degree}{^{\circ}}

\newcommand*{\semla}{\textsc{\textit{ALTO21}}}

\begin{document}

\begin{frontmatter}

\title{
\boldmath{Sensitivity to point-like sources of the ALTO atmospheric particle detector array, designed for $\rm 200\,GeV$--$\rm 50\,TeV$ \\ $\gamma$-ray astronomy}
}

\author[a]{M.~Punch\fnref{corresponding1}}
\author[b]{M.~Senniappan}
\author[a]{Y.~Becherini}
\author[b]{G.~Kukec~Mezek}
\author[c]{S.~Thoudam}
\author[b]{T.~Bylund}
\author[d]{J.-P.~Ernenwein}

\affiliation[a]{organization={Universit\'e Paris Cit\'e, CNRS/IN2P3, AstroParticule et Cosmologie (APC)}, city={Paris~F-75013}, country={France}}

\affiliation[b]{organization={Department of Physics and Electrical Engineering, Linnaeus University}, postcode={35195}, city={V\"axj\"o}, country={Sweden}}

\affiliation[c]{organization={Department of Physics, Khalifa University}, addressline={PO Box 127788}, city={Abu Dhabi}, country={United Arab Emirates}}

\affiliation[d]{organization={Aix Marseille Univ, CNRS/IN2P3, CPPM}, city={Marseille}, country={France}}

\fntext[corresponding1]{Corresponding author:  Michael Punch, %
\href{mailto:punch@in2p3.fr}{punch@in2p3.fr}}

\begin{abstract}
In the context of atmospheric shower arrays designed for $\gamma$-ray astronomy and in the context of the ALTO project, we present: a study of the impact of heavier nuclei in the cosmic-ray background on the estimated $\gamma$-ray detection performance on the basis of dedicated Monte Carlo simulations, a method to calculate the sensitivity to a point-like source, and finally the required observation times to reach a firm detection on a list of known point-like sources. 
\end{abstract}


\end{frontmatter}

\section{Introduction}
\label{sec:intro}

ALTO is a R\&D project that aims to design a wide field-of-view very high energy (VHE; >~\SI{200}{GeV}) gamma-ray observatory \citep{semla_jinst, ALTO1_ICRC2017, ALTO2_ICRC2017, ALTO3_ICRC2021} through prototype operation and simulation studies. 
The ALTO detectors take advantage of the water Cherenkov technique, and are optimized for the detection of VHE gamma rays at high altitudes. 

In contrast to most current and future initiatives in wide-field-of-view ground-based gamma-ray astronomy, which concentrate on the multi-TeV to multi-PeV energy range (e.g., HAWC, High-Altitude Water Cherenkov Gamma-Ray Observatory, \citealt{HAWC:Crab}; LHAASO, Large High Altitude Air Shower Observatory, \citealt{LHAASO_ScienceBook2021}; or the proposed SWGO, Southern Wide-field Gamma-ray Observatory, \citealt{SGSO:WhitePaper}), the primary science goal of the ALTO project is to achieve a low threshold, of the order of \SI{200}{GeV}, for the detection, the monitoring, and the characterization of soft-spectrum VHE gamma-ray sources, which include active galactic nuclei (AGN), and gamma-ray bursts (GRB). 
Therefore, the goal of ALTO is to optimize and improve the existing particle sampling technique in the challenging domain of the energy range from \SI{200}{GeV} to \SI{10}{TeV}, while keeping the costs reasonably low\footnote{We estimate a $\sim 20\rm\, M$\texteuro~  materials construction cost on the basis of an extrapolation from the prototype cost.  The prototype is described in Section \ref{subsec:prototype}.}.

Detecting gamma rays with wide-field-of-view ground-based particle sampling detectors in particular in the \SI{200}{GeV}--\SI{10}{TeV} energy range requires both the capability of sampling ``small'' gamma-ray showers originated from low-energy gamma rays interacting at very high altitude in the atmosphere and an excellent gamma-ray over cosmic-ray separation.

Signal over background (S/B) classification in this energy range is a challenge: small gamma-ray and cosmic-ray showers look very similar, as they trigger only a small number of particle detectors. We showed in \cite{semla_jinst} -- referred to below as {\semla} -- our limited capability of S/B separation in the lowest-energy bin with the SEMLA event classification, which is based on a standard machine learning procedure using human-crafted features. But we also mentioned that a Deep-Learning-based classification might further enhance the event separation in this energy range, as differences between the two classes which are inconclusive for conventional methods might yet be taken advantage of by Deep Learning. 
In this paper, we evaluate the sensitivity of the simple ALTO design with the SEMLA analysis technique without the use of Deep Learning, so the results shown here can be considered as a lower limit to the ALTO capabilities at the lowest energies.

In {\semla}, we demonstrated that using an advanced analysis strategy like SEMLA, we achieve a good performance in the energy range considered\footnote{Besides the lowest-energy bin.}, and that our results do not improve substantially when we add the information from the signals acquired in the scintillators positioned underneath the primary water Cherenkov tanks. Therefore we concluded that in our energy range, the cost of adding a layer of scintillators was not justified; consequently in this paper, we only use the primary water Cherenkov tanks.

Besides the obvious requirement to operate such detectors at high altitude (\SI{5.1}{\km} in our simulations) to reach a lower threshold, our detector design improvements include a simplification of the particle detector which, in this paper, is a water Cherenkov tank\footnote{We will, in a future paper, present an alternative to the water Cherenkov tank. The water Cherenkov tank has the main advantage of providing an excellent timing accuracy and to use only water as the detector medium, but it comes also with some disadvantages: the large volume of water to be transported at high altitude and the need for water purification to avoid the growth of bio-organisms.}, and the use of signal waveforms which improve the timing accuracy over threshold-passing methods.

Our study therefore aims to determine the requirements and possible hardware choices in order for such a technology to be effective for extragalactic gamma-ray astronomy. Our hardware choices could be implemented as a standalone array, or encapsulated as a dedicated low-energy dense core inside a larger array. 
Possible locations for the ALTO array are all in the Andes (Argentina, Chile, Bolivia, Peru), to have the required high-altitude plateau and to ensure complementarity with LHAASO \citep{LHAASO_ScienceBook2021}, which is operational in the Northern hemisphere.

Finally, in this paper we consider point-like $\gamma$-ray sources only, although ALTO should also be sensitive to diffuse and extended sources, thanks to its very wide Field-of-View.

\subsection{The proposed ALTO detector design}
\label{sec:detector}

\begin{figure*}[t]
\centering
\includegraphics[width=0.42\textwidth] {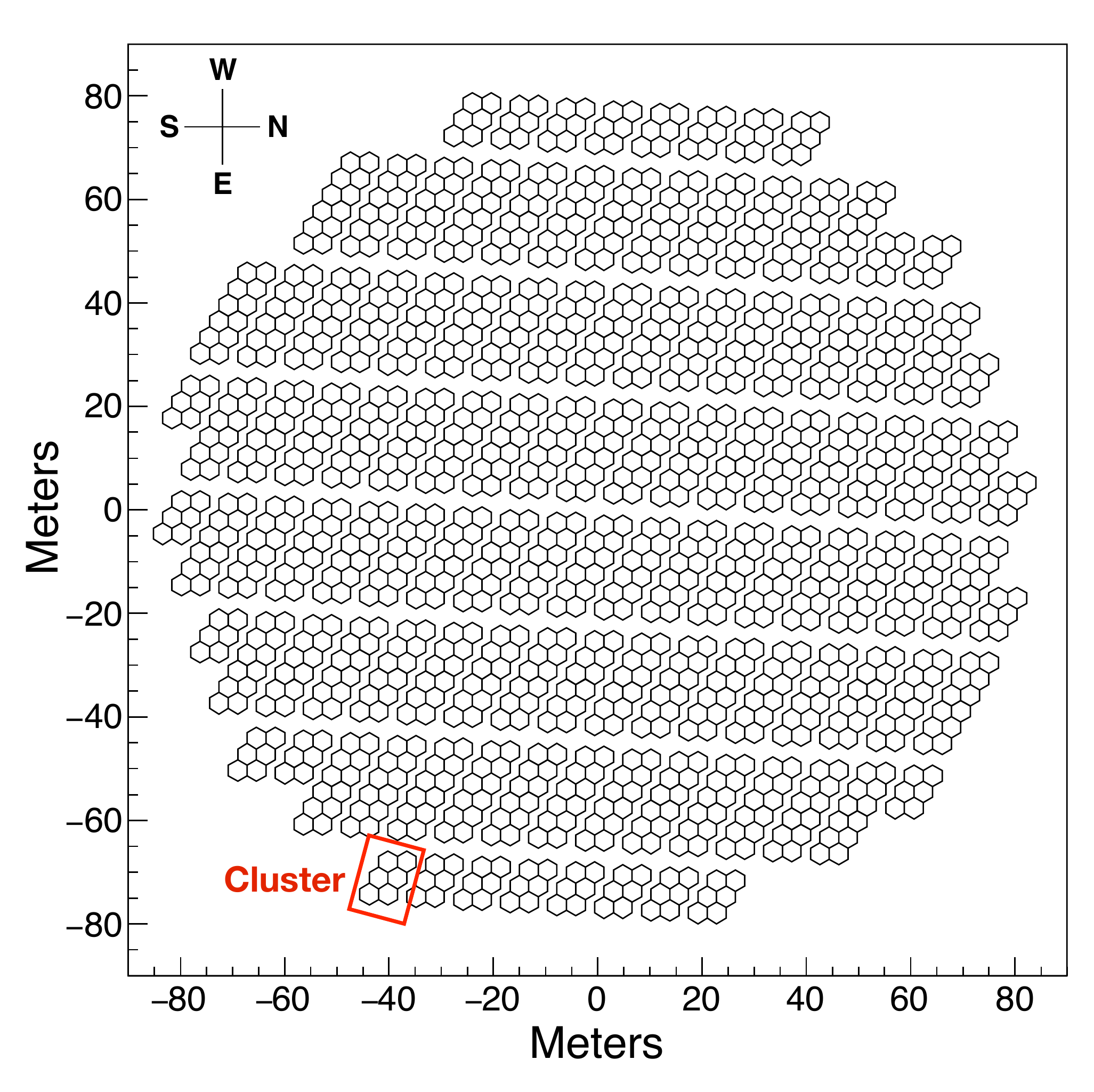}
    \hspace{0.05cm}
    \includegraphics[width=0.54\textwidth] {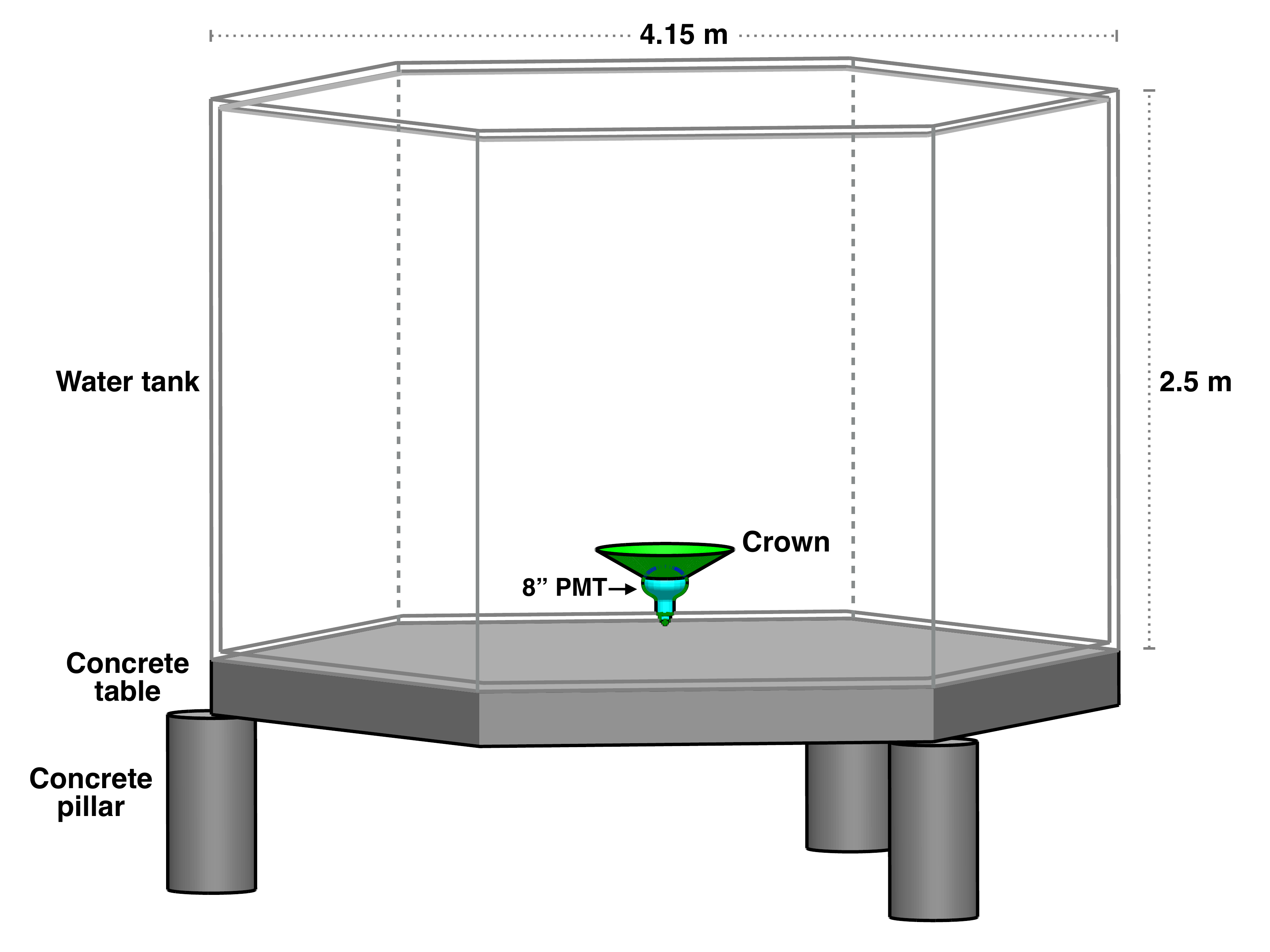}
\caption{\label{fig:alto_array}
\textbf{Proposed ALTO detector array.} \emph{Left panel:} The layout of the proposed ALTO array of \SI{160}{m} diameter. A cluster consists of six ALTO units as highlighted in the red box. \emph{Right panel:} The geometry of an ALTO unit with its dimensions. A unit consists of a water Cherenkov detector (WCD).}

\end{figure*}

The proposed ALTO detector array consists of 1242 detectors in a circular array of \SI{80}{\m} radius as shown in Figure \ref{fig:alto_array}. The full array is segmented into clusters, and each cluster consists of six hexagonal-shaped Water Cherenkov Detectors (WCDs). The WCDs are \SI{2.5}{\m} tall and \SI{4.15}{\m} wide between the opposite corners of the hexagon. Each WCD is filled with \SI{25}{\m^3} of water. An 8" super-bialkali Hamamatsu photo multiplier tube (R5912–100) is placed at the central bottom of the WCDs. A reflective crown is positioned on top of the PMT to increase the water Cherenkov photon collection by $\sim$40\%.  

The initial ALTO design in {\semla} \citep{semla_jinst}  comprised an additional layer of 1242 scintillator detectors, separated from each WCD by \SI{25}{\cm} of concrete. This scintillator layer was planned as a muon detector, aiming to better distinguish gamma rays from cosmic rays.  
As we reported in {\semla} , this scintillator layer contributes somewhat to the background discrimination, yet not significantly enough to justify the cost, at least for the ALTO energy range and detector array size. Hence, only the layer of WCDs is used in this present work.

The simulations used in this paper are carried out for the full proposed ALTO detector, covering an area of \SI{20000}{\m^2}. 
A detailed simulation study comprising shower generation, detector response, trigger, event reconstruction, data analysis, and its subsequent performance study has been presented in {\semla} . 

\subsection{The ALTO prototype at Linnaeus University}
\label{subsec:prototype}

The ALTO R\&D project involves prototype and simulation studies. 
We have operated a prototype of the ALTO detector units at Linnaeus University, Växjö, from February 2019 \citep{ALTO3_ICRC2021} until June 2022.  The prototype consisted of two water Cherenkov detectors, accompanied by two large scintillator detectors and several smaller monitoring scintillator detectors, as well as environmental monitoring devices.  The prototype allowed the verification and the stability of the hardware choices implemented in the simulations, and in particular justifies the final choice to lower the thresholds for triggering and event reconstruction. 
Indeed, the prototype has been operated with a threshold initially safely set at \SI{20}{mV}, for each channel, and then lowered to \SI{10}{mV} while still triggering for $\ge2$~PMTs exceeding this threshold. This allowed to verify the feasibility of the readout at this threshold, which on the other hand, in simulations, has been proven to enhance the sensitivity of a full array at low energy.
 The prototype served to evaluate the readout method based on the WaveCatcher device \citep{7097545} set at \SI{400}{MHz} sampling rate, 128~samples, and its Linux libraries that we used in our real-time data analysis and formatting with ROOT6 \citep{Brun:1997pa,Bellenot_2015}, running on a dedicated Single Board Computer connected to the WaveCatcher via UDP. The sampling rate chosen for simulations is identical, and the number of samples similar (120), to ensure a similar and realistic performance in charge integration and Time-of-Maximum measurement.  The single-photoelectron waveform template implemented in the simulations has been measured on the PMTs used in the prototype, at the operation gain of $10^7$.

\subsection{Aim of current work and evolution with respect to previous work}

The current work presents a refined study compared to {\semla} focusing on the following:
\begin{enumerate}
    \item The aim of {\semla} was to introduce the Monte Carlo data production and analysis procedure and to study the performance in terms of background discrimination, angular resolution, effective area, and resolution on core position and energy.
    We showed that for the purpose of $\gamma$-ray astronomy, the scintillator detector layer as conceived in our initial design gives mostly redundant information, and therefore does not provide significant additional background discrimination. 
    Given this conclusion, in the current paper we present a new study using only one detector layer -- in particular the one giving the better timing information, i.e., the layer of WCDs. 
    \item In {\semla}, a detector ``cluster" is composed by 12 detectors (6 WCDs and 6 scintillation detectors), while in this work, a cluster is composed by only 6 WCDs. Hence, the cluster trigger condition used changes from \textit{2 out of 12} in {\semla} to \textit{2 out of 6} in this work.
    \item From the experience gained in the ALTO prototype activity, we learnt that we could lower the cluster trigger threshold which opens the way to lowering the full detector threshold. In {\semla} , we used a  \SI{20}{\milli\volt} threshold per detector, which corresponds to $\sim3$~photoelectrons (p.e.), while a \SI{10}{\milli\volt} threshold will be used in this work, corresponding to \SI{\sim 1.5}{\pe} Additionally, for the event reconstruction, the minimum level to take into account the WCD signal can be reduced from 5 to \SI{2}{\pe} in integrated signal.
    \item In {\semla} , we assumed that all the cosmic-ray components behave like protons, which we simulated and scaled up to the cosmic-ray spectrum.  This results in an overestimate of the cosmic-ray rate seen by ALTO.  In this work, we study specifically the impact of the helium component on rates, on the background discrimination, and on the sensitivity.
    \item In {\semla} , we provided a performance study in terms of resolutions, biases and effective areas. In this paper, we update this to reflect the new configuration and trigger conditions, and we provide a calculation of the sensitivity.
    We calculate the Instrument Response Functions (IRFs), and use them in the framework of \texttt{Gammapy} (see Section~\ref{sec:sensi}), which allows us to study and predict the detection performance of a number of already-known $\gamma$-ray sources. In this framework, we study the time needed to reach a firm $5\sigma$ detection, as well as their spectra.
\end{enumerate}

\subsection{Structure of paper}

Section \ref{sec:simulation} describes the atmospheric air shower and detector Monte Carlo simulations. The trigger configuration used, and the reconstruction technique, is briefly reviewed in Section \ref{sec:reco}. Section \ref{sec:semla} describes the analysis performed on the reconstructed events, and Section \ref{sec:ana_perf} details the generation of the instrument response functions. Finally, we present the ALTO sensitivity to point-like $\gamma$-ray sources, as well as the detection performance and spectral analysis on a list of VHE $\gamma$-ray sources in section \ref{sec:sensi}. Then, we conclude with a discussion on possible future improvements in section \ref{sec:outlook}.

\section{Shower particles and ALTO detector response simulation}
\label{sec:simulation}

\begin{table*}[t]
    \centering
    \small
    \caption{\small{\textbf{Simulation details.} CORSIKA shower simulation characteristics and number of events simulated.
    $ \theta_{\rm T} $  and $ \phi_{\rm T} $ represent the true zenith and azimuthal angle, respectively.
    The simulated live-time for the background is derived from these characteristics, assuming the proton and helium spectra given in \cite{CTA_MC_KB_2013}.\\ 
    }}
    
    \label{table:simulation}
    
    \begin{tabular}{cScccccS}
        \hline
        \textbf{Primary} & \textbf{Power law} & \textbf{Energy} & \boldmath{$\theta_{\rm T}$}& \boldmath{$\phi_{\rm T}$} & 
        \textbf{Impact} & \textbf{Events} & \textbf{Simulated} \\ 
        
        \textbf{type} & \textbf{spectral} & \textbf{range} & \textbf{[deg]}& \textbf{[deg]} & \textbf{parameter} & $\mathbf{(\times10^6)}$ & \textbf{live-time}\\
        
        & \textbf{index}& \textbf{[TeV]} & & & \textbf{[m]} & & \textbf{[min]} \\ 
        
        \hline
            $\gamma$-ray & -2 & 0.01\rdash100 & 18 & 0 & 0\rdash130& 34 & \\
            proton & -2.7 & 0.06\rdash100 & 15\rdash21 & 0\rdash360 & 0\rdash184 & 224 & 27.6 \\ %
            helium nuclei & -2.64 & 0.06\rdash100 & 15\rdash21 & 0\rdash360 & 0\rdash184 & 199 & 37.6 \\ %
        \hline
    \end{tabular}
\end{table*}
    
For this study, atmospheric air showers are simulated using \texttt{CORSIKA v7.4387} \cite{corsika}. The primary particles simulated include point-like $\gamma$-rays, diffuse protons (\texttt{p}) and helium nuclei (\texttt{He}). The air shower particles are simulated down to \SI{5.1}{\km} above sea level. 
Table \ref{table:simulation} summarises the characteristics of the air shower simulations used in this study, which follow those used in {\semla}, with the addition of the helium background. We use the same characteristics for the proton and helium simulations, with the exception of their power law spectral indices, these being taken from \cite{CTA_MC_KB_2013}.
Given these characteristics, we can calculate the live-times which would be required to produce the numbers of protons and helium nuclei simulated, as 27.6 and \SI{37.6}{minutes}, respectively, as shown in the table.  These live-times are needed to find the corresponding background rates as a function of energy for the IRFs (sections~\ref{sec:semla}, \ref{sec:ana_perf}).

Following the atmospheric air shower simulation, \texttt{Geant4 v4.10.02.p02} \citep{geant4} is used to obtain the ALTO detector response. 
The detector design introduced in Section~\ref{sec:detector} is used for the simulation.  
The number of Cherenkov photons and their arrival time at the PMT level are produced by the detector simulation. Subsequently, the waveform of the signals are generated using our custom PMT response simulation.

\section{Trigger configuration and shower reconstruction}
\label{sec:reco}
In order to optimize the array for the detection of  soft spectrum VHE sources, a loose trigger configuration with a low signal threshold is used: a 6-WCD cluster is considered as triggered if 2 tanks have a signal exceeding \SI{10}{\mV}, which corresponds to about \SI{1.5}{\pe} at the foreseen operation gain of $10^7$ for the photomultipliers. This improvement with respect to the work presented in {\semla} , where the threshold was \SI{20}{\mV}, has been proven to be feasible thanks to the ALTO prototype activity. Indeed, in addition to the fact that operation remains stable and immune to electromagnetic noise, the higher accidental coincidence rate measured at \SI{10}{\mV} can be handled at the level of the cluster, where a single board computer extracts the relevant parameters from the waveform to send them to the central data acquisition, and, when extrapolating, at the level of the central DAQ itself.

Clusters not satisfying the trigger criterion are ignored. For clusters passing the trigger, the parameters extracted from the waveform are the integrated charge in units of photoelectrons and the time of  maximum of the waveform. These two quantities are used in the reconstruction procedure. For an event to be reconstructed, at least eight tanks should have a signal exceeding \SI{2}{\pe} 
This key feature allows to lower the energy threshold of the array, compared to {\semla} , where  \SI{5}{\pe} have been used.

The shower reconstruction is performed in several steps. First, using the integrated charge from each detector, the shower core position is obtained from a fit of the lateral distribution with the NKG function \citep{nkg_1, nkg_2}. Then, the time of maximum and the shower core are used to obtain the arrival direction of the primary particle using a hyperbolic shower front model \citep{hyper_1, hyper_2}. The reconstruction procedure also provides other shower parameters, such as shower size, which are later used in the analysis.
The shower reconstruction and the variables obtained from the procedure are reported in detail in {\semla} . 

\section{Background discrimination and energy reconstruction: the SEMLA analysis}
\label{sec:semla}
The SEMLA (Signal Extraction using Machine Learning for ALTO) analysis involves machine learning methods using artificial neural networks for selecting well-reconstructed events, for gamma over hadron separation, and for energy reconstruction {\semla} . 

The analysis procedure is applied to events for which there are at least eight tanks triggered ($N_{\rm WCD}\geq8$).  It consists of four successive stages: stages A, B, C and D. Stage A is a simple filter to remove the few pathological events for which the reconstruction has clearly failed to converge. 
Stages B and C involve machine learning classification procedures. The classification of the events is performed for three bins in reconstructed shower size to ensure energy-dependent background rejection. After training/testing, the cuts are obtained for a given signal efficiency.
Stage B focuses on removing the poorly reconstructed events.
The crucial part of the SEMLA analysis is to separate the $\gamma$-rays from the cosmic-ray background, which is performed in stage C. 
We noted that if we include helium nuclei in the training, we do not find any improvement in the background rejection performance, so here we use only protons in the training. 
The analysis cuts defined in Stages A, B, and C are then applied to simulated $\gamma$-rays, protons, and helium nuclei leaving a set of well-reconstructed $\gamma$-rays and ``$\gamma$-ray-like'' protons and helium. 
Finally, in stage D, the energy of these selected events is obtained using a machine learning regression method. 

The SEMLA training was done on one half of the simulated events, while the other half was used for the performance plots and the sensitivity calculation in the following sections. 

\begin{figure*}[t]
    \centering
    \includegraphics[width=0.48\textwidth]{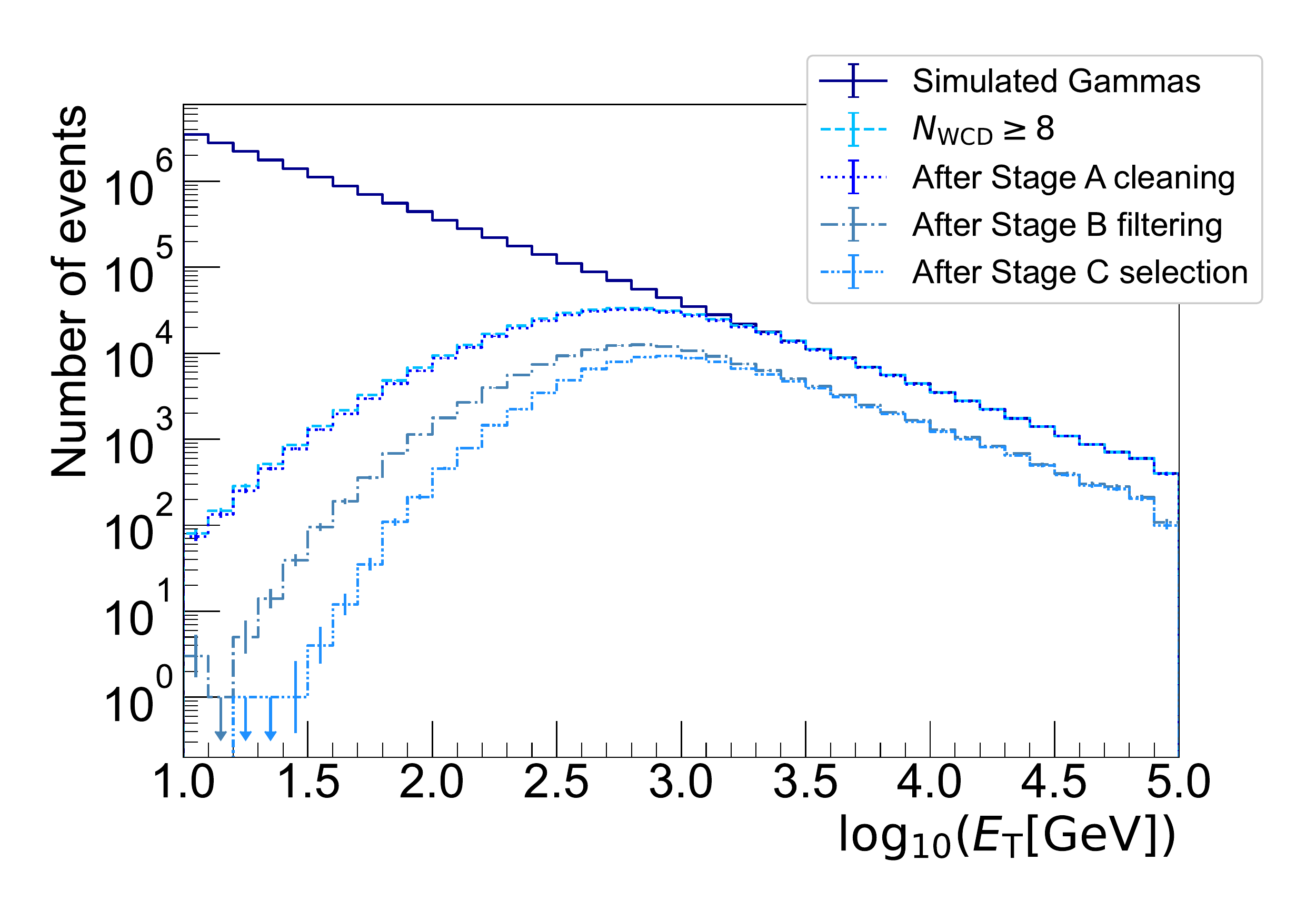}
     \hspace{.05
     cm}
     \\
     \includegraphics[width=0.48\textwidth]{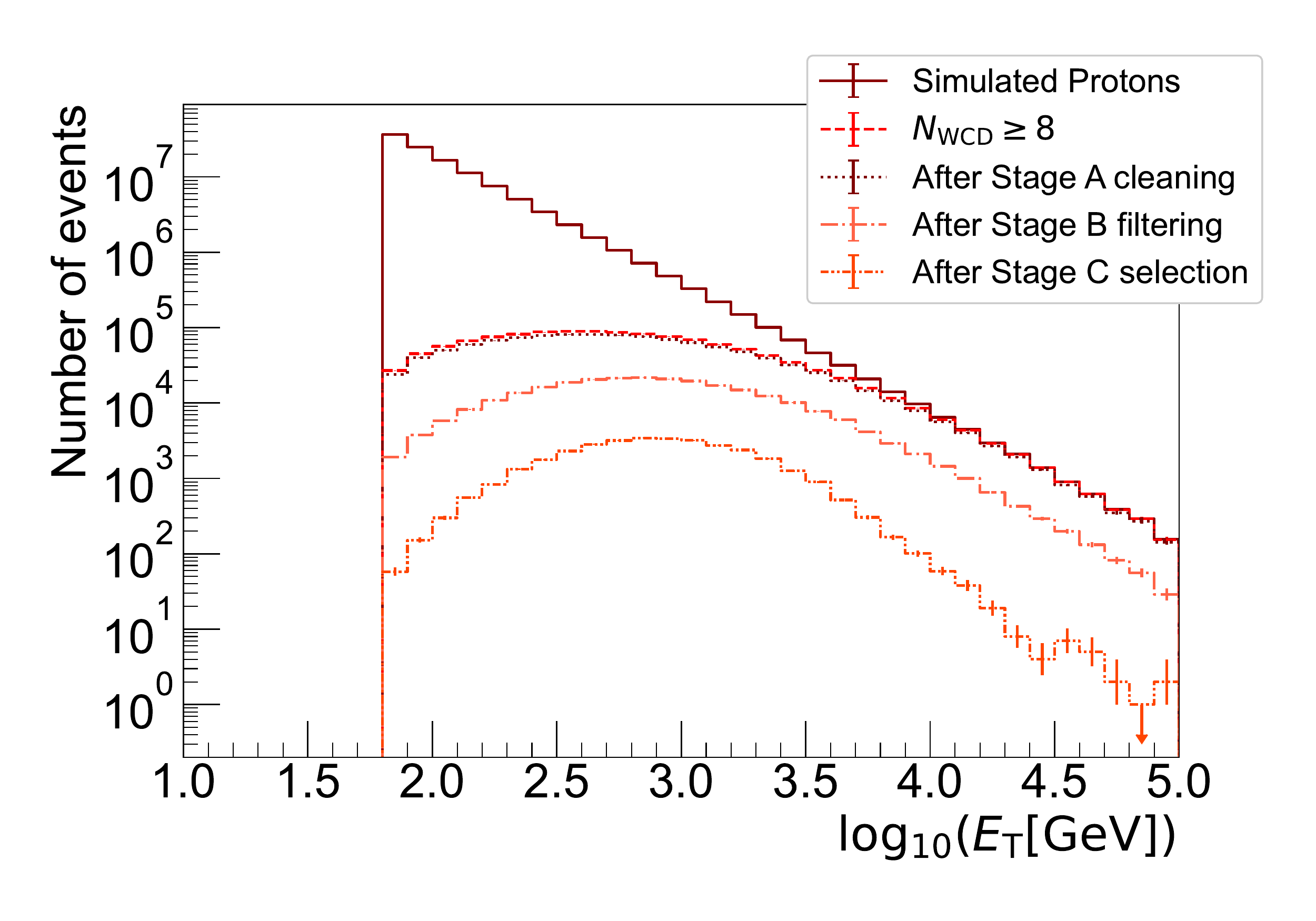}
     \hspace{.05
     cm}
     \includegraphics[width=0.48\textwidth]{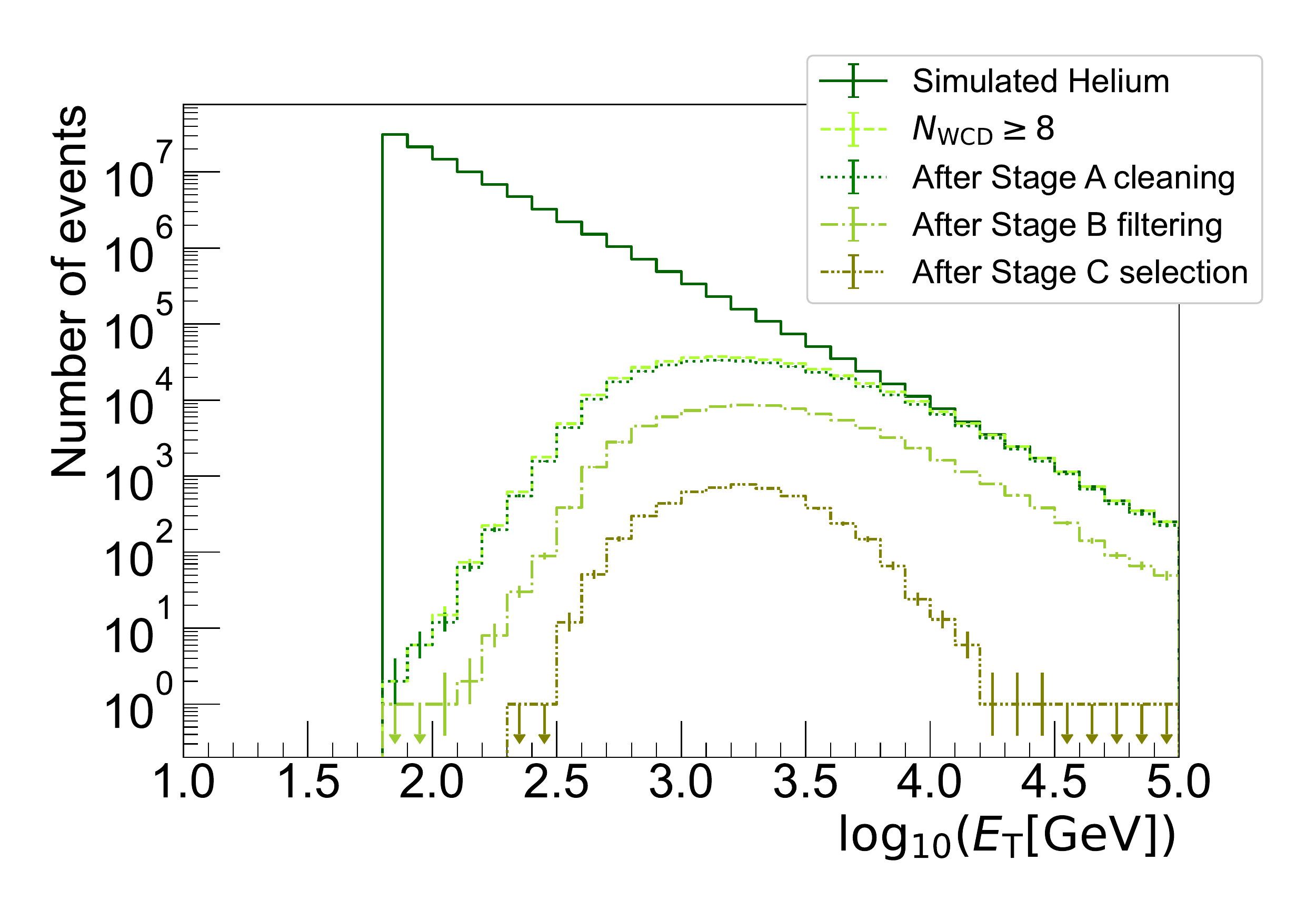}

    \caption{\small{\textbf{SEMLA performance on simulated $\gamma$-rays, protons and helium nuclei.} 
    \emph{Top panel:} True energy distribution of simulated $\gamma$-ray events after each SEMLA analysis stage.
    \emph{Bottom panels:} True energy distribution of simulated protons (\emph{Left}) and helium nuclei (\emph{Right}) after each SEMLA stage. 
    }
    \label{fig:Energy_distributions}}
\end{figure*}

The distributions in true energy of the simulated events after the successive SEMLA cut stages are shown in Figure \ref{fig:Energy_distributions} for gamma rays, protons, and helium.  
Table \ref{table:event_stat} shows the overall efficiency for the selection after the successive SEMLA cuts.  Given the live-times for the simulation derived in Table~\ref{table:simulation}, this can be used to determine the event rates for the background proton and helium cosmic rays.  
For the $\gamma$-ray events, as in {\semla} , we posit a ``pseudo-Crab'' source at $18^\circ$ from Zenith, with the same power law as in \cite{MAGIC_Crab}, but with no exponential cut-off, to have a bright, steady source with a spectral index close to that simulated for the $\gamma$ rays.

\begin{table*}[t]
    \small
    \centering
    \begin{tabular}{c|SSS|SSS}
        \hline

         & \multicolumn{3}{c}{\textbf{Efficiencies}} &         \multicolumn{3}{|c}{\textbf{Rates}} \\
         \cline{2-7} \textbf{Stages} & \textbf{$\gamma$-ray}  & \textbf{Proton} & \textbf{He nuclei} & \textbf{$\gamma$-rays} &  \multicolumn{1}{|c}{\textbf{Protons}} & \textbf{Helium} \\
         \cline{6-7} & & &  & $\mathbf{[min^{-1}]}$ & \multicolumn{2}{|c}{$\mathbf{[deg^{-2}min^{-1}]}$} \\
        \hline
            Stage A & \SI{95.6}{\%} & \SI{90.9}{\%}  & \SI{90.2}{\%} & 2.86 & 121.04 & 27.37 \tabularnewline 
            Stage B & \SI{32.9}{\%} & \SI{21.6}{\%} & \SI{21.7}{\%} & 0.98 & 28.77 & 6.57 \tabularnewline 
            Stages C--D & \SI{23.3}{\%} & \SI{2.7}{\%} & \SI{1.4}{\%} & 0.70 & 3.65 & 0.41 \tabularnewline 
        \hline
    \end{tabular}
    
    \caption{\small{\textbf{SEMLA efficiencies and indicative rates.} 
    Remaining fraction of simulated $\gamma$-rays, protons and helium nuclei relative to the events with $N_{\rm WCD}$ $\geq$ 8 after each SEMLA stage, before the application of the angular selection cut. 
    The $\gamma$-ray rates are estimated for the Pseudo-Crab source described in the text, prior to the angular selection cut.
    }}
    \label{table:event_stat}
    
\end{table*}

In this study, we obtain a proton rate alone of \SI{3.65}{{\!}/ \deg\squared \per\min}, slightly higher than the value of \SI{3.3}{{\!}/ \deg\squared \per\min} reported in {\semla}  for which a scaling factor had been applied on protons to account for all species in Cosmic Rays.
This higher background rate -- even without accounting for heavier species as below in section~\ref{sec:CR_background} -- is due to the lower trigger and reconstruction thresholds of the ALTO detector used in this work.  
This is somewhat compensated for by the higher rate for the pseudo-Crab $\gamma$-ray source of \SI{0.70}{\!/\min}, thanks to the lower threshold, compared to the \SI{0.56}{\!/\min} in {\semla} . 
For soft-spectrum sources such as the AGN and GRBs which are a major target for ALTO, this gain will be even greater.

\begin{figure}[h]
\centering
{\includegraphics[width=0.48\textwidth]{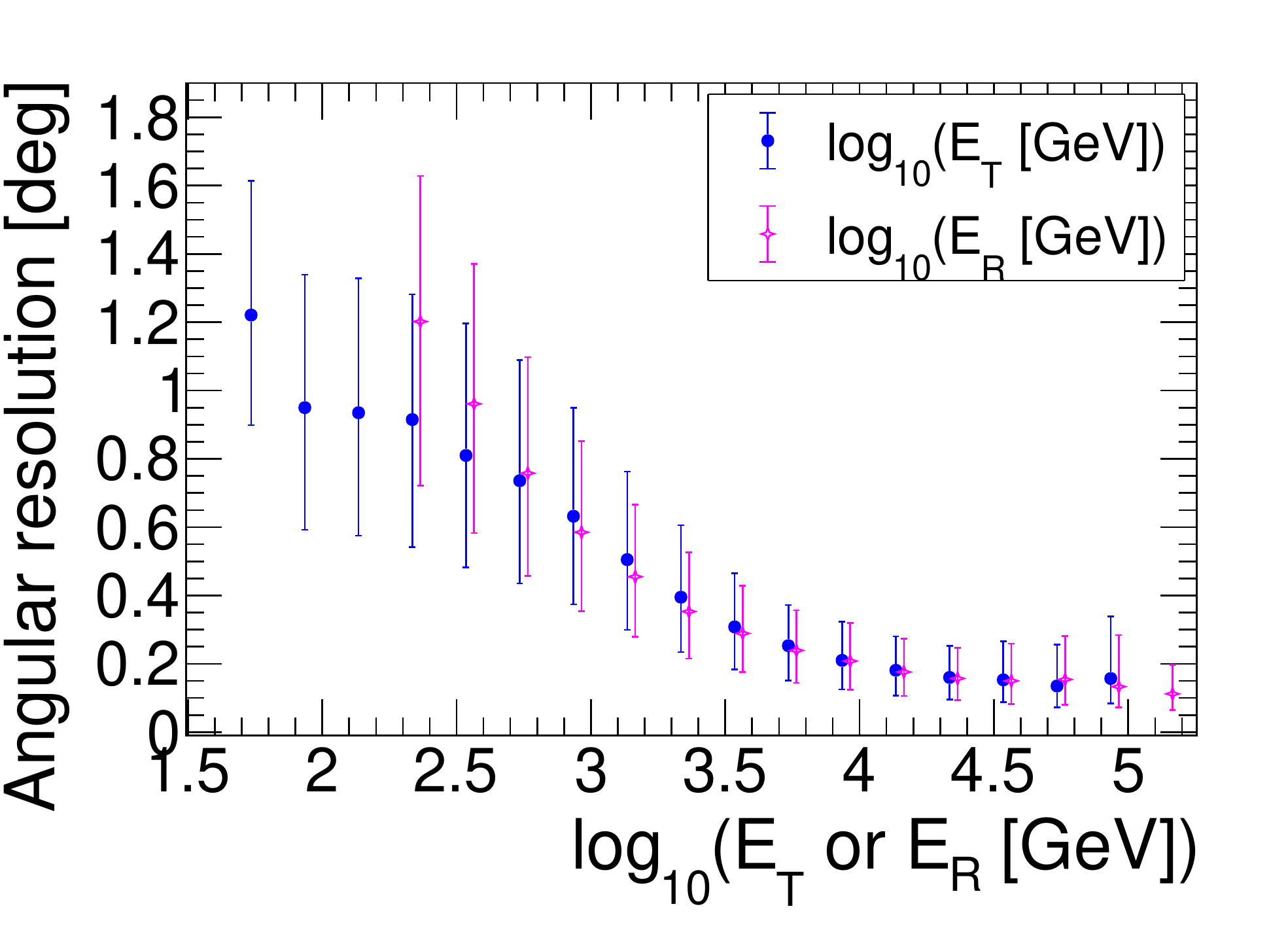}}
{\caption{\small{\textbf{Angular resolution of simulated $\gamma$-rays selected by SEMLA.} 
    The points represent the 68.3\% containment of the distribution of
    the angle between each event’s reconstructed shower direction and the true direction.
    The error bars indicate the 38.3\% to 86.6\% containment range. For clarity, the points in $E_{\rm T}$ and $E_{\rm R}$ are slightly shifted left and right respectively in energy
    }}
    \label{fig:StageC_selected_angular_resolution}}
\end{figure}

Figure~\ref{fig:StageC_selected_angular_resolution} presents the angular resolution (68.3\% containment) for the $\gamma$-rays, shown as a function of both true energy $\log_{10}(E_{\rm T})$ and reconstructed energy $\log_{10}(E_{\rm R})$.  This angular resolution as a function of reconstructed energy is then used (Sections~\ref{sec:ana_perf}, \ref{sec:sensi}) as an angular cut on the point-source $\gamma$-rays, decreasing from $\sim 1.2^\circ$ near the ALTO threshold to below $0.2^\circ$ at the highest energies, corresponding to solid angles of \SI{\sim4}{{\!}\deg\squared} down to \SI{0.1}{{\!}\deg\squared}.

\begin{figure*}[t] 
    \centering
    \includegraphics[width=0.475\textwidth]{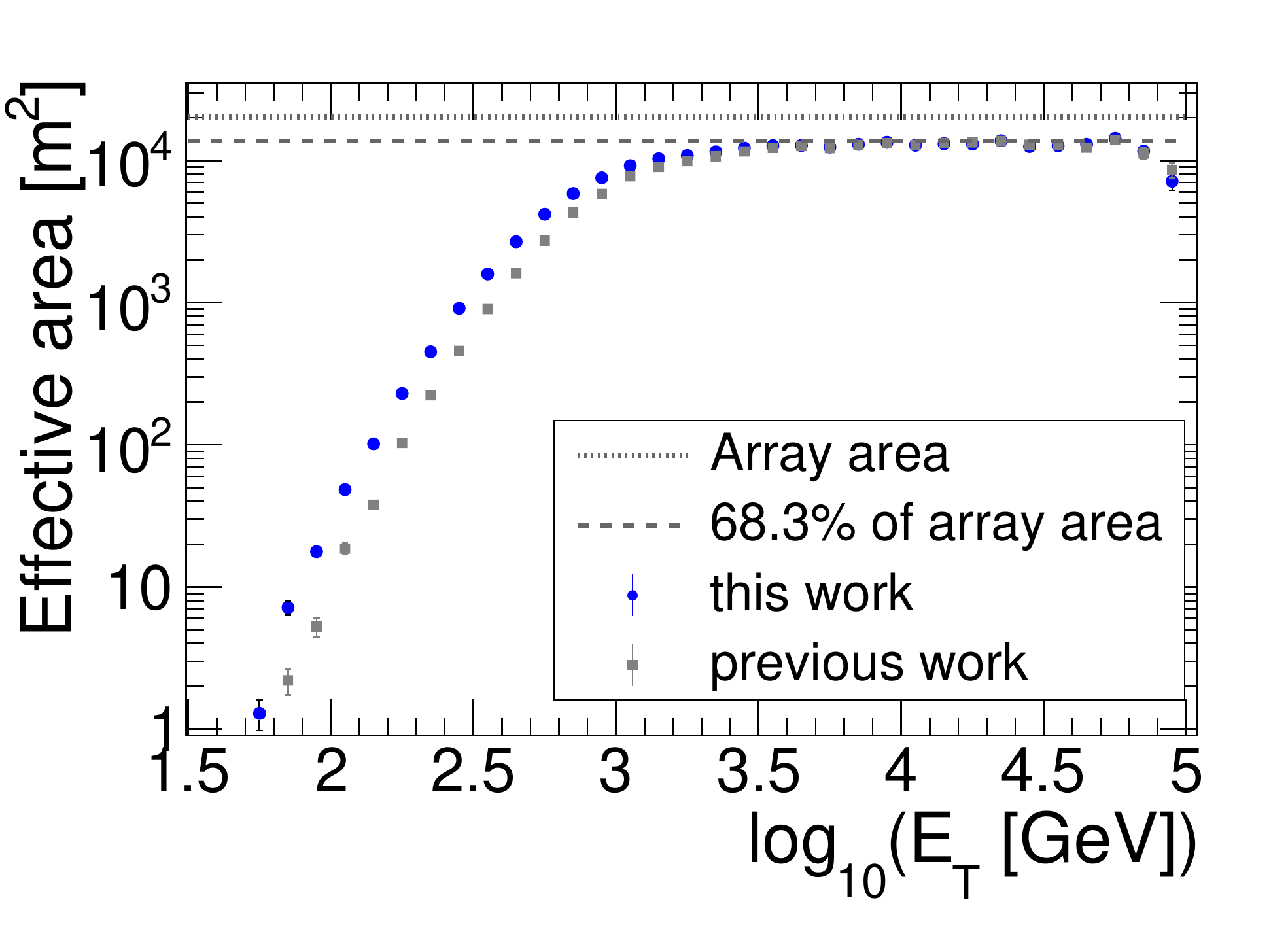}
    \hspace{0.4cm}%
    \includegraphics[width=0.475\textwidth]{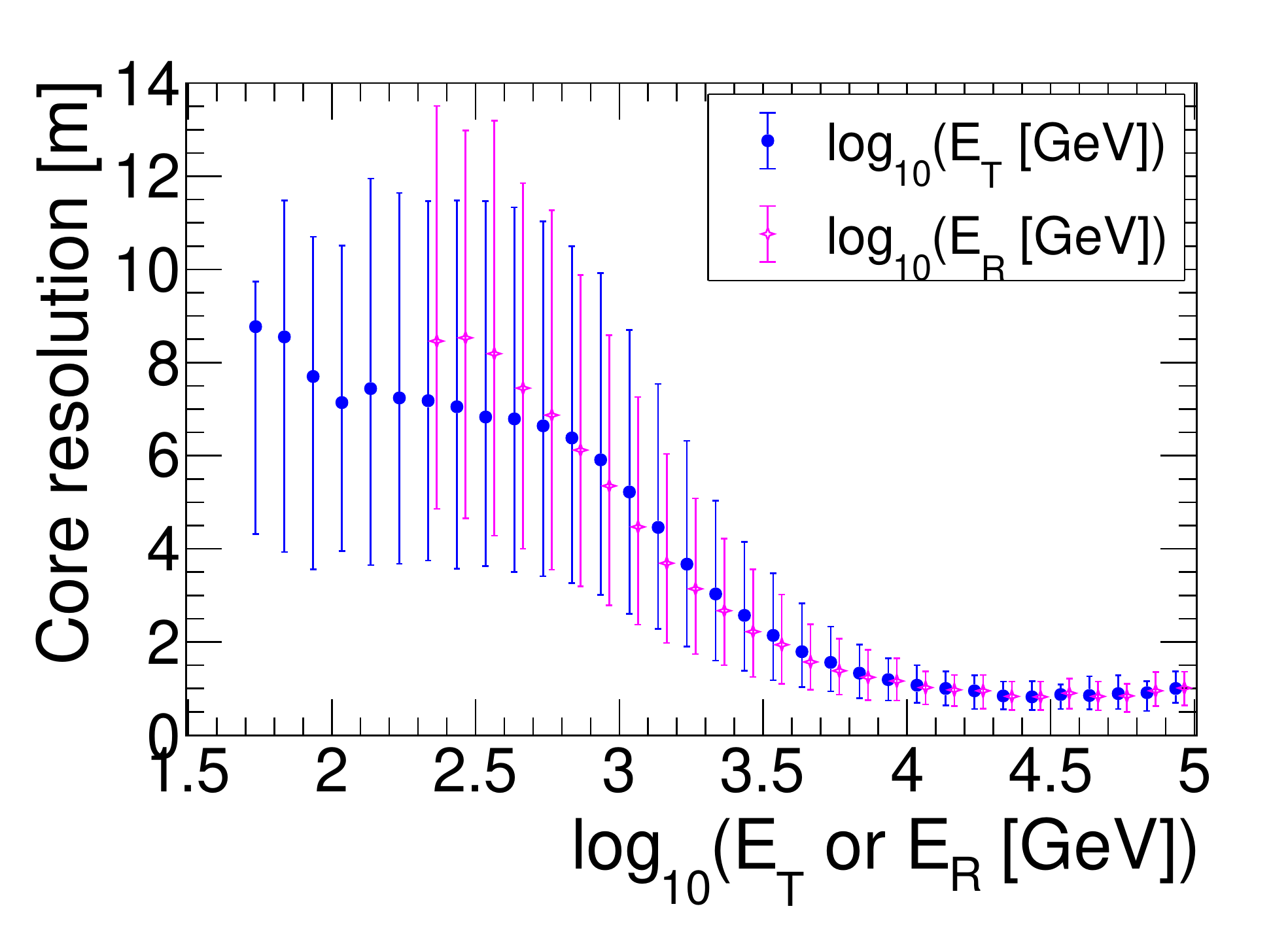}
    
    \includegraphics[width=0.475\textwidth]{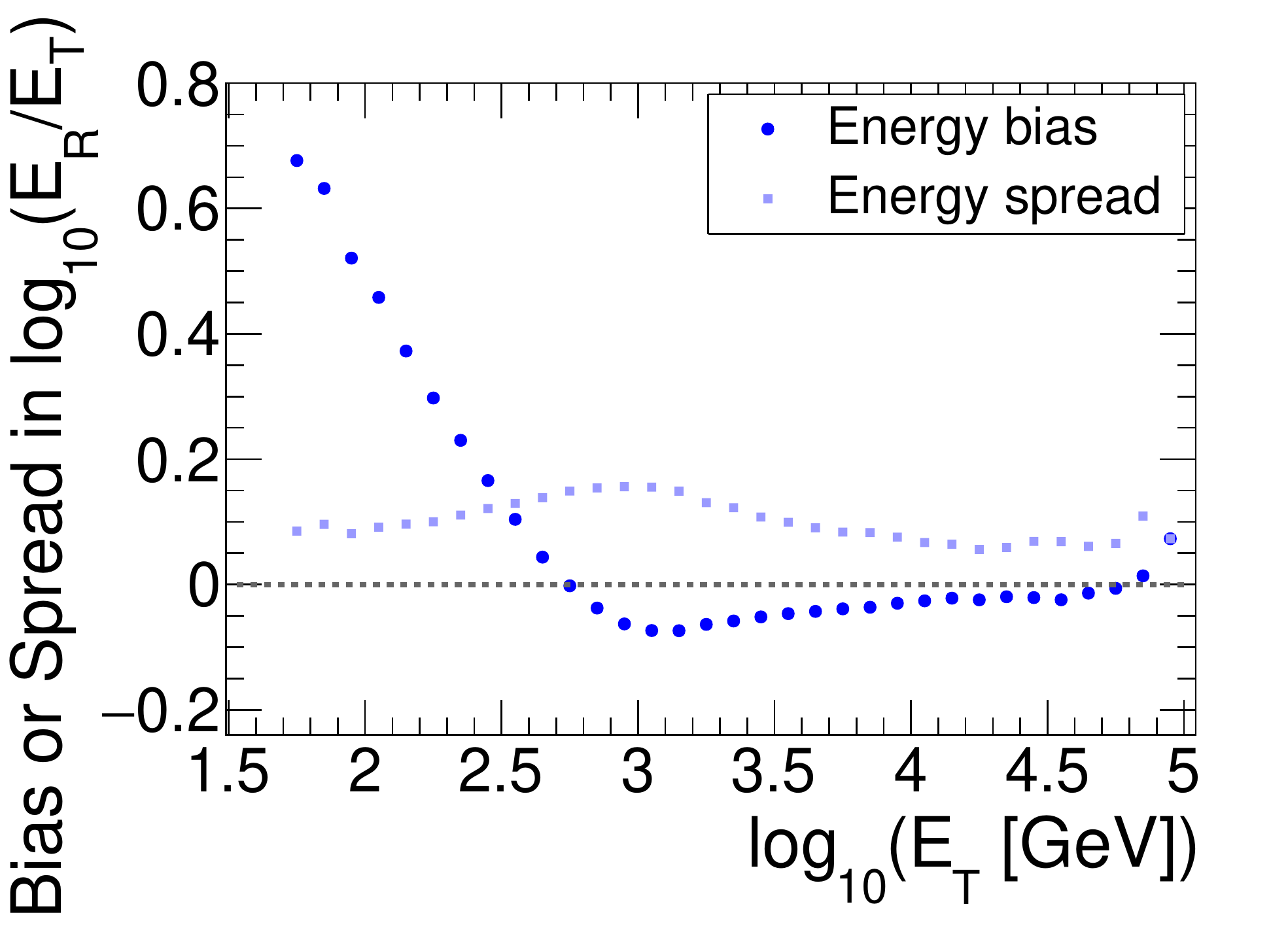}
    \hspace{0.4cm}%
    \includegraphics[width=0.475\textwidth]{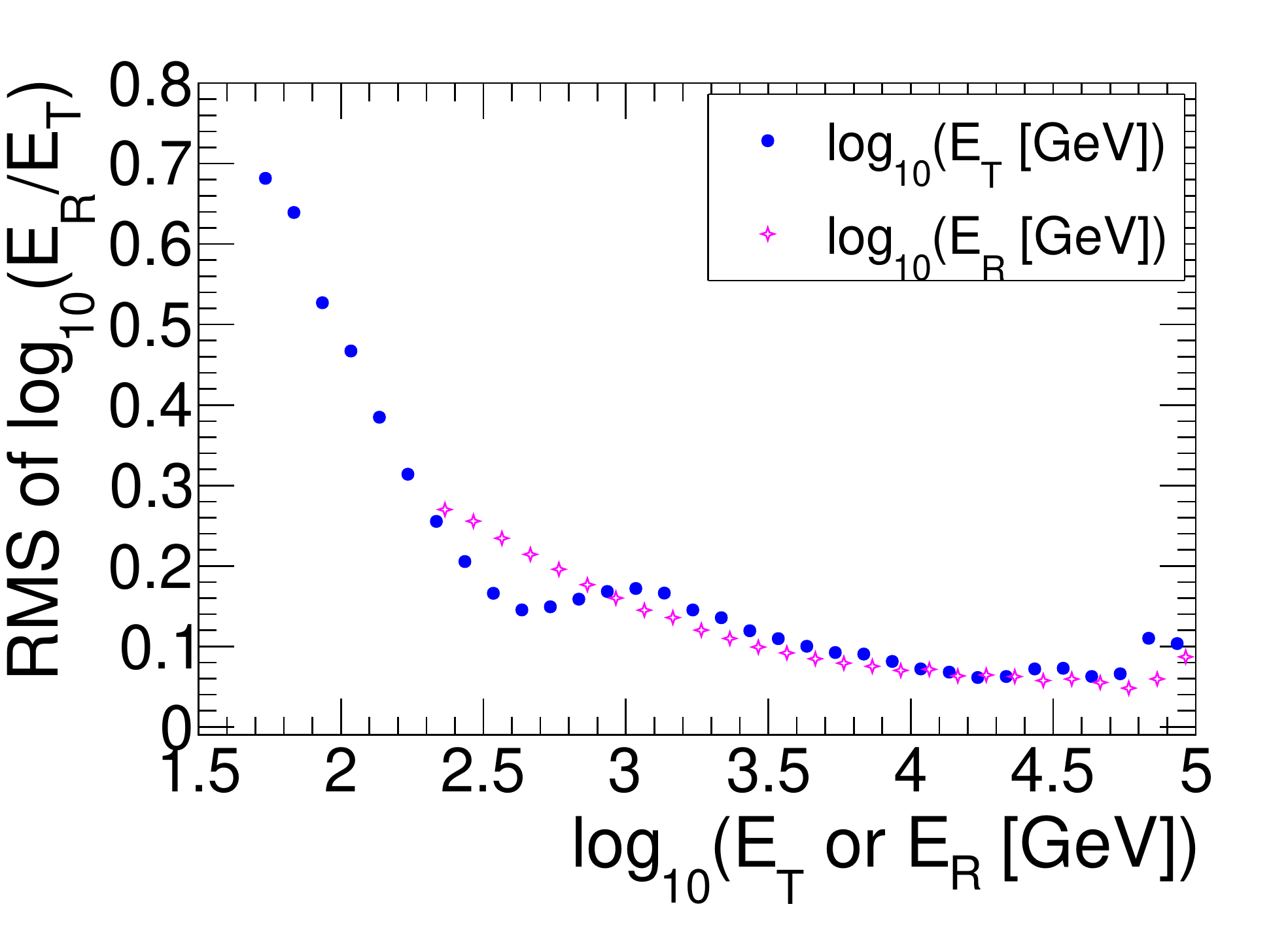}
    
    {\caption{\small{\textbf{SEMLA performance on simulated $\gamma$-rays after the angular selection cut.} 
    \emph{Upper Panel:} The effective area in $\rm m^2$ (left) and the core resolution in metres (right). \emph{Lower Panel:} The energy bias and its spread (left), and the energy resolution (right) as a function of energy. In the right panels, the points in $E_{\rm T}$ and $E_{\rm R}$ are slightly shifted left and right respectively in energy for clarity.
    For details of the definitions of Bias, Spread, and RMS, see Section~8 of the previous work {\semla} .
    }}
    \label{fig:semla_perf_plots_angcut}}
    
\end{figure*}

In Figure~\ref{fig:semla_perf_plots_angcut}, we show the performance of the ALTO detector optimized for soft-spectrum VHE sources. Comparing with Figure~14 from {\semla} , we note that the performance is practically identical, except for the effective area for $\gamma$-rays, where the previous performance is shown for comparison on the figure.  The gain in effective area (or equivalently a lowering of energy by $\sim20$\% for the same effective area at sub-TeV energies) is the consequence of the lowering of the trigger and reconstruction thresholds.

\section{Generation of Instrument Response Functions}
\label{sec:ana_perf}

\subsection{Generation of ALTO Instrument Response Functions after SEMLA}
\label{sec:IRFs}

Instrument Response Functions (IRFs) were generated from the above Monte Carlo simulations using the \texttt{PyIRF v0.6.0} package being developed for CTA \citep{pyirf_060}.
The relevant IRFs are produced after all SEMLA selection cuts, as defined in {\semla} , are applied. 

\subsection{ALTO response to point-like gamma-ray sources (IRFs)}
\label{sec:point-like}

Here, we consider only point-like $\gamma$-ray sources, so in addition to the SEMLA cuts we apply a selection on the reconstructed $\gamma$-ray direction to generate ``point-like'' IRFs. 
This angular cut is the 68.3\% containment radius for the $\gamma$-ray events, and is defined as a function of reconstructed energy $E_{\rm R}$.  
Since we estimate the sensitivity in uniform bins in $\log_{10}(E_{\rm R})$ (five per decade), the angular cut is defined in the same bins, as shown in Figure \ref{fig:StageC_selected_angular_resolution}.

For the gamma rays, the effective area for detection and the energy dispersion matrix are generated with 100 bins per decade of true energy, $\log_{10}(E_{\rm T})$, and 5 bins per decade of reconstructed energy, $\log_{10}(E_{\rm R})$, the latter to match the energy bins used for the sensitivity estimation.  

\subsubsection{Response to a Crab-like Gamma-ray source}
\label{sec:Pseudo_Crab}

The pseudo-Crab source described above would be detected by ALTO with a rate of $0.70$~$\gamma$-rays per minute after all SEMLA cuts and prior to the angular resolution cut, or $0.48$~$\gamma$-rays per minute including the angular selection cut.  The ALTO response to the pseudo-Crab source after SEMLA prior to the angular selection cut as a function of reconstructed energy is shown in Figure~\ref{fig:ALTO_SEMLA_pCrab_backgrounds} \emph{(left)}, 
which can be compared to the response to the proton and helium background on the right panel, discussed below.

\subsection{ALTO response to background}

For the background estimate, we define a $\pm 3^\circ$  band in zenith angles encompassing the source (between 15--$21^\circ$), over all azimuths.  
The background rate across this band being flat at a few \% level, we use the rate averaged over the band to estimate the background event rate, in 5 bins per decade for $\log_{10}(E_{\rm R})$.

\subsubsection{Inclusion of the Cosmic-ray background beyond protons}
\label{sec:CR_background}

For our preceding paper in {\semla}, due to computing-power constraints, we had generated only proton Monte Carlo simulations for the background, with these being scaled to the all-particle cosmic-ray flux to give estimates of the background cosmic-ray rate for ALTO.  This was a conservative approach, as we expected the heavier nuclei to be more easily rejected than protons.  
Note that in the approach initially followed by CTA \citep{CTA_MC_KB_2013} which simulated multiple species of cosmic rays, 
it was found that the heavier nuclei are more easily rejected by Imaging Atmospheric Cherenkov Telescope (IACT) arrays.  Thus, in more recent CTA work, only protons are considered, unscaled to the all-particle flux.  
For particle detector arrays, however, the effect of helium and heavier cosmic-ray backgrounds are generally taken into account based on the most recent abundance measurements (e.g., for HAWC, \citealt{HAWC:GRB_2012, HAWC:DM_2014}, and LHAASO, \citealt{LHAASO:Performance_crab}).
However, the differences in response to the different species are generally not detailed.

For this paper, we have now additionally simulated the helium cosmic-ray background in order to test the assumption that heavier species beyond protons are more easily rejected, even if not as unequivocally as for IACTs.  This also provides a more accurate estimate of the background rate than in our previous paper, needing fewer assumptions.  
 
\begin{figure*}[t]
\centering
{\includegraphics[width=0.485\textwidth]{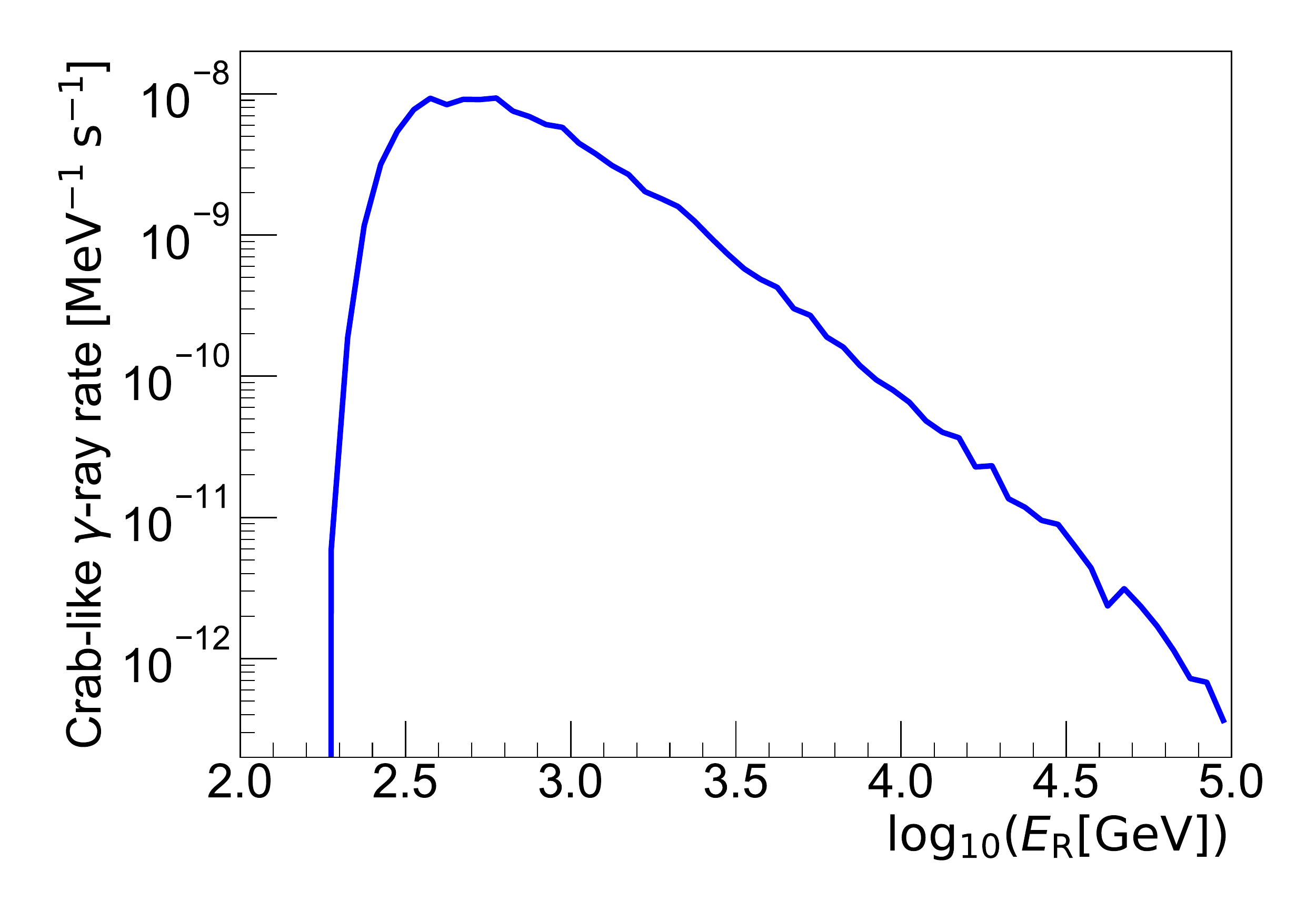}}
{\includegraphics[width=0.485\textwidth]{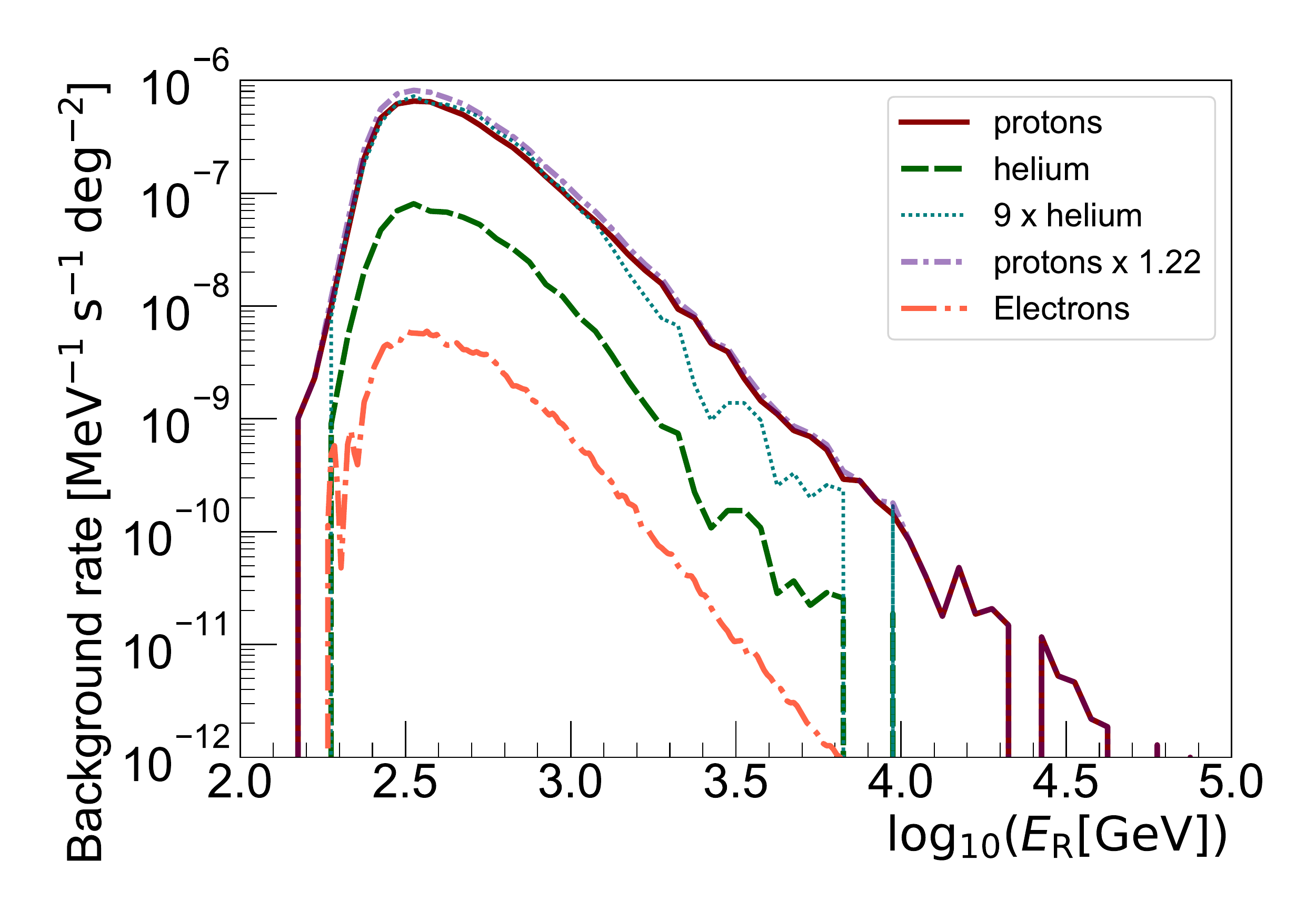}}
{\caption{\small{\textbf{Response to Pseudo-Crab, and to Background from proton, helium and electron fluxes.}
    {\emph{Left panel:}}
    Gamma-ray rate expected from a Crab-like source for the ALTO detector as a function of reconstructed energy. 
    {\emph{Right panel:}}
    Background rates for the ALTO detector, for simulated protons (solid dark-red line), helium (dashed dark-green line), and electrons (dot-dash red line), where the electrons response is inferred from the $\gamma$-ray simulations.  We show also the scaling $\times9$ of helium (dotted teal line) matching the protons over much of the energy range, and the effect of scaling the protons $\times 1.22$ (dot-dashed indigo line) to match the Cosmic Ray rate.  The gaps in the curves at the highest energies are due to low statistics.}}
    \label{fig:ALTO_SEMLA_pCrab_backgrounds}}
\end{figure*}

These background rates, Figure~\ref{fig:ALTO_SEMLA_pCrab_backgrounds}\emph{(right)}, show clearly that in reconstructed energy, after SEMLA cuts, the proton background rate highly dominates over the helium rate, despite the fact that before cuts the helium spectrum is only $\sim25$\% lower than the proton spectrum given in \cite{CTA_MC_KB_2013} (see Table~3).  
We find the ratio of the integral rate of protons over helium to be $\sim 9$, as shown in the figure, 
so taking 0.11 times the proton rate could be used as a proxy for the helium rate.
We also noted that the sum of the fluxes of the species heavier than helium given in  \cite{CTA_MC_KB_2013}, is closely comparable with the helium flux itself, such that taking the proton flux plus twice the helium flux happens to reproduce the all-particle Cosmic-ray flux to within $\pm2$\% in the energy range from \SI{100}{\GeV} to \SI{100}{\TeV}.  

Therefore, we can estimate the background rate taking the ALTO response to Cosmic Rays to be at the level of the response to protons plus twice that to helium.  Given the intrinsic imprecisions in the simulation of the detector response, which can only be fully explored after the construction of a full detector, this approximation is justifiable.
In fact, given that the helium rate is close to 0.11 times the proton rate, these could be combined to give an approximation of the background being 1.22 times the basic proton rate.  Both of these scenarios are considered in the estimation of the sensitivity.

Finally, adding the ALTO detector response to proton showers and the showers from helium nuclei and doubling the latter as a proxy for the responses to heavier species gives the estimate of the background rate, both overall 
(\SI{4.48}{\! \per \deg \squared \per \min})
and as a function of reconstructed energy.  
This summed estimate (i.e., protons plus twice the helium rates) for the background as a function of reconstructed energy, $\log_{10}(E_{\rm R})$, is used in the following section to estimate the sensitivity of ALTO and its response to several example point-like $\gamma$-ray sources.

The question of the contribution of the electron CR background may be posed, especially since electron-initiated showers are practically indistinguishable from those initiated by $\gamma$-rays.  
The electron flux is a tenth of a percent of that of protons at \SI{100}{GeV}, and decreases rapidly with increasing energy, but the collection area for these is designed to be maximal -- as required for the $\gamma$-rays -- whereas the collection area for protons is minimized by their rejection factor after SEMLA cuts.
Given the IRFs of the $\gamma$-rays, and assuming that the electron collection area is the same as that for the $\gamma$-rays, we can evaluate the ALTO response to electrons (again, as given by \citealt{CTA_MC_KB_2013}), simply by multiplying the electron flux by the $\gamma$-ray collection area (corrected to be that prior to angular selection cuts) and convolving with the energy dispersion.  This gives the response in reconstructed energy, $\log_{10}(E_{\rm R})$, shown in Figure~\ref{fig:ALTO_SEMLA_pCrab_backgrounds}\emph{(Right)}.  It can be seen that the irreducible background rate from electrons is still less than 1\% of the proton rate, both after the SEMLA cuts, and so may be neglected.

Furthermore, from the angular cut defined above in Section~\ref{sec:point-like}, also given in bins in $\log_{10}(E_{\rm R})$, we estimate the background rate in each energy bin in the region around the point-source, the sum of which over all energies is found to be \SI{8.28}{\per \minute}.

We conclude that contribution of helium, while not negligible for particle detector arrays, is greatly suppressed with respect to that of the protons for the same reconstructed energy, and so we can make a more accurate estimate of the background rate --- and therefore of the ALTO sensitivity and response to sources than in our previous work.

\section{Sensitivity and Spectral Response calculation}
\label{sec:sensi}
\subsection{Point-source Sensitivity}

For $\gamma$-ray experiments, the sensitivity is generally quoted as a differential flux level in bins in $\log_{10}(E_{\rm reco})$, generally with 5 bins per decade. 
This flux level is required to fulfil several criteria, as follows.  
The flux level must result in  a sufficiently large number of $\gamma$-rays, $N_\gamma$, against the background rate, $N_{\rm BG}$, to give a $5\sigma$ significance in that bin.  
The flux must give a minimum $N_\gamma$, where for ALTO we require $\geq 10$, and we also require $N_\gamma > $ 1\% of the background estimation. 
The latter is much less than commonly used for IACTs, 
since for WCDs there are fewer perturbative effects, for example, there is no night-sky light or stars.

For IACT experiments, an observation time of \SI{50}{\hour} is usually used as a basis,
since this is a reasonable time within a calendar year for pointed observation of a source during Moon-free night-time with good weather conditions.  These observing constraints do not apply to particle detector arrays such as WCDs.
For ALTO, taking the sensitivity and threshold to be reasonably flat up to $30^\circ$ from Zenith,
a source in the declination range for observation 
would be observable for approximately %
\SI{4}{h} per transit (day or night, independent of Moonlight), or \SI{1461}{h} per year.
This range may be extended to higher zenith angles, but with the penalty of a higher threshold which must be taken into account with extensive Monte Carlo simulations which are beyond the scope of this paper.

We use the \texttt{Gammapy 1.0} package \citep{gammapy:2017,gammapy:1.0} for the estimation of the differential sensitivity of ALTO, and for the response to typical known sources, assuming these transit above the ALTO site.  This has been cross-checked with custom codes developed for the previous paper, with good agreement.

For the background estimation, we assume that when observing with a full array the rate can be estimated from circular areas at the same distance from Zenith but different azimuths.  
Given the low background rates after SEMLA cuts -- especially at the highest energies -- we take the background \emph{OFF}-source region to consist of 12 such areas with a radius of \SI{3}{\deg} (so \SI{\sim0.1}{\steradian} in total),  
out of 18 such areas possible in this zenith band. 

However, Gammapy does not yet allow such a background region as standard, being optimized for pointing instruments with small fields-of-view.  
So, we calculate the background counts in the \emph{OFF}-region directly from the background rate IRF and the supposed observation time, and provide this to Gammapy.  Together with \emph{ON} and \emph{OFF} acceptances (which are just the corresponding solid angles), this allows us to use Gammapy to compute the sensitivity, i.e., the flux required in each energy bin to fulfil the criteria listed above.

\begin{figure*}[t]
\centering
{\includegraphics[width=0.49\textwidth]{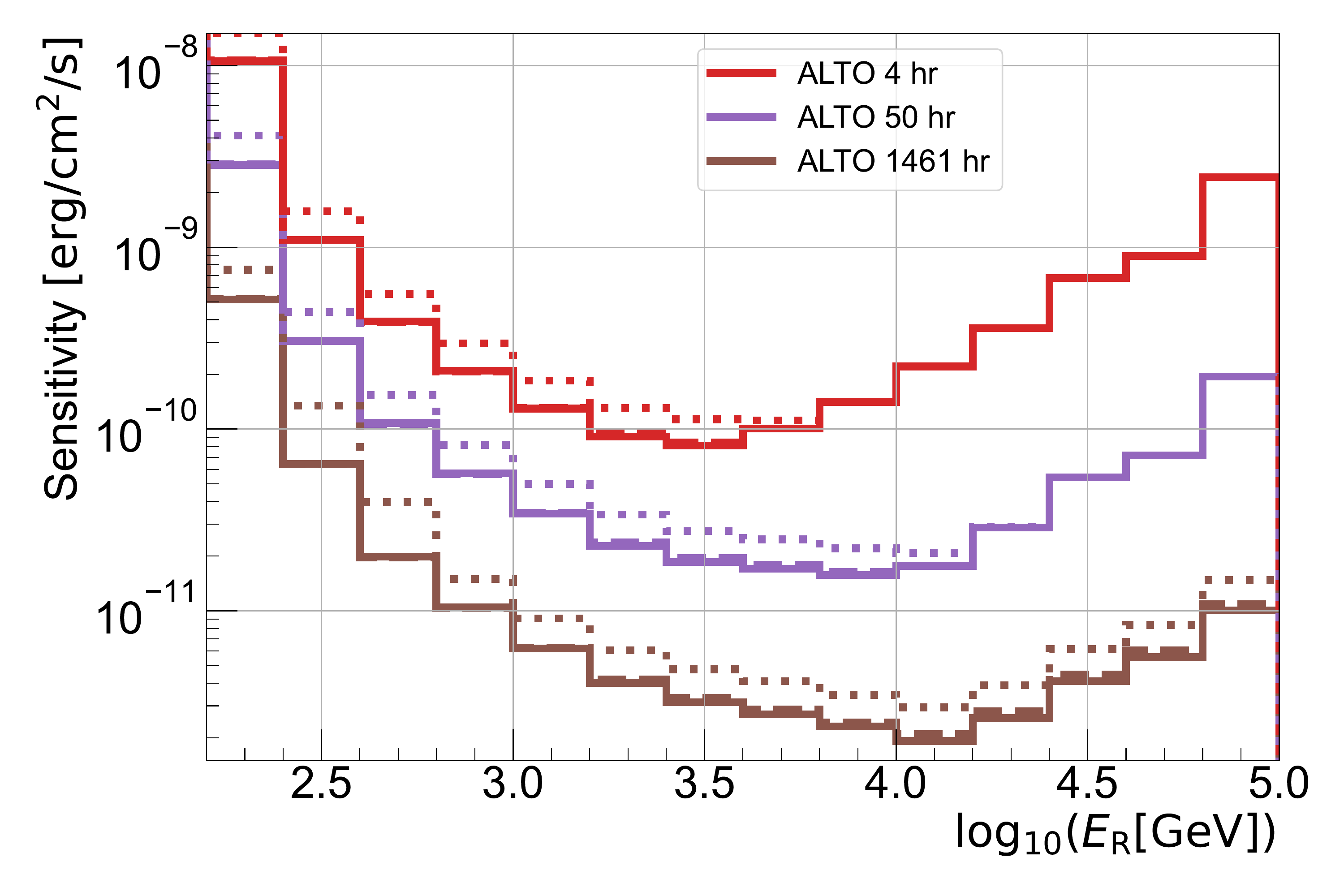}}
\hspace{.05
cm}
{\includegraphics[width=0.49\textwidth]{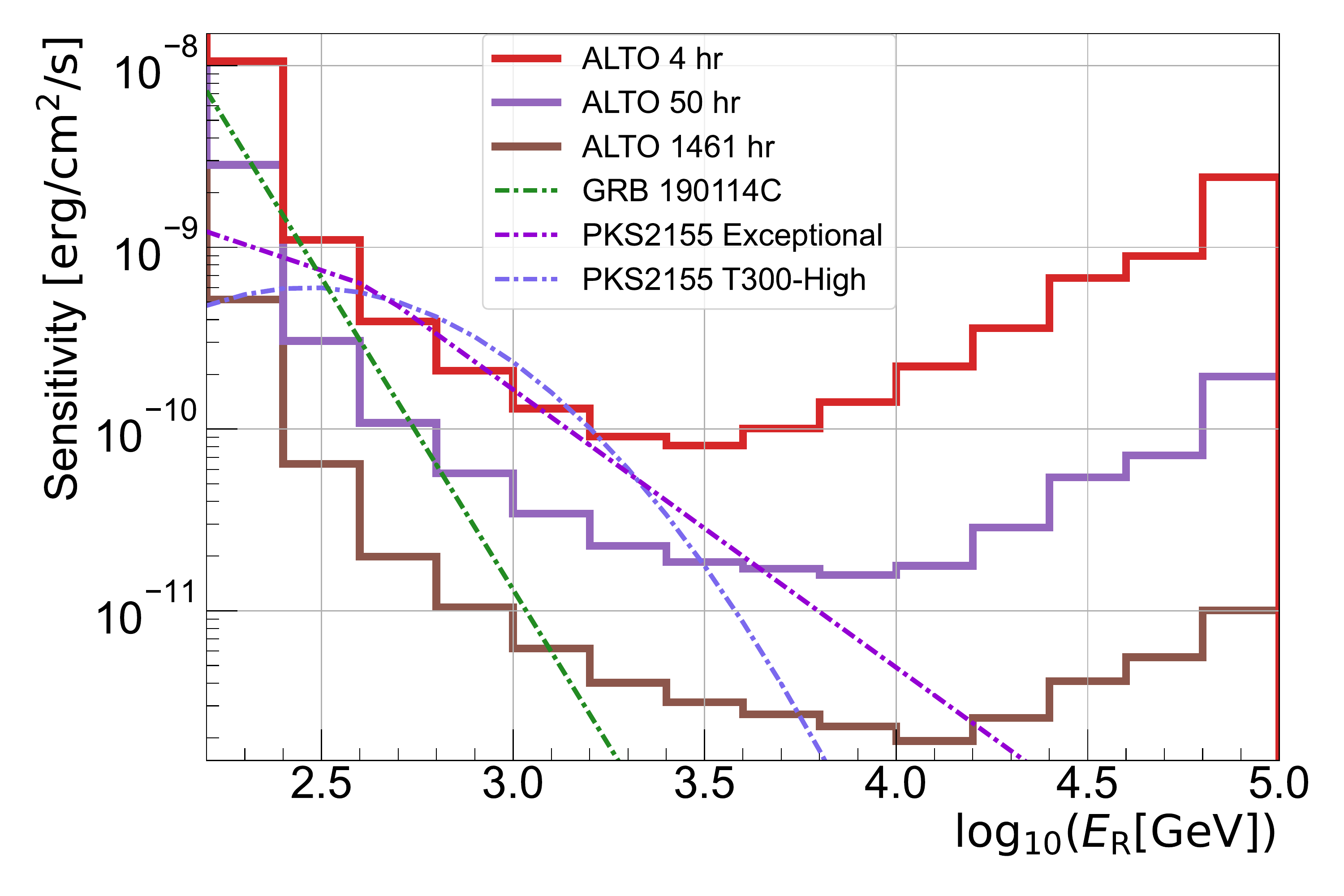}}
\hspace{.05
cm}
\\
{\includegraphics[width=0.49\textwidth]{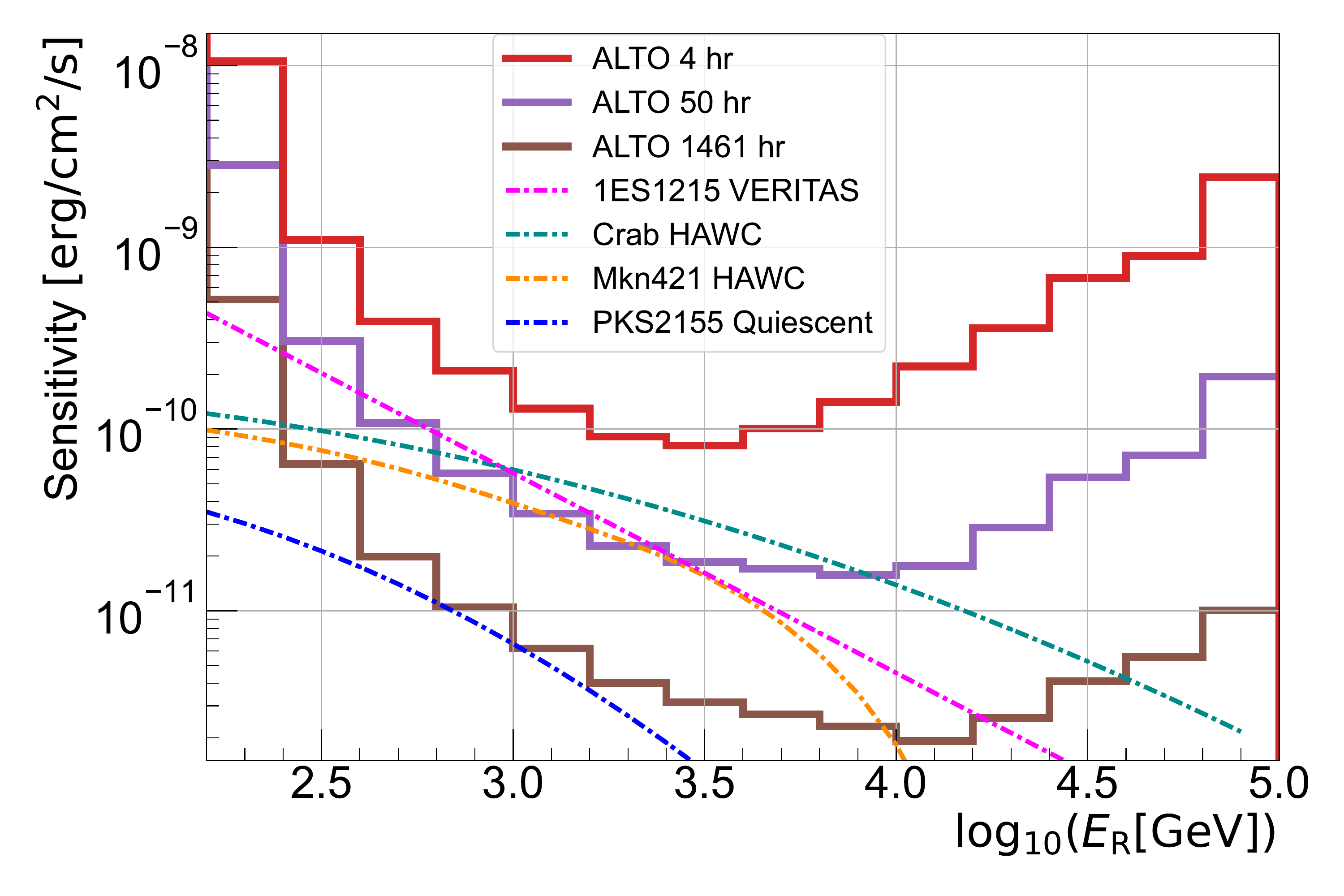}}
{\includegraphics[width=0.49\textwidth]{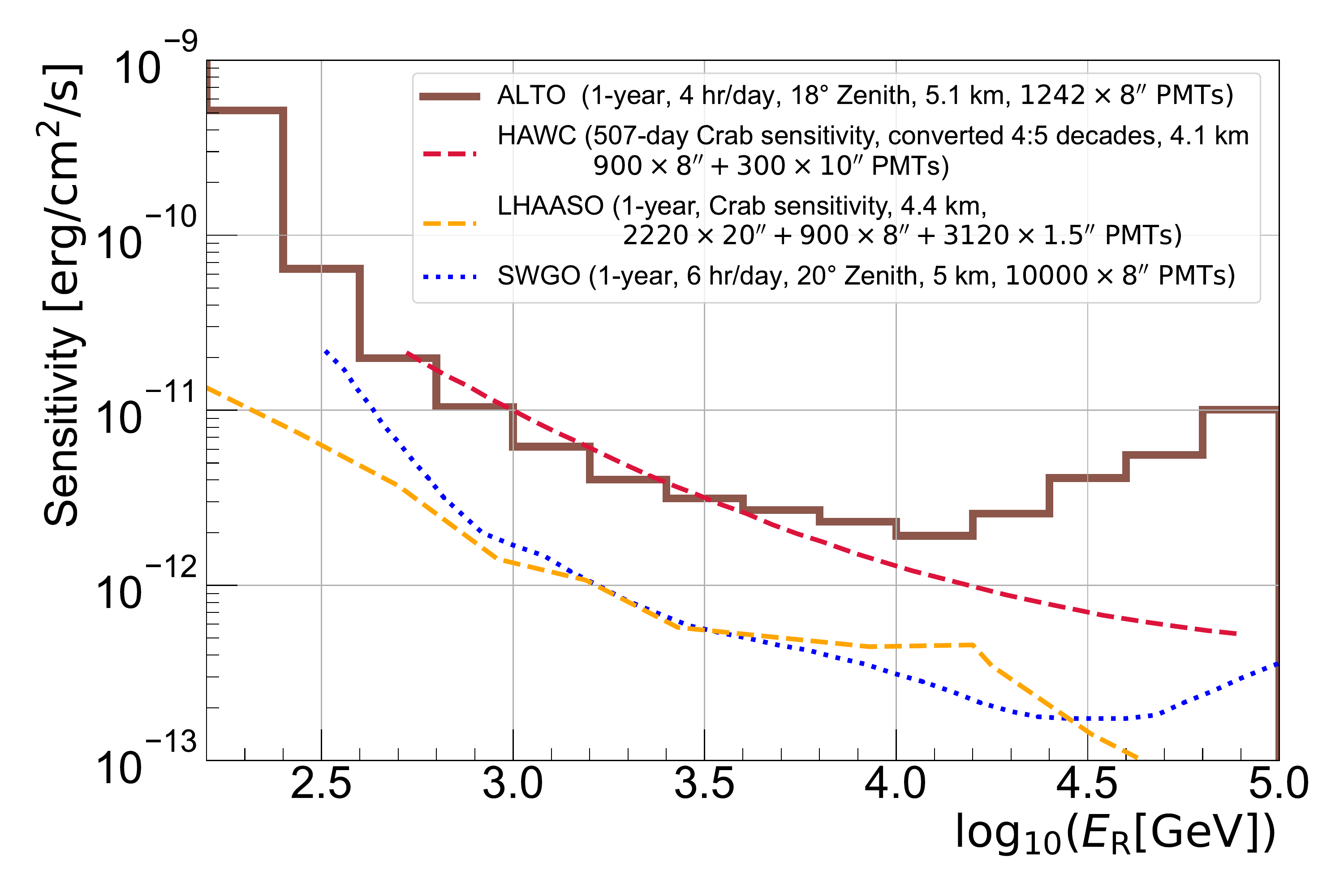}}
{\caption{\small{\textbf{Sensitivity to point-like $\gamma$-ray sources.} 
    Sensitivity for the ALTO detector, shown as step plots:  {\emph{Upper-Left}} for observation times for 4hr (one transit), 50hr, and 1461hr (for a calendar year of observation), with full, dashed, and dotted line steps are for different background estimations, as explained in the text; {\emph{Upper-Right}} and {\emph{Lower-Left}} the same, for the most realistic background alone, compared to the spectra of typical sources which would be targets for ALTO (dot-dashed curves given in the legend in order from the high to low flux in the lowest-energy bin), assuming they were observed continuously at \SI{18}{\degree} from Zenith, for soft and hard sources respectively; {\emph{Lower-Right}} comparison with some current and proposed particle detector array sensitivity curves, giving the assumptions under which these are produced (observation time in a calendar year, Zenith constraints, altitude, number of PMTs) which make such comparisons delicate.
    }}
    \label{fig:ALTO_SEMLA_sensitivity}}
\end{figure*}

The differential sensitivity of the full ALTO detector with the SEMLA analysis, in this estimation, is shown in Figure \ref{fig:ALTO_SEMLA_sensitivity}{\emph{(Upper-Left)}}.
Each step plot corresponds to a given observation time, from \SI{4}{\hour} (one transit) through to the \SI{1461}{\hour} which would be achieved in one calendar year of operation.
The solid-line steps correspond to the estimate of the Cosmic Ray background using the proton background combined with twice the helium background, while almost-indistinguishable dashed line is for the protons scaled by a factor 1.22, and finally the dotted line is the too-conservative proton background alone scaled to the cosmic-ray background (by a factor $\sim 2.5$), as assumed in our previous paper, {\semla} .
This therefore demonstrates, at least for ALTO, that using only proton simulations and scaling by the appropriate factor is equivalent to including scaled helium simulations, and likely equivalent to simulating the full range of cosmic-ray species. 
Figure \ref{fig:ALTO_SEMLA_sensitivity} also shows the ALTO differential sensitivity for one calendar year of operation superimposed with  the spectra from several sources of interest, described in the following section, for soft- and hard-spectrum sources respectively on the {\emph{Upper-Right}} and {\emph{Lower-Left}} plots.  In the next section we also estimate the ALTO spectral response to these sources of interest.

\subsection{Spectral response}

The spectral response is estimated similarly to the procedure for the point-source sensitivity.  From the IRFs, for a given supposed source $\gamma$-ray spectrum in true energy and a given observation time, the signal counts in each bin in reconstructed energy can be calculated, along with the background counts in the \emph{OFF}-source region estimated as above.
We then produce a number of realizations of a typical observation by including Poisson fluctuations in the \emph{ON}- and \emph{OFF}-source counts in each energy bin \footnote{Gammapy's \emph{`fake'} method is used for injecting these Poisson fluctuations.}, to give realistic simulations of binned spectral data.

From each set of realizations of the spectral response for a given source, we can find the average excess counts which would be measured in the {ON}-region 
(where this is an estimate of the $\gamma$-ray counts) 
and its error, compared to the background counts in the {ON}-region along with their error which is estimated from the much larger {OFF}-region.  This gives the overall average detection significance which would be achieved for a source if it were at \SI{18}{\degree} from Zenith.

In Table \ref{table:SensiSources}, we present the performance of ALTO to a number of notable examples of point-like sources, for which the details are given in Section~\ref{subsubsec:sources}.
For transient or outburst episodes, we choose the observation times considered as equal to the flare time actually observed, while for the steady or non-outburst cases we consider longer times needed to achieve a significant signal.  
The `Background' counts can be seen to scale with the observation time according to the \SI{\sim8.28}{\per \minute} rate within the angular cut given above.

The performance was checked on a larger number of sources, but for most of the brighter AGNs (besides Markarian~501) the observation times required for detection of the steady state were found to be several years.
For the other GRBs detected at VHE, even extrapolating back to the onset of the exponential-decay phase, the detectability on the GRB time-scales was not achieved, although there may be other GRBs which were missed by VHE instruments due to observing constraints to which ALTO would not be subject.  
Note that the significances shown in Table \ref{table:SensiSources} are pre-trials, whereas for a blind search for detection of a transient, trials would apply depending on the choice of detection time-window.  However, in the cases where a detection from another instrument or messenger provides a time-window, such trials would be minimized, while if ALTO is used to provide alerts to other observatories, then those observatories can choose which false-alarm-rate is acceptable.

\begin{table*}
    \centering
    \small
    \sisetup{separate-uncertainty,table-align-uncertainty}%
    \begin{tabular}{l|c|SccSSS}
    \hline
    \hline
         \textbf{Source} & \textbf{State}  & \textbf{Obs.} & \textbf{Excess} & \textbf{Background} & \textbf{Signif.} & \textbf{$\mathbf{5\sigma}$-time} \\
         & & \textbf{time (h)}  &  &  & \textbf{($\sigma$)} & \textbf{(h)} \\
         \hline
         \hline
         \multirow{2}{*}{Crab Nebula}  & HAWC & 3349 & $(92.8\pm1.4)\times10^3$ & $1664.1\times10^3$ & 71.1 & {\multirow{2}{*}{17.1}} \\
         & spectrum & 50 & $(1.39\pm0.17)\times10^3$ & $24.8\times10^3$ & 8.7 & \\
         \hline
         \multirow{2}{*}{PKS 2155-304} & Big Flare &  1.32 &  $169\pm31$ & 656  & 6.3 &  0.86 \\
         \multirow{2}{*}{$\,\,\,\,$ (H.E.S.S.)} & \texttt{T300-High} & 1.36 & $158\pm28$ & 676 & 5.8 & 0.86 \\
                      & Quiescent   & 1461 & $(5.87\pm0.91)\times10^3$ & $726\times10^3$ & 6.9 & 759 \\
         \hline
         Mkn 421 & HAWC 3yr & 50 & $940\pm168$ & 24845 & 5.9 & 38 \\
         \hline
         \multirow{2}{*}{1ES~1215$+$30} & VERITAS & 0.75 & $28\pm23$ & 373 & 1.44 & {\multirow{2}{*}{7.9}} \\
         & flare & 4 & $156\pm53$ & 1988 & 3.44 & \\
         \hline
         GRB 190114C & MAGIC & 0.66 & $219\pm25$ & 330 & 11.0 & 0.144  \\
         \hline
         \hline
    \end{tabular}
    \caption{\small{\textbf{Detection significances expected on certain point-like sources with the ALTO array.}  
    The sources are all supposed at \SI{18}{\degree} from Zenith.  
    For the flaring sources or episodes, the observation time corresponds to the burst duration; for the other cases, observations of $\sim$\SI{4}{\hour} per night around transit should be considered.  
    The confidence interval on the excess is the result of the statistical errors on signal and background, for the 100 realisations of each source's spectrum.  The significances are given pre-trials.
    See Section~\ref{subsubsec:sources} for spectral references.}}
    \label{table:SensiSources}
\end{table*}

In addition, given the excesses and background in each bin in reconstructed energy, spectral models can be fitted to each realization.
Given a number of realizations, the average spectrum which would be measured can be found and its parameters determined, and they are found to be in accordance with the injected model.  Examples of single realizations of fitted spectra are shown in Figure~\ref{fig:spectra_results} for the same sources, where the blue points and error bars correspond to the flux point estimation in true (Monte Carlo) energy.

\subsubsection{Sources considered}
\label{subsubsec:sources}

As example notable sources, we consider the brightest AGN and GRBs, but also the Crab nebula which is a standard candle for the domain of VHE astrophysics.  In all cases, we give the response as if the source were at \SI{18}{\degree} from Zenith for \SI{4}{\hour} per transit, which in actuality would only be the case for the AGN PKS~2155$-$304, with ALTO located in the Southern hemisphere.

Comparisons with source detections from existing detectors are only indicative, as they follow sources for longer transit times, while for ALTO we apply a cut-off of \SI{4}{\hour}, or \SI{30}{\degree} for a source at Zenith.  Also, for existing detectors, few of these sources transit at Zenith, in fact.  The estimations for an operational ALTO would be pessimistic in the former case,  while over-optimistic in the latter.

\begin{figure*}[b] 
    \centering
    \includegraphics[width=0.475\textwidth]{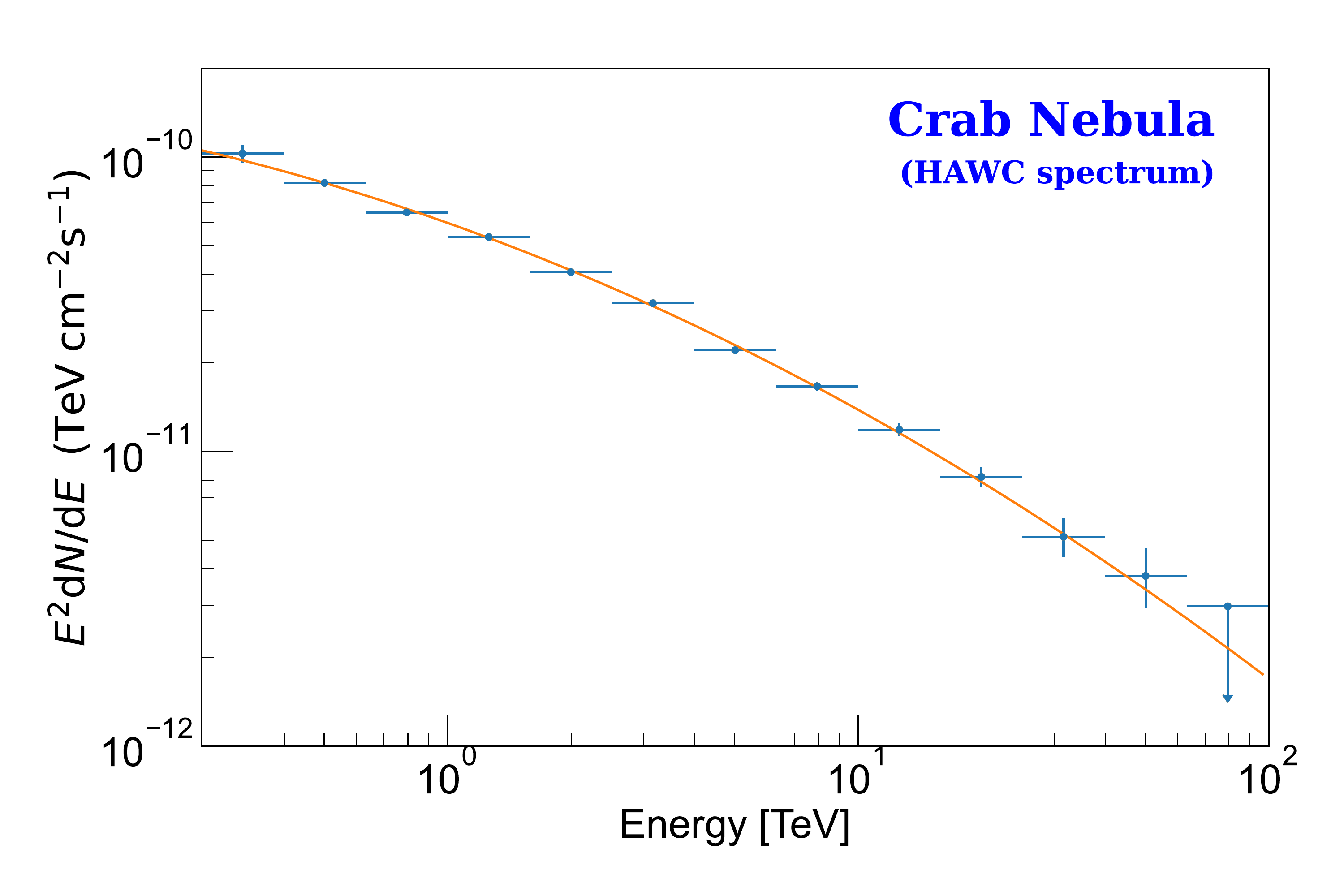}
    \hspace{0.3cm}
    \includegraphics[width=0.475\textwidth]{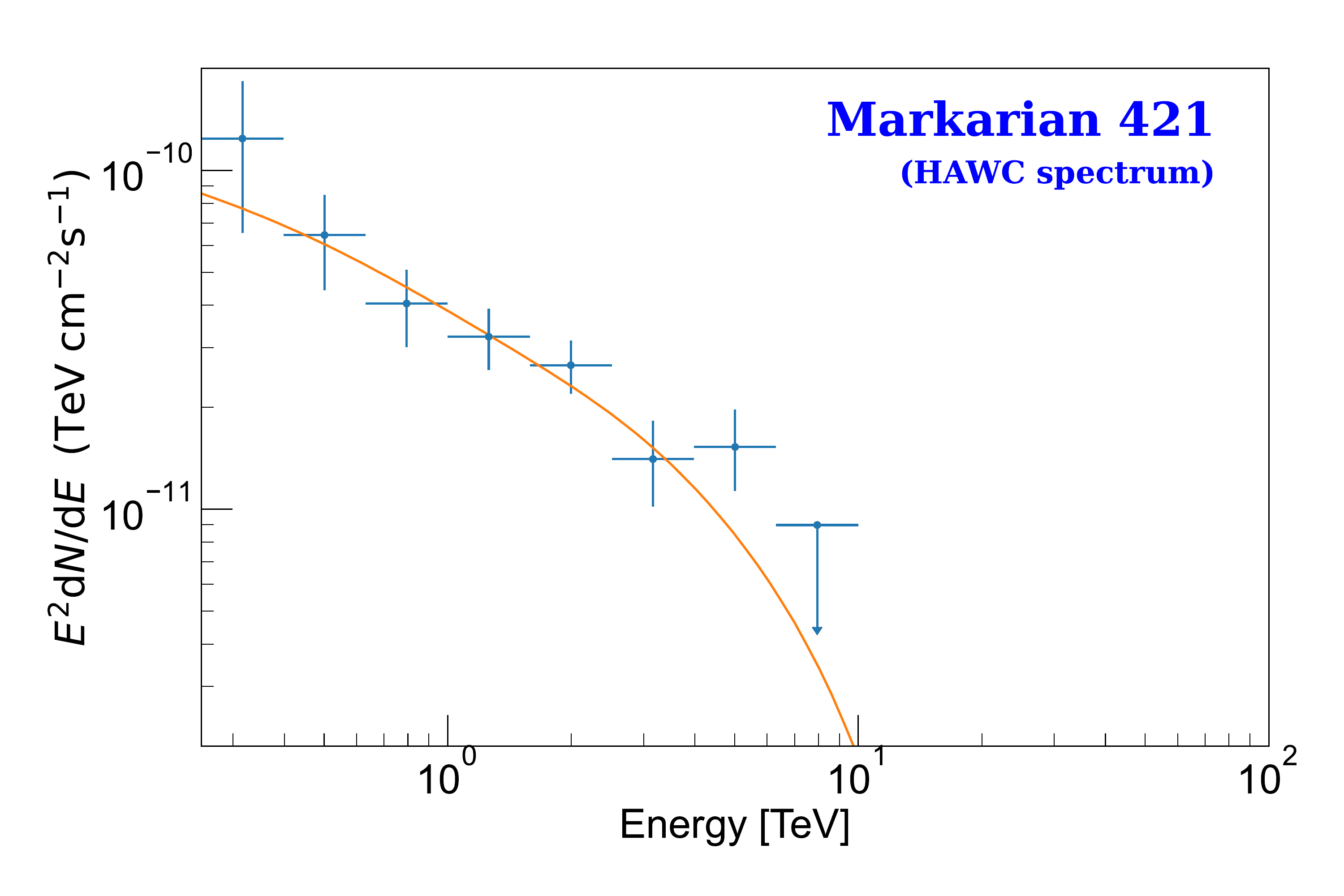}
    
    \includegraphics[width=0.475\textwidth]{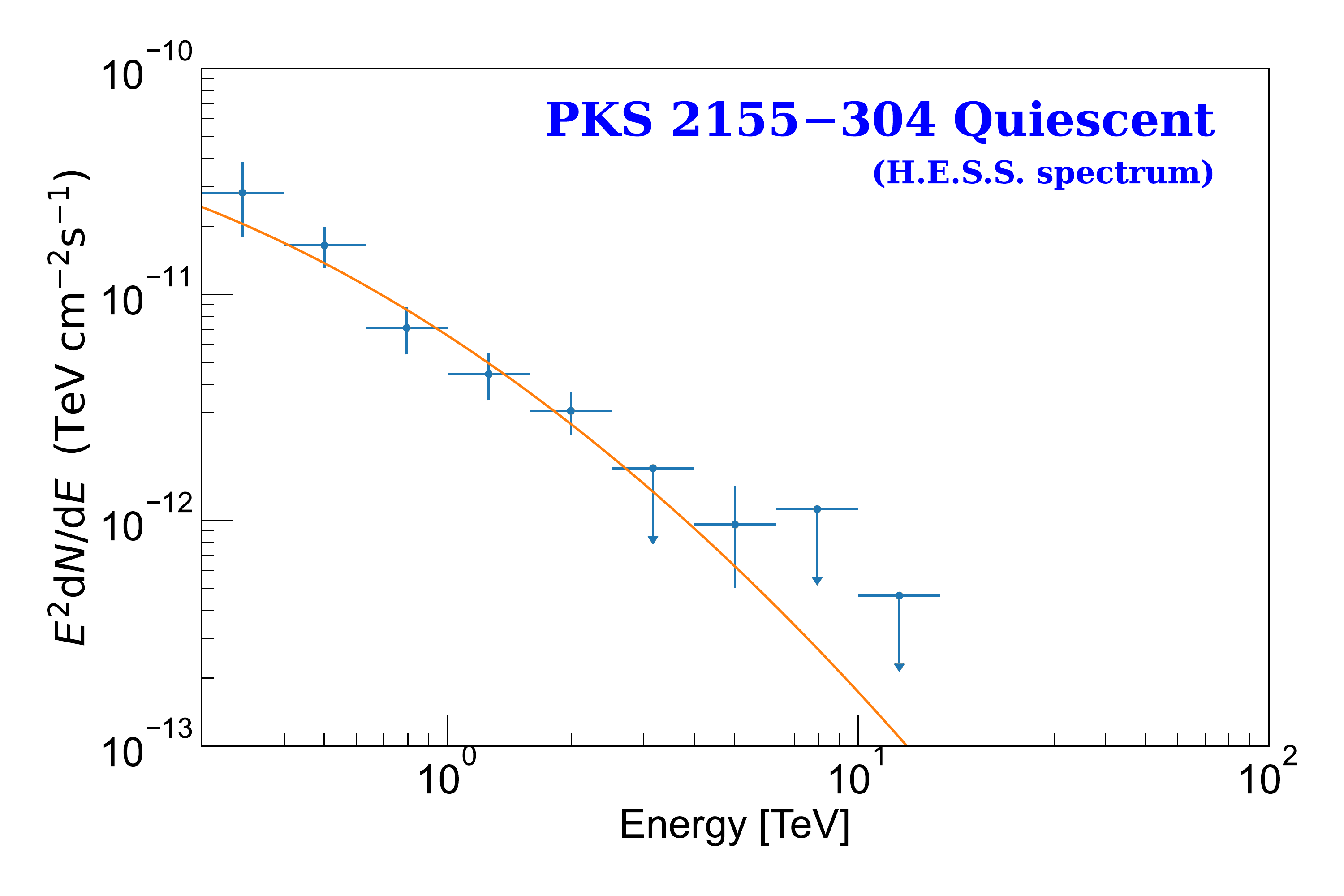}
    \hspace{0.3cm}
    \includegraphics[width=0.475\textwidth]{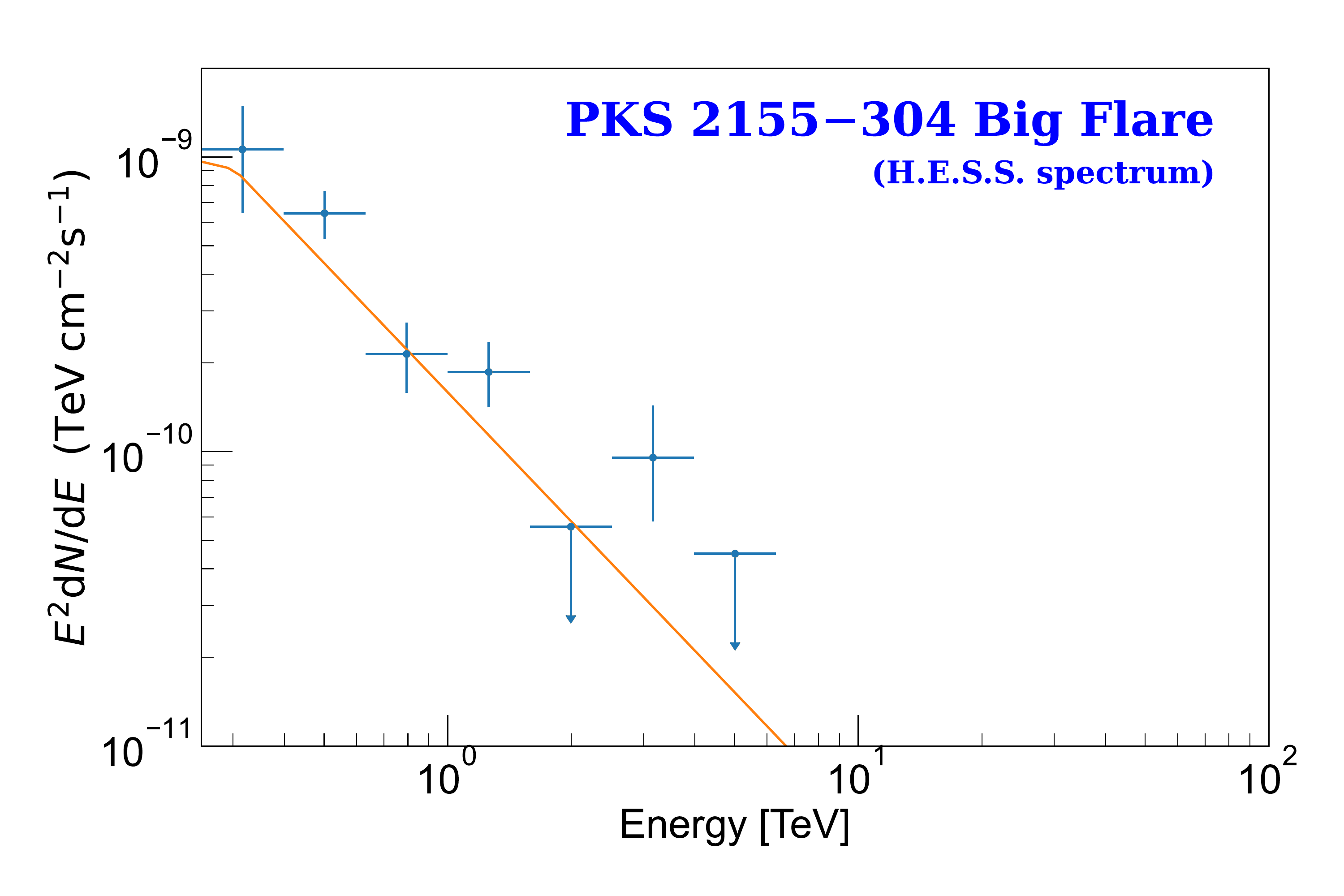}
    
    \includegraphics[width=0.475\textwidth]{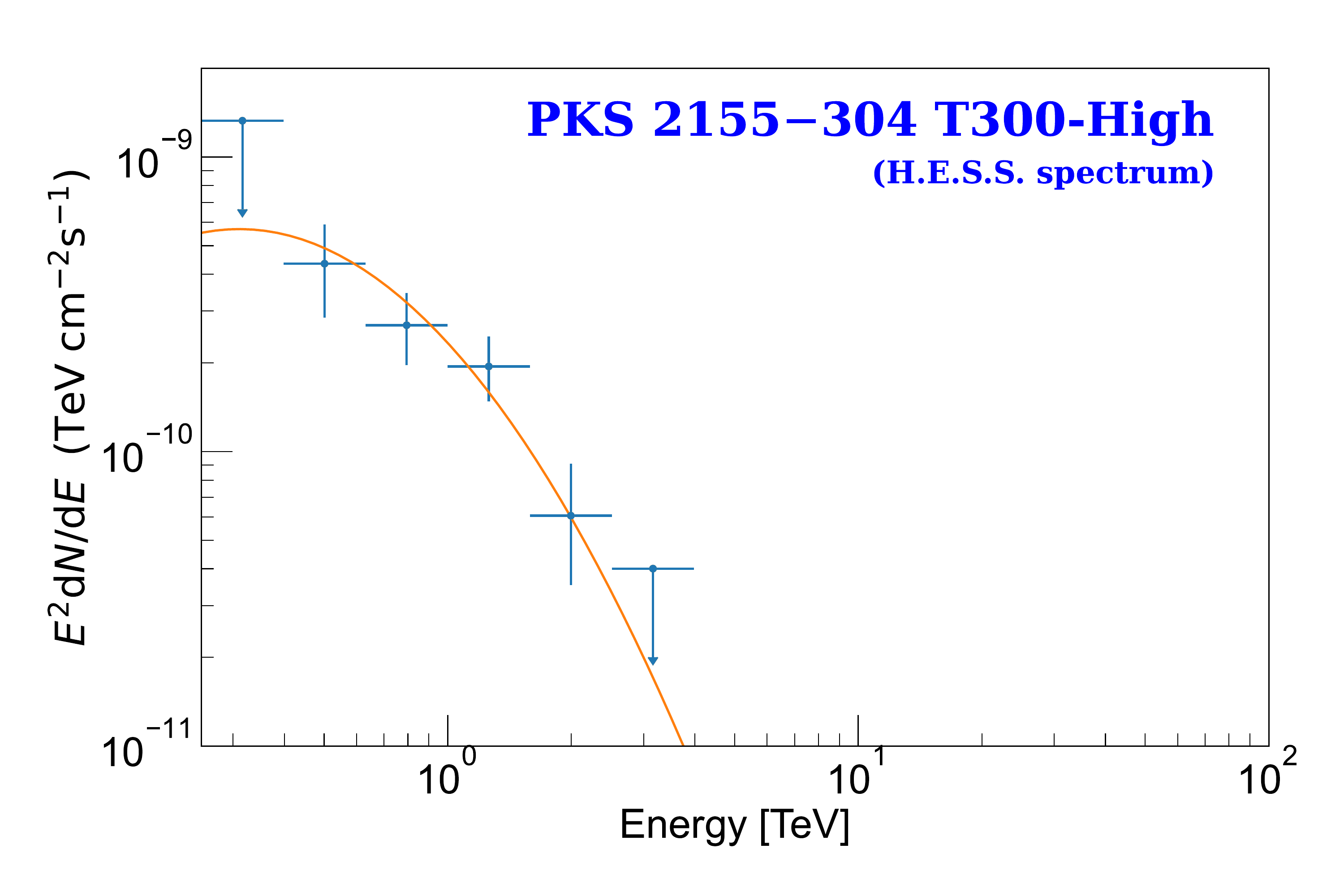}
    \hspace{0.3cm}
    \includegraphics[width=0.475\textwidth]{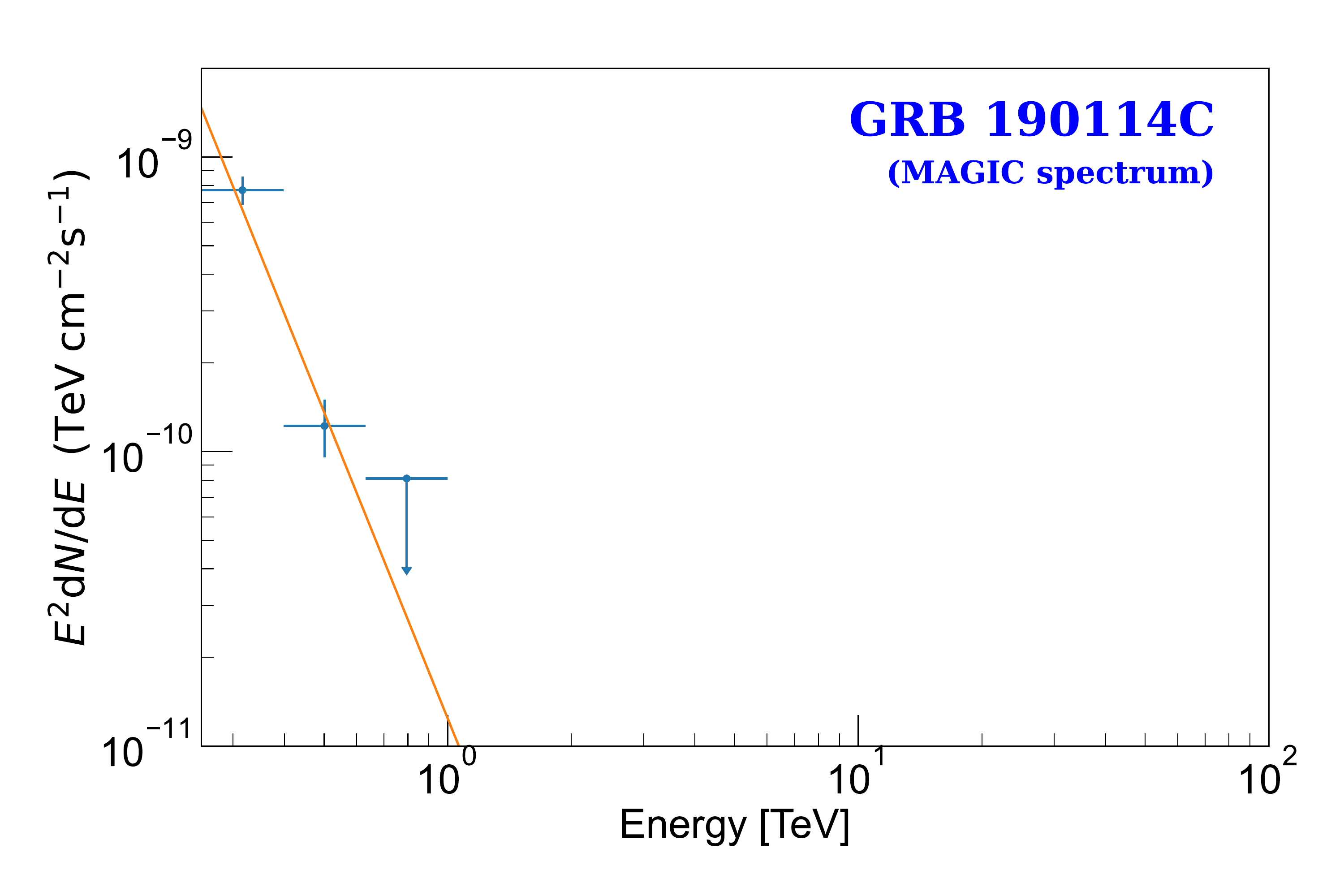}
    
    \caption{\small{{\textbf{Spectral responses for selected sources.}} Crab Nebula \SI{3348.8}{\hour} (to compare with HAWC 837.2 calendar days); Markarian 421 \SI{50}{\hour} (taking HAWC's 3-year average spectrum); PKS~2155$-$304: quiescent for 1 calendar year (\SI{1461}{\hour} live-time), Big Flare \SI{1.32}{\hour}, ``Chandra night'' \texttt{T300-High} \SI{1.36}{\hour}; GRB~190114C for \SI{40}{\minute}. On the spectra displayed, $2\sigma$ upper limits are shown.}}
    
    \label{fig:spectra_results}
\end{figure*}

\begin{itemize}
    \item Crab Nebula: For this `standard candle' we take the Crab spectrum as measured by HAWC (NN-method) \citep{HAWC:Crab}, since this covers a similar energy range.  In 837.2 calendar days, HAWC measured a significance of $139\sigma$, whereas for ALTO in the same time ($837.2 \times 4\,$\si{\hour} $=$ \SI{3348.8}{\hour}), the significance would be $71\sigma$; in \SI{50}{\hour}, a still-significant result of $8.7\sigma$ can be achieved, see Table~\ref{table:SensiSources}. The spectrum shown in Figure~\ref{fig:spectra_results} is for the longer time, for comparison with the HAWC result.
    \item PKS~2155$-$304: This Southern-hemisphere AGN, discovered at VHE by the Durham Mk-VI telescope in 1999 \citep{DurhamMkVI:PKS2155}, is notable in that with the H.E.S.S.\ IACT array it is detected in all states, from quiescent to exceptional outbursts.  Here we consider three states:
    \begin{itemize}
        \item The exceptional flare (or `Big Flare') on July 28, 2006 (MJD~53,944), consisted of several fast bursts observed over a duration of \SI{1.32}{\hour} live-time \citep{PKS2155_bigflare}.  We consider the broken power-law for the full live-time given there.  While for H.E.S.S., an average $\gamma$-ray rate of \SI{2.5}{Hz} and a significance of $168\sigma$ was found, for ALTO this would be detected with \SI{2.1}{\per\minute} at the $6.4\sigma$ level.  However, we note that the H.E.S.S. observations began \SI{\sim2}{\hour} after the start of the observability window of \SI{30}{\degree} from Zenith, so with an instrument such as ALTO a wider time-coverage could be achieved.
        \item \texttt{T300-High}: The second-highest flare occurred on the subsequent night, also known as the ``Chandra night'' given the multi-wavelength observations triggered by the `big flare', as described in \cite{PKS2155_Chandra_night}.  From this, we use the spectrum for the high state \texttt{T300-High} part of the night's observations, with duration \SI{1.36}{\hour}.  The flux level here was similar to the average of the preceding night, with the spectrum described by a log-parabola, giving a similar ALTO response as for the `big flare' night.
        \item Quiescent state: since this is the most powerful nearby AGN at VHE, this source is detectable nightly even in its quiescent state by H.E.S.S., for example in \cite{PKS2155_quiescent}.  For ALTO, this quiescent state would give a highly-significant detection at $6.9\sigma$ in one calendar year of observation (\SI{1461}{\hour} on the source).
\end{itemize}
    \item Mkn 421: This source, detected in 1992 by the Whipple Observatory $\gamma$-ray telescope \citep{Whipple:421} is the most powerful nearby Northern AGN (\SI{+39.2}{\degree} declination). It has been detected by several VHE observatories since then, even at large zenith angles by the Southern-located H.E.S.S.\ Observatory.  Here, we use the 3-year spectrum as measured by HAWC between June 2015 and July 2018 \citep{HAWC:421}, ($\sim1038$ days of exposure) to estimate detectability.  We model this using the intrinsic power-law with exponential cut-off spectrum given there, absorbed on the extragalactic background light, which results in a significant $5.9\sigma$ detection with ALTO in \SI{50}{\hour}.
    With HAWC, where the source transits at \SI{20}{\degree} from Zenith, the detection in the three-year exposure was at the $48\sigma$ level; for ALTO, if the source were Southern, 
    similarly, a $51.1\sigma$ detection would be achieved with 1038 transits of \SI{4}{\hour}.
    \item 1ES~1215$+$30: This blazar with a red-shift above $\sim 0.1$ was first detected at VHE by MAGIC in 2011 \citep{MAGIC:1215}. It has been extensively followed by VERITAS, which has observed several outbursts, notably at the $\sim 2$-Crab level on February 8, 2014, in a 45-minute observation window \citep{VERITAS:1215_flare}.  
    Based on the average flux and Log-Parabolic spectrum during that window, it would hardly be detectable by ALTO in that time-lapse.  However, if that burst (which was compatible with steady emission) continued for a \SI{4}{\hour} transit, ALTO could give an alert at a $> 3\sigma$ level, as shown in Table~\ref{table:SensiSources}.  We note that HAWC observed a $3\sigma$ excess from this source (between November 2014 and June 2019) in their AGN survey \citep{HAWC:AGN_survey}, unfortunately not overlapping the VERITAS outburst.
    \item GRB~190114C: This is one of the most intense GRBs seen at VHE, detected by MAGIC \citep{MAGIC:GRB190114C}.  We take the observed spectrum as seen between $T_0\!+\!62\,\rm s$ and $T_0\!+\!2453\,\rm s$ (from Figure 2 in \citealt{MAGIC:GRB190114C}), i.e., for \SI{\sim 40}{\minute}.  We note (Table~\ref{table:SensiSources}) that the excess is of the same order of the background, so ALTO would significantly detect such a bright GRB.  For this high red-shift source, the spectral index measured by MAGIC is $5.43\!\pm\!0.22$, while for ALTO, based on the procedure above with this injected index, the fitted spectrum in 100 realizations was found to be  $5.49\!\pm\!0.37$.

\end{itemize}

\subsection{Comparison with Current and Future Major Particle Detector Arrays}
\label{subsec:comparison}

In Figure \ref{fig:ALTO_SEMLA_sensitivity}{\emph{(Lower-Right)}}, we also show the ALTO differential sensitivity for one calendar year of operation as compared with major current or proposed future particle detector arrays, although we would place strong caveats on such a comparison given the different assumptions under which the curves have been produced.  
For the comparison, as a proxy for a cost estimate, we give the number and size of the PMTs used in the arrays, since the PMT photocathode area is one major cost driver.  For HAWC \citep{HAWC:Crab} and LHAASO \citep{LHAASO:Performance_crab}, the sensitivity estimation is given for tracking the Crab Nebula over all Zenith angles up to a certain limit.  
The HAWC sensitivity curve was also adjusted to take into account the conversion from 4 bins per decade in that reference to the 5 bins per decade used here.  
For LHAASO, the numbers of PMTs given concern the WCDA component (water Cherenkov detector array), which provides the sensitivity below the break at \SI{16}{\TeV}.
Concerning ALTO (this paper) and the ``SWGO straw-man'' sensitivity \citep{SGSO:WhitePaper}, the estimates are made based on fixed Zenith angle simulations (\SI{18}{\degree} and \SI{20}{\degree} respectively), while assuming the source transits overhead within $\pm30\degree$ and $\pm45\degree$ of Zenith, respectively, so assuming \SI{4}{hr/day} for ALTO versus \SI{6}{hr/day} for SWGO.  
Evidently, such comparisons are indicative only.

Furthermore, we note that, when comparing differential sensitivities, this comparison only carries partial information, as the improvements in energy resolution are essentially independent of the sensitivity. The energy resolution is an important parameter, for instance, for the determination of the spectral cutoffs. 

When comparing the ALTO sensitivity with respect to HAWC, it is difficult to disentangle the effect of the altitude from the hardware and design choices, as the number of PMTs is similar. 

As compared to LHAASO,  it should be noted that while the reported sensitivity extends lower in energy than ALTO, this is at the expense of over ten times more photocathode area for the ensemble of PMTs, and the costly choice of using covered pools for the detector medium.  

Finally, for the scientific goal of the ALTO project which is to enlarge the $\gamma$-ray horizon of high-altitude detector arrays by studying the detection capability for extragalactic soft-spectrum sources, and in particular, the capability of detecting transient phenomena with a wide field-of-view, the ALTO design and hardware choices represent an efficient, cost-effective solution, since with just an eighth of number of similar-sized PMTs than proposed by SWGO, ALTO has a threshold $\sim 2$ times lower than SWGO, while having only $\sim 2$--3 times less sensitivity at \SI{300}{GeV}. 
This implies that ALTO could operate as a standalone array, or as a high-density core of a future larger initiative such as SWGO.

\section{Summary and outlook}
\label{sec:outlook}

In this paper, we show the point-like source detection capability of a ground-based $\gamma$-ray detector array such as ALTO, located at an altitude of \SI{5.1}{\kilo\metre}. 
Until now, in the extragalactic field  similar types of array have detected only the closest hard-spectrum AGNs (Mkn~421 and Mkn~501, by HAWC, \citealt{HAWC:421}) or the recently-detected ultra-bright GRB (GRB~221009A, by LHAASO, \citealt{LHAASO:GRB221009A})\footnote{The ALTO response to GRB~221009A has not been considered here, as spectral parameters are not yet available.}.  Therefore, the $\gamma$-ray horizon of this technique is, at present, quite limited.

The ALTO results presented in this paper were obtained after an effort comprising prototyping, full Monte Carlo simulations, a full reconstruction and analysis strategy and the use of the Gammapy software tool \citep{gammapy:1.0}, which is becoming the standard tool in $\gamma$-ray astronomy for high-level analysis. 
We show here the detection significances for a list of sources of interest, and their reconstructed spectra, focussing our study on extragalactic sources and transient phenomena. 

We find that for short and bright transient phenomena, ALTO is capable of significantly detecting an excess of gamma rays, which makes it interesting for the monitoring of VHE $\gamma$-ray sources, given its large field-of-view.

As a VHE monitor, it can be used to give instant alerts to other pointing observatories, thus optimizing the observation strategies on transient phenomena. 

However, the capability on steady soft-spectrum sources remains quite limited, indicating that the technique still needs to be improved. In a forthcoming publication, we will show how we could possibly improve the sensitivity of wide-field-of-view ground-based gamma-ray observations during darkness, by coupling the ALTO particle detectors with sensitive atmospheric Cherenkov light collectors. 

\section*{Acknowledgements}

The ALTO R\&D project is primarily supported by Linnaeus University and the following Swedish private foundations or public institutes: the Crafoord Foundation, the Foundation in memory of Lars Hierta, the Magnus Bergvall’s Foundation, the Crafoord stipendium of the Royal Swedish Academy of Sciences (KVA), the Märta and Erik Holmberg Endowment of the Royal Physiographic Society in Lund, the Längmanska kulturfonden, the Helge Ax:son Johnson’s Foundation and Linnaeus University. 
We also thank the Swedish National Infrastructure for Computing (SNIC) at Lunarc (Lund, Sweden). 
The authors would like to thank the \emph{Data Intensive Sciences and Applications} (DISA) research centre at Linnaeus University for their valuable support.
We also acknowledge funding from the Abu Dhabi Award for Research Excellence (AARE-2019) under the project number AARE19-224.
MP would like to thank the Gammapy and PyIRF teams for helpful feedback on inclusion of ALTO, especially Maximilian Linhoff, Axel Donath, R\'egis Terrier, and Bruno Kh\'elifi.

\bibliographystyle{elsarticle-num-names} 
\bibliography{ALTO_Sensitivity_Performance}

\begin{thebibliography}{35}
\expandafter\ifx\csname natexlab\endcsname\relax\def\natexlab#1{#1}\fi
\providecommand{\url}[1]{\texttt{#1}}
\providecommand{\href}[2]{#2}
\providecommand{\path}[1]{#1}
\providecommand{\DOIprefix}{doi:}
\providecommand{\ArXivprefix}{arXiv:}
\providecommand{\URLprefix}{URL: }
\providecommand{\Pubmedprefix}{pmid:}
\providecommand{\doi}[1]{\href{http://dx.doi.org/#1}{\path{#1}}}
\providecommand{\Pubmed}[1]{\href{pmid:#1}{\path{#1}}}
\providecommand{\bibinfo}[2]{#2}
\ifx\xfnm\relax \def\xfnm[#1]{\unskip,\space#1}\fi
%Type = Article
\bibitem[{Senniappan et~al.(2021)Senniappan, Becherini, Punch, Thoudam, Bylund,
  Mezek, and Ernenwein}]{semla_jinst}
\bibinfo{author}{M.~Senniappan}, \bibinfo{author}{Y.~Becherini},
  \bibinfo{author}{M.~Punch}, \bibinfo{author}{S.~Thoudam},
  \bibinfo{author}{T.~Bylund}, \bibinfo{author}{G.~K. Mezek},
  \bibinfo{author}{J.-P. Ernenwein},
\newblock \bibinfo{title}{Signal extraction in atmospheric shower arrays
  designed for 200 {GeV}{\textendash}50 {TeV} $\gamma$-ray astronomy},
\newblock \bibinfo{journal}{JInst} \bibinfo{volume}{16} (\bibinfo{year}{2021})
  \bibinfo{pages}{P07050}. \URLprefix
  \url{https://doi.org/10.1088/1748-0221/16/07/p07050}.
  \DOIprefix\doi{10.1088/1748-0221/16/07/p07050}.
%Type = Article
\bibitem[{Becherini et~al.(2017)Becherini, Thoudam, Punch, and
  Ernenwein}]{ALTO1_ICRC2017}
\bibinfo{author}{Y.~Becherini}, \bibinfo{author}{S.~Thoudam},
  \bibinfo{author}{M.~Punch}, \bibinfo{author}{J.-P. Ernenwein},
\newblock \bibinfo{title}{{Very-High-Energy gamma-ray astronomy with the ALTO
  observatory}},
\newblock \bibinfo{journal}{PoS} \bibinfo{volume}{ICRC2017}
  (\bibinfo{year}{2017}) \bibinfo{pages}{782}.
  \DOIprefix\doi{10.22323/1.301.0782}.
%Type = Article
\bibitem[{Thoudam et~al.(2017)Thoudam, Becherini, and Punch}]{ALTO2_ICRC2017}
\bibinfo{author}{S.~Thoudam}, \bibinfo{author}{Y.~Becherini},
  \bibinfo{author}{M.~Punch},
\newblock \bibinfo{title}{{Simulation study for the proposed wide field-of-view
  gamma-ray detector array ALTO}},
\newblock \bibinfo{journal}{PoS} \bibinfo{volume}{ICRC2017}
  (\bibinfo{year}{2017}) \bibinfo{pages}{780}.
  \DOIprefix\doi{10.22323/1.301.0780}.
%Type = Article
\bibitem[{Senniappan et~al.(2021)Senniappan, Becherini, Punch, Thoudam, Bylund,
  Kukec~Mezek, and Ernenwein}]{ALTO3_ICRC2021}
\bibinfo{author}{M.~Senniappan}, \bibinfo{author}{Y.~Becherini},
  \bibinfo{author}{M.~Punch}, \bibinfo{author}{S.~Thoudam},
  \bibinfo{author}{T.~Bylund}, \bibinfo{author}{G.~Kukec~Mezek},
  \bibinfo{author}{J.-P. Ernenwein},
\newblock \bibinfo{title}{{Expected performance of the ALTO particle detector
  array designed for 200 GeV - 50 TeV gamma-ray astronomy}},
\newblock \bibinfo{journal}{PoS} \bibinfo{volume}{ICRC2021}
  (\bibinfo{year}{2021}) \bibinfo{pages}{761}.
  \DOIprefix\doi{10.22323/1.395.0761}.
%Type = Article
\bibitem[{{Abeysekara} et~al.(2019){Abeysekara}, {Albert}, {Alfaro}, {Alvarez},
  {{\'A}lvarez}, {Camacho}, {Arceo}, {Arteaga-Vel{\'a}zquez}, {Arunbabu},
  {Avila Rojas}, {Ayala Solares}, {Baghmanyan}, {Belmont-Moreno}, {BenZvi},
  {Brisbois}, {Caballero-Mora}, {Capistr{\'a}n}, {Carrami{\~n}ana}, {Casanova},
  {Cotti}, {Cotzomi}, {Couti{\~n}o de Le{\'o}n}, {De la Fuente}, {de Le{\'o}n},
  {Dichiara}, {Dingus}, {DuVernois}, {D{\'\i}az-V{\'e}lez}, {Ellsworth},
  {Engel}, {Espinoza}, {Fick}, {Fleischhack}, {Fraija}, {Galv{\'a}n-G{\'a}mez},
  {Garc{\'\i}a-Gonz{\'a}lez}, {Garfias}, {Gonz{\'a}lez}, {Goodman}, {Harding},
  {Hernandez}, {Hinton}, {Hona}, {Hueyotl-Zahuantitla}, {Hui},
  {H{\"u}ntemeyer}, {Iriarte}, {Jardin-Blicq}, {Joshi}, {Kaufmann}, {Kieda},
  {Lara}, {Lee}, {Le{\'o}n Vargas}, {Linnemann}, {Longinotti}, {Luis-Raya},
  {Lundeen}, {Malone}, {Marinelli}, {Martinez}, {Martinez-Castellanos},
  {Mart{\'\i}nez-Castro}, {Mart{\'\i}nez-Huerta}, {Matthews},
  {Miranda-Romagnoli}, {Morales-Soto}, {Moreno}, {Mostaf{\'a}}, {Nayerhoda},
  {Nellen}, {Newbold}, {Nisa}, {Noriega-Papaqui}, {Peisker},
  {P{\'e}rez-P{\'e}rez}, {Pretz}, {Ren}, {Rho}, {Rivi{\`e}re},
  {Rosa-Gonz{\'a}lez}, {Rosenberg}, {Ruiz-Velasco}, {Salazar}, {Salesa Greus},
  {Sandoval}, {Schneider}, {Schoorlemmer}, {Seglar Arroyo}, {Sinnis}, {Smith},
  {Springer}, {Surajbali}, {Tabachnick}, {Tanner}, {Tibolla}, {Tollefson},
  {Torres}, {Weisgarber}, {Westerhoff}, {Wood}, {Yapici}, {Zepeda}, {Zhou}, and
  {HAWC Collaboration}}]{HAWC:Crab}
\bibinfo{author}{A.~U. {Abeysekara}}, \bibinfo{author}{A.~{Albert}},
  \bibinfo{author}{R.~{Alfaro}}, \bibinfo{author}{C.~{Alvarez}},
  \bibinfo{author}{J.~D. {{\'A}lvarez}}, \bibinfo{author}{J.~R.~A. {Camacho}},
  \bibinfo{author}{R.~{Arceo}}, \bibinfo{author}{J.~C.
  {Arteaga-Vel{\'a}zquez}}, \bibinfo{author}{K.~P. {Arunbabu}},
  \bibinfo{author}{D.~{Avila Rojas}}, \bibinfo{author}{H.~A. {Ayala Solares}},
  \bibinfo{author}{V.~{Baghmanyan}}, \bibinfo{author}{E.~{Belmont-Moreno}},
  \bibinfo{author}{S.~Y. {BenZvi}}, \bibinfo{author}{C.~{Brisbois}},
  \bibinfo{author}{K.~S. {Caballero-Mora}},
  \bibinfo{author}{T.~{Capistr{\'a}n}}, \bibinfo{author}{A.~{Carrami{\~n}ana}},
  \bibinfo{author}{S.~{Casanova}}, \bibinfo{author}{U.~{Cotti}},
  \bibinfo{author}{J.~{Cotzomi}}, \bibinfo{author}{S.~{Couti{\~n}o de
  Le{\'o}n}}, \bibinfo{author}{E.~{De la Fuente}}, \bibinfo{author}{C.~{de
  Le{\'o}n}}, \bibinfo{author}{S.~{Dichiara}}, \bibinfo{author}{B.~L.
  {Dingus}}, \bibinfo{author}{M.~A. {DuVernois}}, \bibinfo{author}{J.~C.
  {D{\'\i}az-V{\'e}lez}}, \bibinfo{author}{R.~W. {Ellsworth}},
  \bibinfo{author}{K.~{Engel}}, \bibinfo{author}{C.~{Espinoza}},
  \bibinfo{author}{B.~{Fick}}, \bibinfo{author}{H.~{Fleischhack}},
  \bibinfo{author}{N.~{Fraija}}, \bibinfo{author}{A.~{Galv{\'a}n-G{\'a}mez}},
  \bibinfo{author}{J.~A. {Garc{\'\i}a-Gonz{\'a}lez}},
  \bibinfo{author}{F.~{Garfias}}, \bibinfo{author}{M.~M. {Gonz{\'a}lez}},
  \bibinfo{author}{J.~A. {Goodman}}, \bibinfo{author}{J.~P. {Harding}},
  \bibinfo{author}{S.~{Hernandez}}, \bibinfo{author}{J.~{Hinton}},
  \bibinfo{author}{B.~{Hona}}, \bibinfo{author}{F.~{Hueyotl-Zahuantitla}},
  \bibinfo{author}{C.~M. {Hui}}, \bibinfo{author}{P.~{H{\"u}ntemeyer}},
  \bibinfo{author}{A.~{Iriarte}}, \bibinfo{author}{A.~{Jardin-Blicq}},
  \bibinfo{author}{V.~{Joshi}}, \bibinfo{author}{S.~{Kaufmann}},
  \bibinfo{author}{D.~{Kieda}}, \bibinfo{author}{A.~{Lara}},
  \bibinfo{author}{W.~H. {Lee}}, \bibinfo{author}{H.~{Le{\'o}n Vargas}},
  \bibinfo{author}{J.~T. {Linnemann}}, \bibinfo{author}{A.~L. {Longinotti}},
  \bibinfo{author}{G.~{Luis-Raya}}, \bibinfo{author}{J.~{Lundeen}},
  \bibinfo{author}{K.~{Malone}}, \bibinfo{author}{S.~S. {Marinelli}},
  \bibinfo{author}{O.~{Martinez}}, \bibinfo{author}{I.~{Martinez-Castellanos}},
  \bibinfo{author}{J.~{Mart{\'\i}nez-Castro}},
  \bibinfo{author}{H.~{Mart{\'\i}nez-Huerta}}, \bibinfo{author}{J.~A.
  {Matthews}}, \bibinfo{author}{P.~{Miranda-Romagnoli}}, \bibinfo{author}{J.~A.
  {Morales-Soto}}, \bibinfo{author}{E.~{Moreno}},
  \bibinfo{author}{M.~{Mostaf{\'a}}}, \bibinfo{author}{A.~{Nayerhoda}},
  \bibinfo{author}{L.~{Nellen}}, \bibinfo{author}{M.~{Newbold}},
  \bibinfo{author}{M.~U. {Nisa}}, \bibinfo{author}{R.~{Noriega-Papaqui}},
  \bibinfo{author}{A.~{Peisker}}, \bibinfo{author}{E.~G.
  {P{\'e}rez-P{\'e}rez}}, \bibinfo{author}{J.~{Pretz}},
  \bibinfo{author}{Z.~{Ren}}, \bibinfo{author}{C.~D. {Rho}},
  \bibinfo{author}{C.~{Rivi{\`e}re}}, \bibinfo{author}{D.~{Rosa-Gonz{\'a}lez}},
  \bibinfo{author}{M.~{Rosenberg}}, \bibinfo{author}{E.~{Ruiz-Velasco}},
  \bibinfo{author}{H.~{Salazar}}, \bibinfo{author}{F.~{Salesa Greus}},
  \bibinfo{author}{A.~{Sandoval}}, \bibinfo{author}{M.~{Schneider}},
  \bibinfo{author}{H.~{Schoorlemmer}}, \bibinfo{author}{M.~{Seglar Arroyo}},
  \bibinfo{author}{G.~{Sinnis}}, \bibinfo{author}{A.~J. {Smith}},
  \bibinfo{author}{R.~W. {Springer}}, \bibinfo{author}{P.~{Surajbali}},
  \bibinfo{author}{E.~{Tabachnick}}, \bibinfo{author}{M.~{Tanner}},
  \bibinfo{author}{O.~{Tibolla}}, \bibinfo{author}{K.~{Tollefson}},
  \bibinfo{author}{I.~{Torres}}, \bibinfo{author}{T.~{Weisgarber}},
  \bibinfo{author}{S.~{Westerhoff}}, \bibinfo{author}{J.~{Wood}},
  \bibinfo{author}{T.~{Yapici}}, \bibinfo{author}{A.~{Zepeda}},
  \bibinfo{author}{H.~{Zhou}}, \bibinfo{author}{{HAWC Collaboration}},
\newblock \bibinfo{title}{{Measurement of the Crab Nebula Spectrum Past 100 TeV
  with HAWC}},
\newblock \bibinfo{journal}{\apj} \bibinfo{volume}{881} (\bibinfo{year}{2019})
  \bibinfo{pages}{134}. \DOIprefix\doi{10.3847/1538-4357/ab2f7d}.
  \href{http://arxiv.org/abs/1905.12518}{{\tt arXiv:1905.12518}}.
%Type = Article
\bibitem[{{Cao} et~al.(2019){Cao}, {della Volpe}, {Liu}, {Editors}, {:}, {Bi},
  {Chen}, {D'Ettorre Piazzoli}, {Feng}, {Jia}, {Li}, {Ma}, {Wang}, {Zhang},
  {Referees}, {:}, {Qie}, {Hu}, {Referees}, {:}, {S{\'a}iz}, {Yang},
  {Contributors}, {:}, {Addazi}, {Belotsky}, {Beylin}, {Bi}, {Che}, {Chen},
  {Cheng}, {Chiavassa}, {Cirelli}, {Di Sciascio}, {Esmaili}, {Fang},
  {Fornengo}, {Gou}, {Guo}, {Gan}, {Gong}, {Gu}, {He}, {He}, {Hou}, {Huang},
  {Huang}, {Kachekriess}, {Khlopov}, {Korchagin}, {Korochkin}, {Kuksa},
  {Ksenofontov}, {Liu}, {Liu}, {Liu}, {Marciano}, {Martineau-Huynh},
  {Martraire}, {Ma}, {Neronov}, {Panci}, {Pasechnick}, {Ruffolo}, {Sakharov},
  {Sala}, {Semikoz}, {Shchegolev}, {Serpico}, {Sheng}, {Stenkin}, {Tam},
  {Vernetto}, {Vallania}, {Volchanskiy}, {Wang}, {Wang}, {Wang}, {Wu}, {Wu},
  {Wu}, {Xiao}, {Yang}, {Yan}, {Yao}, {Yin}, {Yuan}, {Zhang}, {Zeng}, {Zhang},
  {Zhang}, {Zhou}, {Zhu}, and {Zuo}}]{LHAASO_ScienceBook2021}
\bibinfo{author}{Z.~{Cao}}, \bibinfo{author}{D.~{della Volpe}},
  \bibinfo{author}{S.~{Liu}}, \bibinfo{author}{{Editors}},
  \bibinfo{author}{{:}}, \bibinfo{author}{X.~{Bi}},
  \bibinfo{author}{Y.~{Chen}}, \bibinfo{author}{B.~{D'Ettorre Piazzoli}},
  \bibinfo{author}{L.~{Feng}}, \bibinfo{author}{H.~{Jia}},
  \bibinfo{author}{Z.~{Li}}, \bibinfo{author}{X.~{Ma}},
  \bibinfo{author}{X.~{Wang}}, \bibinfo{author}{X.~{Zhang}},
  \bibinfo{author}{E.~{Referees}}, \bibinfo{author}{{:}},
  \bibinfo{author}{X.~{Qie}}, \bibinfo{author}{H.~{Hu}},
  \bibinfo{author}{I.~{Referees}}, \bibinfo{author}{{:}},
  \bibinfo{author}{A.~{S{\'a}iz}}, \bibinfo{author}{R.~{Yang}},
  \bibinfo{author}{{Contributors}}, \bibinfo{author}{{:}},
  \bibinfo{author}{A.~{Addazi}}, \bibinfo{author}{K.~{Belotsky}},
  \bibinfo{author}{V.~{Beylin}}, \bibinfo{author}{Y.-J. {Bi}},
  \bibinfo{author}{M.-J. {Che}}, \bibinfo{author}{S.-Z. {Chen}},
  \bibinfo{author}{Y.-D. {Cheng}}, \bibinfo{author}{A.~{Chiavassa}},
  \bibinfo{author}{M.~{Cirelli}}, \bibinfo{author}{G.~{Di Sciascio}},
  \bibinfo{author}{A.~{Esmaili}}, \bibinfo{author}{K.~{Fang}},
  \bibinfo{author}{N.~{Fornengo}}, \bibinfo{author}{Q.~{Gou}},
  \bibinfo{author}{Y.-Q. {Guo}}, \bibinfo{author}{Q.~{Gan}},
  \bibinfo{author}{G.-H. {Gong}}, \bibinfo{author}{M.-H. {Gu}},
  \bibinfo{author}{H.~{He}}, \bibinfo{author}{H.-H. {He}},
  \bibinfo{author}{C.~{Hou}}, \bibinfo{author}{X.-T. {Huang}},
  \bibinfo{author}{W.-H. {Huang}}, \bibinfo{author}{M.~{Kachekriess}},
  \bibinfo{author}{M.~{Khlopov}}, \bibinfo{author}{V.~{Korchagin}},
  \bibinfo{author}{A.~{Korochkin}}, \bibinfo{author}{V.~{Kuksa}},
  \bibinfo{author}{L.~T. {Ksenofontov}}, \bibinfo{author}{Y.~{Liu}},
  \bibinfo{author}{R.-Y. {Liu}}, \bibinfo{author}{C.~{Liu}},
  \bibinfo{author}{A.~{Marciano}}, \bibinfo{author}{O.~{Martineau-Huynh}},
  \bibinfo{author}{D.~{Martraire}}, \bibinfo{author}{L.~{Ma}},
  \bibinfo{author}{A.~{Neronov}}, \bibinfo{author}{P.~{Panci}},
  \bibinfo{author}{R.~{Pasechnick}}, \bibinfo{author}{D.~{Ruffolo}},
  \bibinfo{author}{A.~{Sakharov}}, \bibinfo{author}{F.~{Sala}},
  \bibinfo{author}{D.~{Semikoz}}, \bibinfo{author}{O.~{Shchegolev}},
  \bibinfo{author}{P.~D. {Serpico}}, \bibinfo{author}{X.-D. {Sheng}},
  \bibinfo{author}{Y.~V. {Stenkin}}, \bibinfo{author}{P.~H.~T. {Tam}},
  \bibinfo{author}{S.~{Vernetto}}, \bibinfo{author}{P.~{Vallania}},
  \bibinfo{author}{N.~{Volchanskiy}}, \bibinfo{author}{Z.~{Wang}},
  \bibinfo{author}{K.~{Wang}}, \bibinfo{author}{X.-Y. {Wang}},
  \bibinfo{author}{H.-R. {Wu}}, \bibinfo{author}{C.-Y. {Wu}},
  \bibinfo{author}{S.~{Wu}}, \bibinfo{author}{G.~{Xiao}},
  \bibinfo{author}{R.-z. {Yang}}, \bibinfo{author}{D.~{Yan}},
  \bibinfo{author}{Z.-G. {Yao}}, \bibinfo{author}{P.~{Yin}},
  \bibinfo{author}{Q.~{Yuan}}, \bibinfo{author}{X.~{Zhang}},
  \bibinfo{author}{H.~{Zeng}}, \bibinfo{author}{S.-S. {Zhang}},
  \bibinfo{author}{Y.~{Zhang}}, \bibinfo{author}{X.~{Zhou}},
  \bibinfo{author}{H.~{Zhu}}, \bibinfo{author}{X.~{Zuo}},
\newblock \bibinfo{title}{{The Large High Altitude Air Shower Observatory
  (LHAASO) Science Book (2021 Edition)}},
\newblock \bibinfo{journal}{arXiv e-prints}  (\bibinfo{year}{2019})
  \bibinfo{pages}{arXiv:1905.02773}. \DOIprefix\doi{10.48550/arXiv.1905.02773}.
  \href{http://arxiv.org/abs/1905.02773}{{\tt arXiv:1905.02773}}.
%Type = Article
\bibitem[{{Albert} et~al.(2019){Albert}, {Alfaro}, {Ashkar}, {Alvarez},
  {{\'A}lvarez}, {Arteaga-Vel{\'a}zquez}, {Ayala Solares}, {Arceo}, {Bellido},
  {BenZvi}, {Bretz}, {Brisbois}, {Brown}, {Brun}, {Caballero-Mora}, {Carosi},
  {Carrami{\~n}ana}, {Casanova}, {Chadwick}, {Cotter}, {Couti{\~n}o De
  Le{\'o}n}, {Cristofari}, {Dasso}, {de la Fuente}, {Dingus}, {Desiati},
  {Salles}, {de Souza}, {Dorner}, {D{\'\i}az-V{\'e}lez},
  {Garc{\'\i}a-Gonz{\'a}lez}, {DuVernois}, {Di Sciascio}, {Engel},
  {Fleischhack}, {Fraija}, {Funk}, {Glicenstein}, {Gonzalez}, {Gonz{\'a}lez},
  {Goodman}, {Harding}, {Haungs}, {Hinton}, {Hona}, {Hoyos}, {Huentemeyer},
  {Iriarte}, {Jardin-Blicq}, {Joshi}, {Kaufmann}, {Kawata}, {Kunwar},
  {Lefaucheur}, {Lenain}, {Link}, {L{\'o}pez-Coto}, {Marandon}, {Mariotti},
  {Mart{\'\i}nez-Castro}, {Mart{\'\i}nez-Huerta}, {Mostaf{\'a}}, {Nayerhoda},
  {Nellen}, {de O{\~n}a Wilhelmi}, {Parsons}, {Patricelli}, {Pichel}, {Piel},
  {Prandini}, {Pueschel}, {Procureur}, {Reisenegger}, {Rivi{\`e}re},
  {Rodriguez}, {Rovero}, {Rowell}, {Ruiz-Velasco}, {Sandoval}, {Santander},
  {Sako}, {Sako}, {Satalecka}, {Schoorlemmer}, {Sch{\"u}ssler},
  {Seglar-Arroyo}, {Smith}, {Spencer}, {Surajbali}, {Tabachnick}, {Taylor},
  {Tibolla}, {Torres}, {Vallage}, {Viana}, {Watson}, {Weisgarber}, {Werner},
  {White}, {Wischnewski}, {Yang}, {Zepeda}, and {Zhou}}]{SGSO:WhitePaper}
\bibinfo{author}{A.~{Albert}}, \bibinfo{author}{R.~{Alfaro}},
  \bibinfo{author}{H.~{Ashkar}}, \bibinfo{author}{C.~{Alvarez}},
  \bibinfo{author}{J.~{{\'A}lvarez}}, \bibinfo{author}{J.~C.
  {Arteaga-Vel{\'a}zquez}}, \bibinfo{author}{H.~A. {Ayala Solares}},
  \bibinfo{author}{R.~{Arceo}}, \bibinfo{author}{J.~A. {Bellido}},
  \bibinfo{author}{S.~{BenZvi}}, \bibinfo{author}{T.~{Bretz}},
  \bibinfo{author}{C.~A. {Brisbois}}, \bibinfo{author}{A.~M. {Brown}},
  \bibinfo{author}{F.~{Brun}}, \bibinfo{author}{K.~S. {Caballero-Mora}},
  \bibinfo{author}{A.~{Carosi}}, \bibinfo{author}{A.~{Carrami{\~n}ana}},
  \bibinfo{author}{S.~{Casanova}}, \bibinfo{author}{P.~M. {Chadwick}},
  \bibinfo{author}{G.~{Cotter}}, \bibinfo{author}{S.~{Couti{\~n}o De
  Le{\'o}n}}, \bibinfo{author}{P.~{Cristofari}}, \bibinfo{author}{S.~{Dasso}},
  \bibinfo{author}{E.~{de la Fuente}}, \bibinfo{author}{B.~L. {Dingus}},
  \bibinfo{author}{P.~{Desiati}}, \bibinfo{author}{F.~d.~O. {Salles}},
  \bibinfo{author}{V.~{de Souza}}, \bibinfo{author}{D.~{Dorner}},
  \bibinfo{author}{J.~C. {D{\'\i}az-V{\'e}lez}}, \bibinfo{author}{J.~A.
  {Garc{\'\i}a-Gonz{\'a}lez}}, \bibinfo{author}{M.~A. {DuVernois}},
  \bibinfo{author}{G.~{Di Sciascio}}, \bibinfo{author}{K.~{Engel}},
  \bibinfo{author}{H.~{Fleischhack}}, \bibinfo{author}{N.~{Fraija}},
  \bibinfo{author}{S.~{Funk}}, \bibinfo{author}{J.-F. {Glicenstein}},
  \bibinfo{author}{J.~{Gonzalez}}, \bibinfo{author}{M.~M. {Gonz{\'a}lez}},
  \bibinfo{author}{J.~A. {Goodman}}, \bibinfo{author}{J.~P. {Harding}},
  \bibinfo{author}{A.~{Haungs}}, \bibinfo{author}{J.~{Hinton}},
  \bibinfo{author}{B.~{Hona}}, \bibinfo{author}{D.~{Hoyos}},
  \bibinfo{author}{P.~{Huentemeyer}}, \bibinfo{author}{A.~{Iriarte}},
  \bibinfo{author}{A.~{Jardin-Blicq}}, \bibinfo{author}{V.~{Joshi}},
  \bibinfo{author}{S.~{Kaufmann}}, \bibinfo{author}{K.~{Kawata}},
  \bibinfo{author}{S.~{Kunwar}}, \bibinfo{author}{J.~{Lefaucheur}},
  \bibinfo{author}{J.~P. {Lenain}}, \bibinfo{author}{K.~{Link}},
  \bibinfo{author}{R.~{L{\'o}pez-Coto}}, \bibinfo{author}{V.~{Marandon}},
  \bibinfo{author}{M.~{Mariotti}}, \bibinfo{author}{J.~{Mart{\'\i}nez-Castro}},
  \bibinfo{author}{H.~{Mart{\'\i}nez-Huerta}},
  \bibinfo{author}{M.~{Mostaf{\'a}}}, \bibinfo{author}{A.~{Nayerhoda}},
  \bibinfo{author}{L.~{Nellen}}, \bibinfo{author}{E.~{de O{\~n}a Wilhelmi}},
  \bibinfo{author}{R.~D. {Parsons}}, \bibinfo{author}{B.~{Patricelli}},
  \bibinfo{author}{A.~{Pichel}}, \bibinfo{author}{Q.~{Piel}},
  \bibinfo{author}{E.~{Prandini}}, \bibinfo{author}{E.~{Pueschel}},
  \bibinfo{author}{S.~{Procureur}}, \bibinfo{author}{A.~{Reisenegger}},
  \bibinfo{author}{C.~{Rivi{\`e}re}}, \bibinfo{author}{J.~{Rodriguez}},
  \bibinfo{author}{A.~C. {Rovero}}, \bibinfo{author}{G.~{Rowell}},
  \bibinfo{author}{E.~L. {Ruiz-Velasco}}, \bibinfo{author}{A.~{Sandoval}},
  \bibinfo{author}{M.~{Santander}}, \bibinfo{author}{T.~{Sako}},
  \bibinfo{author}{T.~K. {Sako}}, \bibinfo{author}{K.~{Satalecka}},
  \bibinfo{author}{H.~{Schoorlemmer}}, \bibinfo{author}{F.~{Sch{\"u}ssler}},
  \bibinfo{author}{M.~{Seglar-Arroyo}}, \bibinfo{author}{A.~J. {Smith}},
  \bibinfo{author}{S.~{Spencer}}, \bibinfo{author}{P.~{Surajbali}},
  \bibinfo{author}{E.~{Tabachnick}}, \bibinfo{author}{A.~M. {Taylor}},
  \bibinfo{author}{O.~{Tibolla}}, \bibinfo{author}{I.~{Torres}},
  \bibinfo{author}{B.~{Vallage}}, \bibinfo{author}{A.~{Viana}},
  \bibinfo{author}{J.~J. {Watson}}, \bibinfo{author}{T.~{Weisgarber}},
  \bibinfo{author}{F.~{Werner}}, \bibinfo{author}{R.~{White}},
  \bibinfo{author}{R.~{Wischnewski}}, \bibinfo{author}{R.~{Yang}},
  \bibinfo{author}{A.~{Zepeda}}, \bibinfo{author}{H.~{Zhou}},
\newblock \bibinfo{title}{{Science Case for a Wide Field-of-View
  Very-High-Energy Gamma-Ray Observatory in the Southern Hemisphere}},
\newblock \bibinfo{journal}{arXiv e-prints}  (\bibinfo{year}{2019})
  \bibinfo{pages}{arXiv:1902.08429}.
  \href{http://arxiv.org/abs/1902.08429}{{\tt arXiv:1902.08429}}.
%Type = Inproceedings
\bibitem[{Breton et~al.(2014)Breton, Delagnes, Maalmi, and Rusquart}]{7097545}
\bibinfo{author}{D.~Breton}, \bibinfo{author}{E.~Delagnes},
  \bibinfo{author}{J.~Maalmi}, \bibinfo{author}{P.~Rusquart},
\newblock \bibinfo{title}{The wavecatcher family of sca-based 12-bit 3.2-gs/s
  fast digitizers},
\newblock in: \bibinfo{booktitle}{2014 19th IEEE-NPSS Real Time Conference},
  \bibinfo{year}{2014}, pp. \bibinfo{pages}{1--8}.
  \DOIprefix\doi{10.1109/RTC.2014.7097545}.
%Type = Article
\bibitem[{Brun and Rademakers(1997)}]{Brun:1997pa}
\bibinfo{author}{R.~Brun}, \bibinfo{author}{F.~Rademakers},
\newblock \bibinfo{title}{{ROOT: An object oriented data analysis framework}},
\newblock \bibinfo{journal}{Nucl. Instrum. Meth. A} \bibinfo{volume}{389}
  (\bibinfo{year}{1997}) \bibinfo{pages}{81--86}.
  \DOIprefix\doi{10.1016/S0168-9002(97)00048-X}.
%Type = Article
\bibitem[{Bellenot et~al.(2015)Bellenot, Canal, Couet, Ganis, Mato, Moneta,
  Naumann, and Piparo}]{Bellenot_2015}
\bibinfo{author}{B.~Bellenot}, \bibinfo{author}{P.~Canal},
  \bibinfo{author}{O.~Couet}, \bibinfo{author}{G.~Ganis},
  \bibinfo{author}{P.~Mato}, \bibinfo{author}{L.~Moneta},
  \bibinfo{author}{A.~Naumann}, \bibinfo{author}{D.~Piparo},
\newblock \bibinfo{title}{Root 6 and beyond: Tobject, c++14 and many cores.},
\newblock \bibinfo{journal}{Journal of Physics: Conference Series}
  \bibinfo{volume}{664} (\bibinfo{year}{2015}) \bibinfo{pages}{062006}.
  \URLprefix \url{https://dx.doi.org/10.1088/1742-6596/664/6/062006}.
  \DOIprefix\doi{10.1088/1742-6596/664/6/062006}.
%Type = Article
\bibitem[{Bernl\"ohr et~al.(2013)Bernl\"ohr, Barnacka, Becherini, {Blanch
  Bigas}, Carmona, Colin, Decerprit, {Di Pierro}, Dubois, Farnier, Funk,
  Hermann, Hinton, Humensky, Khélifi, Kihm, Komin, Lenain, Maier, Mazin,
  Medina, Moralejo, Nolan, Ohm, {de Oña Wilhelmi}, Parsons, {Paz Arribas},
  Pedaletti, Pita, Prokoph, Rulten, Schwanke, Shayduk, Stamatescu, Vallania,
  Vorobiov, Wischnewski, Yoshikoshi, and Zech}]{CTA_MC_KB_2013}
\bibinfo{author}{K.~Bernl\"ohr}, \bibinfo{author}{A.~Barnacka},
  \bibinfo{author}{Y.~Becherini}, \bibinfo{author}{O.~{Blanch Bigas}},
  \bibinfo{author}{E.~Carmona}, \bibinfo{author}{P.~Colin},
  \bibinfo{author}{G.~Decerprit}, \bibinfo{author}{F.~{Di Pierro}},
  \bibinfo{author}{F.~Dubois}, \bibinfo{author}{C.~Farnier},
  \bibinfo{author}{S.~Funk}, \bibinfo{author}{G.~Hermann},
  \bibinfo{author}{J.~Hinton}, \bibinfo{author}{T.~Humensky},
  \bibinfo{author}{B.~Khélifi}, \bibinfo{author}{T.~Kihm},
  \bibinfo{author}{N.~Komin}, \bibinfo{author}{J.-P. Lenain},
  \bibinfo{author}{G.~Maier}, \bibinfo{author}{D.~Mazin},
  \bibinfo{author}{M.~Medina}, \bibinfo{author}{A.~Moralejo},
  \bibinfo{author}{S.~Nolan}, \bibinfo{author}{S.~Ohm}, \bibinfo{author}{E.~{de
  Oña Wilhelmi}}, \bibinfo{author}{R.~Parsons}, \bibinfo{author}{M.~{Paz
  Arribas}}, \bibinfo{author}{G.~Pedaletti}, \bibinfo{author}{S.~Pita},
  \bibinfo{author}{H.~Prokoph}, \bibinfo{author}{C.~Rulten},
  \bibinfo{author}{U.~Schwanke}, \bibinfo{author}{M.~Shayduk},
  \bibinfo{author}{V.~Stamatescu}, \bibinfo{author}{P.~Vallania},
  \bibinfo{author}{S.~Vorobiov}, \bibinfo{author}{R.~Wischnewski},
  \bibinfo{author}{T.~Yoshikoshi}, \bibinfo{author}{A.~Zech},
\newblock \bibinfo{title}{"monte carlo design studies for the cherenkov
  telescope array"},
\newblock \bibinfo{journal}{APh} \bibinfo{volume}{43} (\bibinfo{year}{2013})
  \bibinfo{pages}{171--188}. \URLprefix
  \url{https://www.sciencedirect.com/science/article/pii/S0927650512001867}.
  \DOIprefix\doi{https://doi.org/10.1016/j.astropartphys.2012.10.002},
  \bibinfo{note}{seeing the High-Energy Universe with the Cherenkov Telescope
  Array - The Science Explored with the CTA}.
%Type = Book
\bibitem[{{Heck} et~al.(1998){Heck}, {Knapp}, {Capdevielle}, {Schatz}, and
  {Thouw}}]{corsika}
\bibinfo{author}{D.~{Heck}}, \bibinfo{author}{J.~{Knapp}},
  \bibinfo{author}{J.~N. {Capdevielle}}, \bibinfo{author}{G.~{Schatz}},
  \bibinfo{author}{T.~{Thouw}}, \bibinfo{title}{{CORSIKA: a Monte Carlo code to
  simulate extensive air showers.}}, \bibinfo{publisher}{Forschungszentrum
  Karlsruhe}, \bibinfo{year}{1998}.
%Type = Article
\bibitem[{Agostinelli et~al.(2003)}]{geant4}
\bibinfo{author}{S.~Agostinelli}, et~al. (\bibinfo{collaboration}{GEANT4}),
\newblock \bibinfo{title}{{GEANT4--a simulation toolkit}},
\newblock \bibinfo{journal}{NIM A} \bibinfo{volume}{506} (\bibinfo{year}{2003})
  \bibinfo{pages}{250--303}. \DOIprefix\doi{10.1016/S0168-9002(03)01368-8}.
%Type = Article
\bibitem[{Kamata and Nishimura(1958)}]{nkg_1}
\bibinfo{author}{K.~Kamata}, \bibinfo{author}{J.~Nishimura},
\newblock \bibinfo{title}{{The Lateral and the Angular Structure Functions of
  Electron Showers}},
\newblock \bibinfo{journal}{Progress of Theoretical Physics Supplement}
  \bibinfo{volume}{6} (\bibinfo{year}{1958}) \bibinfo{pages}{93--155}.
  \URLprefix \url{https://doi.org/10.1143/PTPS.6.93}.
  \DOIprefix\doi{10.1143/PTPS.6.93}.
%Type = Article
\bibitem[{Greisen(1960)}]{nkg_2}
\bibinfo{author}{K.~Greisen},
\newblock \bibinfo{title}{Cosmic ray showers},
\newblock \bibinfo{journal}{Annual Review of Nuclear Science}
  \bibinfo{volume}{10} (\bibinfo{year}{1960}) \bibinfo{pages}{63--108}.
  \URLprefix \url{https://doi.org/10.1146/annurev.ns.10.120160.000431}.
  \DOIprefix\doi{10.1146/annurev.ns.10.120160.000431}.
%Type = Article
\bibitem[{{Corstanje} et~al.(2015){Corstanje}, {Schellart}, {Nelles},
  {Buitink}, {Enriquez}, {Falcke}, {Frieswijk}, {H{\"o}randel}, {Krause},
  {Rachen}, {Scholten}, {ter Veen}, {Thoudam}, {Trinh}, {van den Akker},
  {Alexov}, {Anderson}, {Avruch}, {Bell}, {Bentum}, {Bernardi}, {Best},
  {Bonafede}, {Breitling}, {Broderick}, {Br{\"u}ggen}, {Butcher}, {Ciardi}, {de
  Gasperin}, {de Geus}, {de Vos}, {Duscha}, {Eisl{\"o}ffel}, {Engels},
  {Fallows}, {Ferrari}, {Garrett}, {Grie{\ss}meier}, {Gunst}, {Hamaker},
  {Hoeft}, {Horneffer}, {Iacobelli}, {Juette}, {Karastergiou}, {Kohler},
  {Kondratiev}, {Kuniyoshi}, {Kuper}, {Maat}, {Mann}, {McFadden},
  {McKay-Bukowski}, {Mevius}, {Munk}, {Norden}, {Orru}, {Paas},
  {Pandey-Pommier}, {Pandey}, {Pizzo}, {Polatidis}, {Reich}, {R{\"o}ttgering},
  {Scaife}, {Schwarz}, {Smirnov}, {Stewart}, {Steinmetz}, {Swinbank}, {Tagger},
  {Tang}, {Tasse}, {Toribio}, {Vermeulen}, {Vocks}, {van Weeren}, {Wijnholds},
  {Wucknitz}, {Yatawatta}, and {Zarka}}]{hyper_1}
\bibinfo{author}{A.~{Corstanje}}, \bibinfo{author}{P.~{Schellart}},
  \bibinfo{author}{A.~{Nelles}}, \bibinfo{author}{S.~{Buitink}},
  \bibinfo{author}{J.~E. {Enriquez}}, \bibinfo{author}{H.~{Falcke}},
  \bibinfo{author}{W.~{Frieswijk}}, \bibinfo{author}{J.~R. {H{\"o}randel}},
  \bibinfo{author}{M.~{Krause}}, \bibinfo{author}{J.~P. {Rachen}},
  \bibinfo{author}{O.~{Scholten}}, \bibinfo{author}{S.~{ter Veen}},
  \bibinfo{author}{S.~{Thoudam}}, \bibinfo{author}{T.~N.~G. {Trinh}},
  \bibinfo{author}{M.~{van den Akker}}, \bibinfo{author}{A.~{Alexov}},
  \bibinfo{author}{J.~{Anderson}}, \bibinfo{author}{I.~M. {Avruch}},
  \bibinfo{author}{M.~E. {Bell}}, \bibinfo{author}{M.~J. {Bentum}},
  \bibinfo{author}{G.~{Bernardi}}, \bibinfo{author}{P.~{Best}},
  \bibinfo{author}{A.~{Bonafede}}, \bibinfo{author}{F.~{Breitling}},
  \bibinfo{author}{J.~{Broderick}}, \bibinfo{author}{M.~{Br{\"u}ggen}},
  \bibinfo{author}{H.~R. {Butcher}}, \bibinfo{author}{B.~{Ciardi}},
  \bibinfo{author}{F.~{de Gasperin}}, \bibinfo{author}{E.~{de Geus}},
  \bibinfo{author}{M.~{de Vos}}, \bibinfo{author}{S.~{Duscha}},
  \bibinfo{author}{J.~{Eisl{\"o}ffel}}, \bibinfo{author}{D.~{Engels}},
  \bibinfo{author}{R.~A. {Fallows}}, \bibinfo{author}{C.~{Ferrari}},
  \bibinfo{author}{M.~A. {Garrett}}, \bibinfo{author}{J.~{Grie{\ss}meier}},
  \bibinfo{author}{A.~W. {Gunst}}, \bibinfo{author}{J.~P. {Hamaker}},
  \bibinfo{author}{M.~{Hoeft}}, \bibinfo{author}{A.~{Horneffer}},
  \bibinfo{author}{M.~{Iacobelli}}, \bibinfo{author}{E.~{Juette}},
  \bibinfo{author}{A.~{Karastergiou}}, \bibinfo{author}{J.~{Kohler}},
  \bibinfo{author}{V.~I. {Kondratiev}}, \bibinfo{author}{M.~{Kuniyoshi}},
  \bibinfo{author}{G.~{Kuper}}, \bibinfo{author}{P.~{Maat}},
  \bibinfo{author}{G.~{Mann}}, \bibinfo{author}{R.~{McFadden}},
  \bibinfo{author}{D.~{McKay-Bukowski}}, \bibinfo{author}{M.~{Mevius}},
  \bibinfo{author}{H.~{Munk}}, \bibinfo{author}{M.~J. {Norden}},
  \bibinfo{author}{E.~{Orru}}, \bibinfo{author}{H.~{Paas}},
  \bibinfo{author}{M.~{Pandey-Pommier}}, \bibinfo{author}{V.~N. {Pandey}},
  \bibinfo{author}{R.~{Pizzo}}, \bibinfo{author}{A.~G. {Polatidis}},
  \bibinfo{author}{W.~{Reich}}, \bibinfo{author}{H.~{R{\"o}ttgering}},
  \bibinfo{author}{A.~M.~M. {Scaife}}, \bibinfo{author}{D.~{Schwarz}},
  \bibinfo{author}{O.~{Smirnov}}, \bibinfo{author}{A.~{Stewart}},
  \bibinfo{author}{M.~{Steinmetz}}, \bibinfo{author}{J.~{Swinbank}},
  \bibinfo{author}{M.~{Tagger}}, \bibinfo{author}{Y.~{Tang}},
  \bibinfo{author}{C.~{Tasse}}, \bibinfo{author}{C.~{Toribio}},
  \bibinfo{author}{R.~{Vermeulen}}, \bibinfo{author}{C.~{Vocks}},
  \bibinfo{author}{R.~J. {van Weeren}}, \bibinfo{author}{S.~J. {Wijnholds}},
  \bibinfo{author}{O.~{Wucknitz}}, \bibinfo{author}{S.~{Yatawatta}},
  \bibinfo{author}{P.~{Zarka}},
\newblock \bibinfo{title}{{The shape of the radio wavefront of extensive air
  showers as measured with LOFAR}},
\newblock \bibinfo{journal}{APh} \bibinfo{volume}{61} (\bibinfo{year}{2015})
  \bibinfo{pages}{22--31}. \DOIprefix\doi{10.1016/j.astropartphys.2014.06.001}.
  \href{http://arxiv.org/abs/1404.3907}{{\tt arXiv:1404.3907}}.
%Type = Article
\bibitem[{{Apel} et~al.(2014){Apel}, {Arteaga-Vel{\'a}zquez}, {B{\"a}hren},
  {Bekk}, {Bertaina}, {Biermann}, {Bl{\"u}mer}, {Bozdog}, {Brancus}, {Cantoni},
  {Chiavassa}, {Daumiller}, {de Souza}, {Di Pierro}, {Doll}, {Engel}, {Falcke},
  {Fuchs}, {Gemmeke}, {Grupen}, {Haungs}, {Heck}, {H{\"o}randel}, {Horneffer},
  {Huber}, {Huege}, {Isar}, {Kampert}, {Kang}, {Kr{\"o}mer}, {Kuijpers},
  {Link}, {{\L}uczak}, {Ludwig}, {Mathes}, {Melissas}, {Morello},
  {Oehlschl{\"a}ger}, {Palmieri}, {Pierog}, {Rautenberg}, {Rebel}, {Roth},
  {R{\"u}hle}, {Saftoiu}, {Schieler}, {Schmidt}, {Schoo}, {Schr{\"o}der},
  {Sima}, {Toma}, {Trinchero}, {Weindl}, {Wochele}, {Zabierowski}, and
  {Zensus}}]{hyper_2}
\bibinfo{author}{W.~D. {Apel}}, \bibinfo{author}{J.~C.
  {Arteaga-Vel{\'a}zquez}}, \bibinfo{author}{L.~{B{\"a}hren}},
  \bibinfo{author}{K.~{Bekk}}, \bibinfo{author}{M.~{Bertaina}},
  \bibinfo{author}{P.~L. {Biermann}}, \bibinfo{author}{J.~{Bl{\"u}mer}},
  \bibinfo{author}{H.~{Bozdog}}, \bibinfo{author}{I.~M. {Brancus}},
  \bibinfo{author}{E.~{Cantoni}}, \bibinfo{author}{A.~{Chiavassa}},
  \bibinfo{author}{K.~{Daumiller}}, \bibinfo{author}{V.~{de Souza}},
  \bibinfo{author}{F.~{Di Pierro}}, \bibinfo{author}{P.~{Doll}},
  \bibinfo{author}{R.~{Engel}}, \bibinfo{author}{H.~{Falcke}},
  \bibinfo{author}{B.~{Fuchs}}, \bibinfo{author}{H.~{Gemmeke}},
  \bibinfo{author}{C.~{Grupen}}, \bibinfo{author}{A.~{Haungs}},
  \bibinfo{author}{D.~{Heck}}, \bibinfo{author}{J.~R. {H{\"o}randel}},
  \bibinfo{author}{A.~{Horneffer}}, \bibinfo{author}{D.~{Huber}},
  \bibinfo{author}{T.~{Huege}}, \bibinfo{author}{P.~G. {Isar}},
  \bibinfo{author}{K.~H. {Kampert}}, \bibinfo{author}{D.~{Kang}},
  \bibinfo{author}{O.~{Kr{\"o}mer}}, \bibinfo{author}{J.~{Kuijpers}},
  \bibinfo{author}{K.~{Link}}, \bibinfo{author}{P.~{{\L}uczak}},
  \bibinfo{author}{M.~{Ludwig}}, \bibinfo{author}{H.~J. {Mathes}},
  \bibinfo{author}{M.~{Melissas}}, \bibinfo{author}{C.~{Morello}},
  \bibinfo{author}{J.~{Oehlschl{\"a}ger}}, \bibinfo{author}{N.~{Palmieri}},
  \bibinfo{author}{T.~{Pierog}}, \bibinfo{author}{J.~{Rautenberg}},
  \bibinfo{author}{H.~{Rebel}}, \bibinfo{author}{M.~{Roth}},
  \bibinfo{author}{C.~{R{\"u}hle}}, \bibinfo{author}{A.~{Saftoiu}},
  \bibinfo{author}{H.~{Schieler}}, \bibinfo{author}{A.~{Schmidt}},
  \bibinfo{author}{S.~{Schoo}}, \bibinfo{author}{F.~G. {Schr{\"o}der}},
  \bibinfo{author}{O.~{Sima}}, \bibinfo{author}{G.~{Toma}},
  \bibinfo{author}{G.~C. {Trinchero}}, \bibinfo{author}{A.~{Weindl}},
  \bibinfo{author}{J.~{Wochele}}, \bibinfo{author}{J.~{Zabierowski}},
  \bibinfo{author}{J.~A. {Zensus}},
\newblock \bibinfo{title}{{The wavefront of the radio signal emitted by cosmic
  ray air showers}},
\newblock \bibinfo{journal}{\jcap} \bibinfo{volume}{2014}
  (\bibinfo{year}{2014}) \bibinfo{pages}{025--025}.
  \DOIprefix\doi{10.1088/1475-7516/2014/09/025}.
  \href{http://arxiv.org/abs/1404.3283}{{\tt arXiv:1404.3283}}.
%Type = Article
\bibitem[{{Aleksi{\'c}} et~al.(2015){Aleksi{\'c}}, {Ansoldi}, {Antonelli},
  {Antoranz}, {Babic}, {Bangale}, {Barrio}, {Becerra Gonz{\'a}lez}, {Bednarek},
  {Bernardini}, {Biasuzzi}, {Biland}, {Blanch}, {Bonnefoy}, {Bonnoli},
  {Borracci}, {Bretz}, {Carmona}, {Carosi}, {Colin}, {Colombo}, {Contreras},
  {Cortina}, {Covino}, {Da Vela}, {Dazzi}, {De Angelis}, {De Caneva}, {De
  Lotto}, {de O{\~n}a Wilhelmi}, {Delgado Mendez}, {Doert}, {Dominis Prester},
  {Dorner}, {Doro}, {Einecke}, {Eisenacher}, {Elsaesser}, {Fonseca}, {Font},
  {Frantzen}, {Fruck}, {Galindo}, {Garc{\'\i}a L{\'o}pez}, {Garczarczyk},
  {Garrido Terrats}, {Gaug}, {Godinovi{\'c}}, {Gonz{\'a}lez Mu{\~n}oz},
  {Gozzini}, {Hadasch}, {Hanabata}, {Hayashida}, {Herrera}, {Hildebrand},
  {Hose}, {Hrupec}, {Idec}, {Kadenius}, {Kellermann}, {Kodani}, {Konno},
  {Krause}, {Kubo}, {Kushida}, {La Barbera}, {Lelas}, {Lewandowska},
  {Lindfors}, {Lombardi}, {L{\'o}pez}, {L{\'o}pez-Coto}, {L{\'o}pez-Oramas},
  {Lorenz}, {Lozano}, {Makariev}, {Mallot}, {Maneva}, {Mankuzhiyil},
  {Mannheim}, {Maraschi}, {Marcote}, {Mariotti}, {Mart{\'\i}nez}, {Mazin},
  {Menzel}, {Miranda}, {Mirzoyan}, {Moralejo}, {Munar-Adrover}, {Nakajima},
  {Niedzwiecki}, {Nilsson}, {Nishijima}, {Noda}, {Nowak}, {Orito},
  {Overkemping}, {Paiano}, {Palatiello}, {Paneque}, {Paoletti}, {Paredes},
  {Paredes-Fortuny}, {Persic}, {Prada Moroni}, {Prandini}, {Preziuso},
  {Puljak}, {Reinthal}, {Rhode}, {Rib{\'o}}, {Rico}, {Rodriguez Garcia},
  {R{\"u}gamer}, {Saggion}, {Saito}, {Saito}, {Satalecka}, {Scalzotto},
  {Scapin}, {Schultz}, {Schweizer}, {Shore}, {Sillanp{\"a}{\"a}}, {Sitarek},
  {Snidaric}, {Sobczynska}, {Spanier}, {Stamatescu}, {Stamerra}, {Steinbring},
  {Storz}, {Strzys}, {Takalo}, {Takami}, {Tavecchio}, {Temnikov}, {Terzi{\'c}},
  {Tescaro}, {Teshima}, {Thaele}, {Tibolla}, {Torres}, {Toyama}, {Treves},
  {Uellenbeck}, {Vogler}, {Wagner}, {Zanin}, {Horns}, {Mart{\'\i}n}, and
  {Meyer}}]{MAGIC_Crab}
\bibinfo{author}{J.~{Aleksi{\'c}}}, \bibinfo{author}{S.~{Ansoldi}},
  \bibinfo{author}{L.~A. {Antonelli}}, \bibinfo{author}{P.~{Antoranz}},
  \bibinfo{author}{A.~{Babic}}, \bibinfo{author}{P.~{Bangale}},
  \bibinfo{author}{J.~A. {Barrio}}, \bibinfo{author}{J.~{Becerra
  Gonz{\'a}lez}}, \bibinfo{author}{W.~{Bednarek}},
  \bibinfo{author}{E.~{Bernardini}}, \bibinfo{author}{B.~{Biasuzzi}},
  \bibinfo{author}{A.~{Biland}}, \bibinfo{author}{O.~{Blanch}},
  \bibinfo{author}{S.~{Bonnefoy}}, \bibinfo{author}{G.~{Bonnoli}},
  \bibinfo{author}{F.~{Borracci}}, \bibinfo{author}{T.~{Bretz}},
  \bibinfo{author}{E.~{Carmona}}, \bibinfo{author}{A.~{Carosi}},
  \bibinfo{author}{P.~{Colin}}, \bibinfo{author}{E.~{Colombo}},
  \bibinfo{author}{J.~L. {Contreras}}, \bibinfo{author}{J.~{Cortina}},
  \bibinfo{author}{S.~{Covino}}, \bibinfo{author}{P.~{Da Vela}},
  \bibinfo{author}{F.~{Dazzi}}, \bibinfo{author}{A.~{De Angelis}},
  \bibinfo{author}{G.~{De Caneva}}, \bibinfo{author}{B.~{De Lotto}},
  \bibinfo{author}{E.~{de O{\~n}a Wilhelmi}}, \bibinfo{author}{C.~{Delgado
  Mendez}}, \bibinfo{author}{M.~{Doert}}, \bibinfo{author}{D.~{Dominis
  Prester}}, \bibinfo{author}{D.~{Dorner}}, \bibinfo{author}{M.~{Doro}},
  \bibinfo{author}{S.~{Einecke}}, \bibinfo{author}{D.~{Eisenacher}},
  \bibinfo{author}{D.~{Elsaesser}}, \bibinfo{author}{M.~V. {Fonseca}},
  \bibinfo{author}{L.~{Font}}, \bibinfo{author}{K.~{Frantzen}},
  \bibinfo{author}{C.~{Fruck}}, \bibinfo{author}{D.~{Galindo}},
  \bibinfo{author}{R.~J. {Garc{\'\i}a L{\'o}pez}},
  \bibinfo{author}{M.~{Garczarczyk}}, \bibinfo{author}{D.~{Garrido Terrats}},
  \bibinfo{author}{M.~{Gaug}}, \bibinfo{author}{N.~{Godinovi{\'c}}},
  \bibinfo{author}{A.~{Gonz{\'a}lez Mu{\~n}oz}}, \bibinfo{author}{S.~R.
  {Gozzini}}, \bibinfo{author}{D.~{Hadasch}}, \bibinfo{author}{Y.~{Hanabata}},
  \bibinfo{author}{M.~{Hayashida}}, \bibinfo{author}{J.~{Herrera}},
  \bibinfo{author}{D.~{Hildebrand}}, \bibinfo{author}{J.~{Hose}},
  \bibinfo{author}{D.~{Hrupec}}, \bibinfo{author}{W.~{Idec}},
  \bibinfo{author}{V.~{Kadenius}}, \bibinfo{author}{H.~{Kellermann}},
  \bibinfo{author}{K.~{Kodani}}, \bibinfo{author}{Y.~{Konno}},
  \bibinfo{author}{J.~{Krause}}, \bibinfo{author}{H.~{Kubo}},
  \bibinfo{author}{J.~{Kushida}}, \bibinfo{author}{A.~{La Barbera}},
  \bibinfo{author}{D.~{Lelas}}, \bibinfo{author}{N.~{Lewandowska}},
  \bibinfo{author}{E.~{Lindfors}}, \bibinfo{author}{S.~{Lombardi}},
  \bibinfo{author}{M.~{L{\'o}pez}}, \bibinfo{author}{R.~{L{\'o}pez-Coto}},
  \bibinfo{author}{A.~{L{\'o}pez-Oramas}}, \bibinfo{author}{E.~{Lorenz}},
  \bibinfo{author}{I.~{Lozano}}, \bibinfo{author}{M.~{Makariev}},
  \bibinfo{author}{K.~{Mallot}}, \bibinfo{author}{G.~{Maneva}},
  \bibinfo{author}{N.~{Mankuzhiyil}}, \bibinfo{author}{K.~{Mannheim}},
  \bibinfo{author}{L.~{Maraschi}}, \bibinfo{author}{B.~{Marcote}},
  \bibinfo{author}{M.~{Mariotti}}, \bibinfo{author}{M.~{Mart{\'\i}nez}},
  \bibinfo{author}{D.~{Mazin}}, \bibinfo{author}{U.~{Menzel}},
  \bibinfo{author}{J.~M. {Miranda}}, \bibinfo{author}{R.~{Mirzoyan}},
  \bibinfo{author}{A.~{Moralejo}}, \bibinfo{author}{P.~{Munar-Adrover}},
  \bibinfo{author}{D.~{Nakajima}}, \bibinfo{author}{A.~{Niedzwiecki}},
  \bibinfo{author}{K.~{Nilsson}}, \bibinfo{author}{K.~{Nishijima}},
  \bibinfo{author}{K.~{Noda}}, \bibinfo{author}{N.~{Nowak}},
  \bibinfo{author}{R.~{Orito}}, \bibinfo{author}{A.~{Overkemping}},
  \bibinfo{author}{S.~{Paiano}}, \bibinfo{author}{M.~{Palatiello}},
  \bibinfo{author}{D.~{Paneque}}, \bibinfo{author}{R.~{Paoletti}},
  \bibinfo{author}{J.~M. {Paredes}}, \bibinfo{author}{X.~{Paredes-Fortuny}},
  \bibinfo{author}{M.~{Persic}}, \bibinfo{author}{P.~G. {Prada Moroni}},
  \bibinfo{author}{E.~{Prandini}}, \bibinfo{author}{S.~{Preziuso}},
  \bibinfo{author}{I.~{Puljak}}, \bibinfo{author}{R.~{Reinthal}},
  \bibinfo{author}{W.~{Rhode}}, \bibinfo{author}{M.~{Rib{\'o}}},
  \bibinfo{author}{J.~{Rico}}, \bibinfo{author}{J.~{Rodriguez Garcia}},
  \bibinfo{author}{S.~{R{\"u}gamer}}, \bibinfo{author}{A.~{Saggion}},
  \bibinfo{author}{T.~{Saito}}, \bibinfo{author}{K.~{Saito}},
  \bibinfo{author}{K.~{Satalecka}}, \bibinfo{author}{V.~{Scalzotto}},
  \bibinfo{author}{V.~{Scapin}}, \bibinfo{author}{C.~{Schultz}},
  \bibinfo{author}{T.~{Schweizer}}, \bibinfo{author}{S.~N. {Shore}},
  \bibinfo{author}{A.~{Sillanp{\"a}{\"a}}}, \bibinfo{author}{J.~{Sitarek}},
  \bibinfo{author}{I.~{Snidaric}}, \bibinfo{author}{D.~{Sobczynska}},
  \bibinfo{author}{F.~{Spanier}}, \bibinfo{author}{V.~{Stamatescu}},
  \bibinfo{author}{A.~{Stamerra}}, \bibinfo{author}{T.~{Steinbring}},
  \bibinfo{author}{J.~{Storz}}, \bibinfo{author}{M.~{Strzys}},
  \bibinfo{author}{L.~{Takalo}}, \bibinfo{author}{H.~{Takami}},
  \bibinfo{author}{F.~{Tavecchio}}, \bibinfo{author}{P.~{Temnikov}},
  \bibinfo{author}{T.~{Terzi{\'c}}}, \bibinfo{author}{D.~{Tescaro}},
  \bibinfo{author}{M.~{Teshima}}, \bibinfo{author}{J.~{Thaele}},
  \bibinfo{author}{O.~{Tibolla}}, \bibinfo{author}{D.~F. {Torres}},
  \bibinfo{author}{T.~{Toyama}}, \bibinfo{author}{A.~{Treves}},
  \bibinfo{author}{M.~{Uellenbeck}}, \bibinfo{author}{P.~{Vogler}},
  \bibinfo{author}{R.~M. {Wagner}}, \bibinfo{author}{R.~{Zanin}},
  \bibinfo{author}{D.~{Horns}}, \bibinfo{author}{J.~{Mart{\'\i}n}},
  \bibinfo{author}{M.~{Meyer}},
\newblock \bibinfo{title}{{Measurement of the Crab Nebula spectrum over three
  decades in energy with the MAGIC telescopes}},
\newblock \bibinfo{journal}{JHEAp} \bibinfo{volume}{5} (\bibinfo{year}{2015})
  \bibinfo{pages}{30--38}. \DOIprefix\doi{10.1016/j.jheap.2015.01.002}.
  \href{http://arxiv.org/abs/1406.6892}{{\tt arXiv:1406.6892}}.
%Type = Misc
\bibitem[{N\"othe et~al.(2022)N\"othe, Peresano, Sitarek, Vuillaume, Nickel,
  Biederbeck, Jouvin, Verna, and Moralejo}]{pyirf_060}
\bibinfo{author}{M.~N\"othe}, \bibinfo{author}{M.~Peresano},
  \bibinfo{author}{J.~Sitarek}, \bibinfo{author}{T.~Vuillaume},
  \bibinfo{author}{L.~Nickel}, \bibinfo{author}{N.~Biederbeck},
  \bibinfo{author}{L.~Jouvin}, \bibinfo{author}{G.~Verna},
  \bibinfo{author}{A.~Moralejo}, \bibinfo{title}{cta-observatory/pyirf: v0.6.0
  – 2022-01-10}, \bibinfo{year}{2022}. \URLprefix
  \url{https://doi.org/10.5281/zenodo.5833284}.
  \DOIprefix\doi{10.5281/zenodo.5833284}.
%Type = Article
\bibitem[{Abeysekara et~al.(2012)Abeysekara, Aguilar, Aguilar, Alfaro, Almaraz,
  {\'{A}}lvarez, de~D.~{\'{A}}lvarez-Romero, {\'{A}}lvarez, Arceo,
  Arteaga-Vel{\'{a}}zquez, Badillo, Barber, Baughman, Bautista-Elivar, Belmont,
  Ben{\'{\i}}tez, BenZvi, Berley, Bernal, Bonamente, Braun, Caballero-Lopez,
  Cabrera, Carrami{\~{n}}ana, Carrasco, Castillo, Chambers, Conde, Condreay,
  Cotti, Cotzomi, D'Olivo, de~la Fuente, Le{\'{o}}n, Delay, Delepine, DeYoung,
  Diaz, Diaz-Cruz, Dingus, Duvernois, Edmunds, Ellsworth, Fick, Fiorino,
  Flandes, Fraija, Galindo, Garc{\i}{\textasciiacute}a-Luna,
  Garc{\i}{\textasciiacute}a-Torales, Garfias, Gonz{\'{a}}lez, Gonz{\'{a}}lez,
  Goodman, Grabski, Gussert, Guzm{\'{a}}n-Ceron, Hampel-Arias, Harris, Hays,
  Hernandez-Cervantes, Hüntemeyer, Imran, Iriarte, Jimenez, Karn,
  Kelley-Hoskins, Kieda, Langarica, Lara, Lauer, Lee, Linares, Linnemann,
  Longo, Luna-Garc{\'{\i}}a, Mart{\i}{\textasciiacute}nez, Mart{\'{\i}}nez,
  Mart{\i}{\textasciiacute}nez, Mart{\i}{\textasciiacute}nez,
  Mart{\i}{\textasciiacute}nez-Castro, Martos, Matthews, McEnery, Medina-Tanco,
  Mendoza-Torres, Miranda-Romagnoli, Montaruli, Moreno, Mostafa, Napsuciale,
  Nava, Nellen, Newbold, Noriega-Papaqui, Oceguera-Becerra, Tapia, Orozco,
  P{\'{e}}rez, P{\'{e}}rez-P{\'{e}}rez, Perkins, Pretz, Ramirez,
  Ram{\'{\i}}rez, Rebello, Renter{\'{\i}}a, Reyes, Rosa-Gonz{\'{a}}lez, Rosado,
  Ryan, Sacahui, Salazar, Salesa, Sandoval, Santos, Schneider, Shoup, Silich,
  Sinnis, Smith, Sparks, Springer, Su{\'{a}}rez, Suarez, Taboada, Tellez,
  Tenorio-Tagle, Tepe, Toale, Tollefson, Torres, Ukwatta, Valdes-Galicia,
  Vanegas, Vasileiou, V{\'{a}}zquez, V{\'{a}}zquez, Villase{\~{n}}or, Wall,
  Walters, Warner, Westerhoff, Wisher, Wood, Yodh, Zaborov, and
  Zepeda}]{HAWC:GRB_2012}
\bibinfo{author}{A.~Abeysekara}, \bibinfo{author}{J.~Aguilar},
  \bibinfo{author}{S.~Aguilar}, \bibinfo{author}{R.~Alfaro},
  \bibinfo{author}{E.~Almaraz}, \bibinfo{author}{C.~{\'{A}}lvarez},
  \bibinfo{author}{J.~de~D.~{\'{A}}lvarez-Romero},
  \bibinfo{author}{M.~{\'{A}}lvarez}, \bibinfo{author}{R.~Arceo},
  \bibinfo{author}{J.~Arteaga-Vel{\'{a}}zquez}, \bibinfo{author}{C.~Badillo},
  \bibinfo{author}{A.~Barber}, \bibinfo{author}{B.~Baughman},
  \bibinfo{author}{N.~Bautista-Elivar}, \bibinfo{author}{E.~Belmont},
  \bibinfo{author}{E.~Ben{\'{\i}}tez}, \bibinfo{author}{S.~BenZvi},
  \bibinfo{author}{D.~Berley}, \bibinfo{author}{A.~Bernal},
  \bibinfo{author}{E.~Bonamente}, \bibinfo{author}{J.~Braun},
  \bibinfo{author}{R.~Caballero-Lopez}, \bibinfo{author}{I.~Cabrera},
  \bibinfo{author}{A.~Carrami{\~{n}}ana}, \bibinfo{author}{L.~Carrasco},
  \bibinfo{author}{M.~Castillo}, \bibinfo{author}{L.~Chambers},
  \bibinfo{author}{R.~Conde}, \bibinfo{author}{P.~Condreay},
  \bibinfo{author}{U.~Cotti}, \bibinfo{author}{J.~Cotzomi},
  \bibinfo{author}{J.~D'Olivo}, \bibinfo{author}{E.~de~la Fuente},
  \bibinfo{author}{C.~D. Le{\'{o}}n}, \bibinfo{author}{S.~Delay},
  \bibinfo{author}{D.~Delepine}, \bibinfo{author}{T.~DeYoung},
  \bibinfo{author}{L.~Diaz}, \bibinfo{author}{L.~Diaz-Cruz},
  \bibinfo{author}{B.~Dingus}, \bibinfo{author}{M.~Duvernois},
  \bibinfo{author}{D.~Edmunds}, \bibinfo{author}{R.~Ellsworth},
  \bibinfo{author}{B.~Fick}, \bibinfo{author}{D.~Fiorino},
  \bibinfo{author}{A.~Flandes}, \bibinfo{author}{N.~Fraija},
  \bibinfo{author}{A.~Galindo},
  \bibinfo{author}{J.~Garc{\i}{\textasciiacute}a-Luna},
  \bibinfo{author}{G.~Garc{\i}{\textasciiacute}a-Torales},
  \bibinfo{author}{F.~Garfias}, \bibinfo{author}{L.~Gonz{\'{a}}lez},
  \bibinfo{author}{M.~Gonz{\'{a}}lez}, \bibinfo{author}{J.~Goodman},
  \bibinfo{author}{V.~Grabski}, \bibinfo{author}{M.~Gussert},
  \bibinfo{author}{C.~Guzm{\'{a}}n-Ceron}, \bibinfo{author}{Z.~Hampel-Arias},
  \bibinfo{author}{T.~Harris}, \bibinfo{author}{E.~Hays},
  \bibinfo{author}{L.~Hernandez-Cervantes}, \bibinfo{author}{P.~Hüntemeyer},
  \bibinfo{author}{A.~Imran}, \bibinfo{author}{A.~Iriarte},
  \bibinfo{author}{J.~Jimenez}, \bibinfo{author}{P.~Karn},
  \bibinfo{author}{N.~Kelley-Hoskins}, \bibinfo{author}{D.~Kieda},
  \bibinfo{author}{R.~Langarica}, \bibinfo{author}{A.~Lara},
  \bibinfo{author}{R.~Lauer}, \bibinfo{author}{W.~Lee},
  \bibinfo{author}{E.~Linares}, \bibinfo{author}{J.~Linnemann},
  \bibinfo{author}{M.~Longo}, \bibinfo{author}{R.~Luna-Garc{\'{\i}}a},
  \bibinfo{author}{H.~Mart{\i}{\textasciiacute}nez},
  \bibinfo{author}{J.~Mart{\'{\i}}nez},
  \bibinfo{author}{L.~Mart{\i}{\textasciiacute}nez},
  \bibinfo{author}{O.~Mart{\i}{\textasciiacute}nez},
  \bibinfo{author}{J.~Mart{\i}{\textasciiacute}nez-Castro},
  \bibinfo{author}{M.~Martos}, \bibinfo{author}{J.~Matthews},
  \bibinfo{author}{J.~McEnery}, \bibinfo{author}{G.~Medina-Tanco},
  \bibinfo{author}{J.~Mendoza-Torres}, \bibinfo{author}{P.~Miranda-Romagnoli},
  \bibinfo{author}{T.~Montaruli}, \bibinfo{author}{E.~Moreno},
  \bibinfo{author}{M.~Mostafa}, \bibinfo{author}{M.~Napsuciale},
  \bibinfo{author}{J.~Nava}, \bibinfo{author}{L.~Nellen},
  \bibinfo{author}{M.~Newbold}, \bibinfo{author}{R.~Noriega-Papaqui},
  \bibinfo{author}{T.~Oceguera-Becerra}, \bibinfo{author}{A.~O. Tapia},
  \bibinfo{author}{V.~Orozco}, \bibinfo{author}{V.~P{\'{e}}rez},
  \bibinfo{author}{E.~P{\'{e}}rez-P{\'{e}}rez}, \bibinfo{author}{J.~Perkins},
  \bibinfo{author}{J.~Pretz}, \bibinfo{author}{C.~Ramirez},
  \bibinfo{author}{I.~Ram{\'{\i}}rez}, \bibinfo{author}{D.~Rebello},
  \bibinfo{author}{A.~Renter{\'{\i}}a}, \bibinfo{author}{J.~Reyes},
  \bibinfo{author}{D.~Rosa-Gonz{\'{a}}lez}, \bibinfo{author}{A.~Rosado},
  \bibinfo{author}{J.~Ryan}, \bibinfo{author}{J.~Sacahui},
  \bibinfo{author}{H.~Salazar}, \bibinfo{author}{F.~Salesa},
  \bibinfo{author}{A.~Sandoval}, \bibinfo{author}{E.~Santos},
  \bibinfo{author}{M.~Schneider}, \bibinfo{author}{A.~Shoup},
  \bibinfo{author}{S.~Silich}, \bibinfo{author}{G.~Sinnis},
  \bibinfo{author}{A.~Smith}, \bibinfo{author}{K.~Sparks},
  \bibinfo{author}{W.~Springer}, \bibinfo{author}{F.~Su{\'{a}}rez},
  \bibinfo{author}{N.~Suarez}, \bibinfo{author}{I.~Taboada},
  \bibinfo{author}{A.~Tellez}, \bibinfo{author}{G.~Tenorio-Tagle},
  \bibinfo{author}{A.~Tepe}, \bibinfo{author}{P.~Toale},
  \bibinfo{author}{K.~Tollefson}, \bibinfo{author}{I.~Torres},
  \bibinfo{author}{T.~Ukwatta}, \bibinfo{author}{J.~Valdes-Galicia},
  \bibinfo{author}{P.~Vanegas}, \bibinfo{author}{V.~Vasileiou},
  \bibinfo{author}{O.~V{\'{a}}zquez}, \bibinfo{author}{X.~V{\'{a}}zquez},
  \bibinfo{author}{L.~Villase{\~{n}}or}, \bibinfo{author}{W.~Wall},
  \bibinfo{author}{J.~Walters}, \bibinfo{author}{D.~Warner},
  \bibinfo{author}{S.~Westerhoff}, \bibinfo{author}{I.~Wisher},
  \bibinfo{author}{J.~Wood}, \bibinfo{author}{G.~Yodh},
  \bibinfo{author}{D.~Zaborov}, \bibinfo{author}{A.~Zepeda},
\newblock \bibinfo{title}{On the sensitivity of the {HAWC} observatory to
  gamma-ray bursts},
\newblock \bibinfo{journal}{APh} \bibinfo{volume}{35} (\bibinfo{year}{2012})
  \bibinfo{pages}{641--650}. \URLprefix
  \url{https://doi.org/10.1016/j.astropartphys.2012.02.001}.
  \DOIprefix\doi{10.1016/j.astropartphys.2012.02.001}.
%Type = Article
\bibitem[{Abeysekara et~al.(2014)Abeysekara, Alfaro, Alvarez, {\'{A}}lvarez,
  Arceo, Arteaga-Vel{\'{a}}zquez, Solares, Barber, Baughman, Bautista-Elivar,
  Gonzalez, Belmont, BenZvi, Berley, Rosales, Braun, Caballero-Lopez,
  Caballero-Mora, Carrami{\~{n}}ana, Castillo, Cotti, Cotzomi, de~la Fuente,
  Le{\'{o}}n, DeYoung, Hernandez, Diaz-Cruz, D{\'{\i}}az-V{\'{e}}lez, Dingus,
  DuVernois, Ellsworth, Fiorino, Fraija, Galindo, Garfias, Gonz{\'{a}}lez,
  Goodman, Grabski, Gussert, Hampel-Arias, Harding, Hui, Hüntemeyer, Imran,
  Iriarte, Karn, Kieda, Kunde, Lara, Lauer, Lee, Lennarz, Vargas, Linares,
  Linnemann, Longo, Luna-Garcia, Marinelli, Martinez, Martinez,
  Mart{\'{\i}}nez-Castro, Matthews, McEnery, Torres, Miranda-Romagnoli, Moreno,
  Mostaf{\'{a}}, Nellen, Newbold, Noriega-Papaqui, Oceguera-Becerra,
  Patricelli, Pelayo, P{\'{e}}rez-P{\'{e}}rez, Pretz, Rivi{\`{e}}re,
  Rosa-Gonz{\'{a}}lez, Ryan, Salazar, Salesa, Sanchez, Sandoval, Schneider,
  Silich, Sinnis, Smith, Woodle, Springer, Taboada, Toale, Tollefson, Torres,
  Ukwatta, Villase{\~{n}}or, Weisgarber, Westerhoff, Wisher, Wood, Yodh, Younk,
  Zaborov, Zepeda, Zhou, and and}]{HAWC:DM_2014}
\bibinfo{author}{A.~Abeysekara}, \bibinfo{author}{R.~Alfaro},
  \bibinfo{author}{C.~Alvarez}, \bibinfo{author}{J.~{\'{A}}lvarez},
  \bibinfo{author}{R.~Arceo}, \bibinfo{author}{J.~Arteaga-Vel{\'{a}}zquez},
  \bibinfo{author}{H.~A. Solares}, \bibinfo{author}{A.~Barber},
  \bibinfo{author}{B.~Baughman}, \bibinfo{author}{N.~Bautista-Elivar},
  \bibinfo{author}{J.~B. Gonzalez}, \bibinfo{author}{E.~Belmont},
  \bibinfo{author}{S.~BenZvi}, \bibinfo{author}{D.~Berley},
  \bibinfo{author}{M.~B. Rosales}, \bibinfo{author}{J.~Braun},
  \bibinfo{author}{R.~Caballero-Lopez}, \bibinfo{author}{K.~Caballero-Mora},
  \bibinfo{author}{A.~Carrami{\~{n}}ana}, \bibinfo{author}{M.~Castillo},
  \bibinfo{author}{U.~Cotti}, \bibinfo{author}{J.~Cotzomi},
  \bibinfo{author}{E.~de~la Fuente}, \bibinfo{author}{C.~D. Le{\'{o}}n},
  \bibinfo{author}{T.~DeYoung}, \bibinfo{author}{R.~D. Hernandez},
  \bibinfo{author}{L.~Diaz-Cruz}, \bibinfo{author}{J.~D{\'{\i}}az-V{\'{e}}lez},
  \bibinfo{author}{B.~Dingus}, \bibinfo{author}{M.~DuVernois},
  \bibinfo{author}{R.~Ellsworth}, \bibinfo{author}{D.~Fiorino},
  \bibinfo{author}{N.~Fraija}, \bibinfo{author}{A.~Galindo},
  \bibinfo{author}{F.~Garfias}, \bibinfo{author}{M.~Gonz{\'{a}}lez},
  \bibinfo{author}{J.~Goodman}, \bibinfo{author}{V.~Grabski},
  \bibinfo{author}{M.~Gussert}, \bibinfo{author}{Z.~Hampel-Arias},
  \bibinfo{author}{J.~Harding}, \bibinfo{author}{C.~Hui},
  \bibinfo{author}{P.~Hüntemeyer}, \bibinfo{author}{A.~Imran},
  \bibinfo{author}{A.~Iriarte}, \bibinfo{author}{P.~Karn},
  \bibinfo{author}{D.~Kieda}, \bibinfo{author}{G.~Kunde},
  \bibinfo{author}{A.~Lara}, \bibinfo{author}{R.~Lauer},
  \bibinfo{author}{W.~Lee}, \bibinfo{author}{D.~Lennarz},
  \bibinfo{author}{H.~L. Vargas}, \bibinfo{author}{E.~Linares},
  \bibinfo{author}{J.~Linnemann}, \bibinfo{author}{M.~Longo},
  \bibinfo{author}{R.~Luna-Garcia}, \bibinfo{author}{A.~Marinelli},
  \bibinfo{author}{H.~Martinez}, \bibinfo{author}{O.~Martinez},
  \bibinfo{author}{J.~Mart{\'{\i}}nez-Castro}, \bibinfo{author}{J.~Matthews},
  \bibinfo{author}{J.~McEnery}, \bibinfo{author}{E.~M. Torres},
  \bibinfo{author}{P.~Miranda-Romagnoli}, \bibinfo{author}{E.~Moreno},
  \bibinfo{author}{M.~Mostaf{\'{a}}}, \bibinfo{author}{L.~Nellen},
  \bibinfo{author}{M.~Newbold}, \bibinfo{author}{R.~Noriega-Papaqui},
  \bibinfo{author}{T.~Oceguera-Becerra}, \bibinfo{author}{B.~Patricelli},
  \bibinfo{author}{R.~Pelayo}, \bibinfo{author}{E.~P{\'{e}}rez-P{\'{e}}rez},
  \bibinfo{author}{J.~Pretz}, \bibinfo{author}{C.~Rivi{\`{e}}re},
  \bibinfo{author}{D.~Rosa-Gonz{\'{a}}lez}, \bibinfo{author}{J.~Ryan},
  \bibinfo{author}{H.~Salazar}, \bibinfo{author}{F.~Salesa},
  \bibinfo{author}{F.~Sanchez}, \bibinfo{author}{A.~Sandoval},
  \bibinfo{author}{M.~Schneider}, \bibinfo{author}{S.~Silich},
  \bibinfo{author}{G.~Sinnis}, \bibinfo{author}{A.~Smith},
  \bibinfo{author}{K.~S. Woodle}, \bibinfo{author}{R.~Springer},
  \bibinfo{author}{I.~Taboada}, \bibinfo{author}{P.~Toale},
  \bibinfo{author}{K.~Tollefson}, \bibinfo{author}{I.~Torres},
  \bibinfo{author}{T.~Ukwatta}, \bibinfo{author}{L.~Villase{\~{n}}or},
  \bibinfo{author}{T.~Weisgarber}, \bibinfo{author}{S.~Westerhoff},
  \bibinfo{author}{I.~Wisher}, \bibinfo{author}{J.~Wood},
  \bibinfo{author}{G.~Yodh}, \bibinfo{author}{P.~Younk},
  \bibinfo{author}{D.~Zaborov}, \bibinfo{author}{A.~Zepeda},
  \bibinfo{author}{H.~Zhou}, \bibinfo{author}{K.~A. and},
\newblock \bibinfo{title}{Sensitivity of {HAWC} to high-mass dark matter
  annihilations},
\newblock \bibinfo{journal}{\prd} \bibinfo{volume}{90} (\bibinfo{year}{2014}).
  \URLprefix \url{https://doi.org/10.1103/physrevd.90.122002}.
  \DOIprefix\doi{10.1103/physrevd.90.122002}.
%Type = Article
\bibitem[{{LHAASO collaboration}(2021)}]{LHAASO:Performance_crab}
\bibinfo{author}{{LHAASO collaboration}},
\newblock \bibinfo{title}{{Performance of LHAASO-WCDA and Observation of Crab
  Nebula as a Standard Candle}},
\newblock \bibinfo{journal}{arXiv e-prints}  (\bibinfo{year}{2021})
  \bibinfo{pages}{arXiv:2101.03508}.
  \href{http://arxiv.org/abs/2101.03508}{{\tt arXiv:2101.03508}}.
%Type = Inproceedings
\bibitem[{{Deil} et~al.(2017){Deil}, {Zanin}, {Lefaucheur}, {Boisson},
  {Khelifi}, {Terrier}, {Wood}, {Mohrmann}, {Chakraborty}, {Watson},
  {Lopez-Coto}, {Klepser}, {Cerruti}, {Lenain}, {Acero}, {Djannati-Ata{\"\i}},
  {Pita}, {Bosnjak}, {Trichard}, {Vuillaume}, {Donath}, {Consortium}, {King},
  {Jouvin}, {Owen}, {Sipocz}, {Lennarz}, {Voruganti}, {Spir-Jacob}, {Ruiz}, and
  {Arribas}}]{gammapy:2017}
\bibinfo{author}{C.~{Deil}}, \bibinfo{author}{R.~{Zanin}},
  \bibinfo{author}{J.~{Lefaucheur}}, \bibinfo{author}{C.~{Boisson}},
  \bibinfo{author}{B.~{Khelifi}}, \bibinfo{author}{R.~{Terrier}},
  \bibinfo{author}{M.~{Wood}}, \bibinfo{author}{L.~{Mohrmann}},
  \bibinfo{author}{N.~{Chakraborty}}, \bibinfo{author}{J.~{Watson}},
  \bibinfo{author}{R.~{Lopez-Coto}}, \bibinfo{author}{S.~{Klepser}},
  \bibinfo{author}{M.~{Cerruti}}, \bibinfo{author}{J.~P. {Lenain}},
  \bibinfo{author}{F.~{Acero}}, \bibinfo{author}{A.~{Djannati-Ata{\"\i}}},
  \bibinfo{author}{S.~{Pita}}, \bibinfo{author}{Z.~{Bosnjak}},
  \bibinfo{author}{C.~{Trichard}}, \bibinfo{author}{T.~{Vuillaume}},
  \bibinfo{author}{A.~{Donath}}, \bibinfo{author}{C.~{Consortium}},
  \bibinfo{author}{J.~{King}}, \bibinfo{author}{L.~{Jouvin}},
  \bibinfo{author}{E.~{Owen}}, \bibinfo{author}{B.~{Sipocz}},
  \bibinfo{author}{D.~{Lennarz}}, \bibinfo{author}{A.~{Voruganti}},
  \bibinfo{author}{M.~{Spir-Jacob}}, \bibinfo{author}{J.~E. {Ruiz}},
  \bibinfo{author}{M.~P. {Arribas}},
\newblock \bibinfo{title}{{Gammapy - A prototype for the CTA science tools}},
\newblock in: \bibinfo{booktitle}{35th International Cosmic Ray Conference
  (ICRC2017)}, volume \bibinfo{volume}{301} of
  \textit{\bibinfo{series}{International Cosmic Ray Conference}},
  \bibinfo{year}{2017}, p. \bibinfo{pages}{766}.
  \href{http://arxiv.org/abs/1709.01751}{{\tt arXiv:1709.01751}}.
%Type = Misc
\bibitem[{Acero et~al.(2022)Acero, Aguasca-Cabot, Buchner, Carreto~Fidalgo,
  Chen, Chromey, Contreras~Gonzalez, de~Bony~de Lavergne, de~Miranda~Cardoso,
  Deil, Donath, Giunti, Hinton, Jouvin, Khélifi, King, Lefaucheur, Lenain,
  Linhoff, López-Coto, Mohrmann, Morcuende, Nakashima, Nigro, Olivera-Nieto,
  Owen, Panny, Papadopoulos~Orfanos, Paz~Arribas, Pintore, Poon, Remy, Ruiz,
  Siejkowski, Sinha, Sipőcz, Spir-Jacob, Terrier, Tibaldo, Unbehaun, van
  Eldik, Vuillaume, Weinstein, and Wood}]{gammapy:1.0}
\bibinfo{author}{F.~Acero}, \bibinfo{author}{A.~Aguasca-Cabot},
  \bibinfo{author}{J.~Buchner}, \bibinfo{author}{D.~Carreto~Fidalgo},
  \bibinfo{author}{A.~Chen}, \bibinfo{author}{A.~Chromey},
  \bibinfo{author}{J.~L. Contreras~Gonzalez}, \bibinfo{author}{M.~de~Bony~de
  Lavergne}, \bibinfo{author}{J.~V. de~Miranda~Cardoso},
  \bibinfo{author}{C.~Deil}, \bibinfo{author}{A.~Donath},
  \bibinfo{author}{L.~Giunti}, \bibinfo{author}{J.~Hinton},
  \bibinfo{author}{L.~Jouvin}, \bibinfo{author}{B.~Khélifi},
  \bibinfo{author}{J.~King}, \bibinfo{author}{J.~Lefaucheur},
  \bibinfo{author}{J.-P. Lenain}, \bibinfo{author}{M.~Linhoff},
  \bibinfo{author}{R.~López-Coto}, \bibinfo{author}{L.~Mohrmann},
  \bibinfo{author}{D.~Morcuende}, \bibinfo{author}{K.~Nakashima},
  \bibinfo{author}{C.~Nigro}, \bibinfo{author}{L.~Olivera-Nieto},
  \bibinfo{author}{E.~Owen}, \bibinfo{author}{S.~Panny},
  \bibinfo{author}{D.~Papadopoulos~Orfanos}, \bibinfo{author}{M.~Paz~Arribas},
  \bibinfo{author}{F.~Pintore}, \bibinfo{author}{H.~Poon},
  \bibinfo{author}{Q.~Remy}, \bibinfo{author}{J.~E. Ruiz},
  \bibinfo{author}{H.~Siejkowski}, \bibinfo{author}{A.~Sinha},
  \bibinfo{author}{B.~M. Sipőcz}, \bibinfo{author}{M.~Spir-Jacob},
  \bibinfo{author}{R.~Terrier}, \bibinfo{author}{L.~Tibaldo},
  \bibinfo{author}{T.~Unbehaun}, \bibinfo{author}{C.~van Eldik},
  \bibinfo{author}{T.~Vuillaume}, \bibinfo{author}{A.~Weinstein},
  \bibinfo{author}{M.~Wood}, \bibinfo{title}{Gammapy: Python toolbox for
  gamma-ray astronomy}, \bibinfo{year}{2022}. \URLprefix
  \url{https://doi.org/10.5281/zenodo.7311399}.
  \DOIprefix\doi{10.5281/zenodo.7311399}, \bibinfo{note}{{If you use this
  software, please cite it using the metadata from this file.}}
%Type = Article
\bibitem[{{Chadwick} et~al.(1999){Chadwick}, {Lyons}, {McComb}, {Orford},
  {Osborne}, {Rayner}, {Shaw}, {Turver}, and {Wieczorek}}]{DurhamMkVI:PKS2155}
\bibinfo{author}{P.~M. {Chadwick}}, \bibinfo{author}{K.~{Lyons}},
  \bibinfo{author}{T.~J.~L. {McComb}}, \bibinfo{author}{K.~J. {Orford}},
  \bibinfo{author}{J.~L. {Osborne}}, \bibinfo{author}{S.~M. {Rayner}},
  \bibinfo{author}{S.~E. {Shaw}}, \bibinfo{author}{K.~E. {Turver}},
  \bibinfo{author}{G.~J. {Wieczorek}},
\newblock \bibinfo{title}{{Very High Energy Gamma Rays from PKS 2155-304}},
\newblock \bibinfo{journal}{\apj} \bibinfo{volume}{513} (\bibinfo{year}{1999})
  \bibinfo{pages}{161--167}. \DOIprefix\doi{10.1086/306862}.
  \href{http://arxiv.org/abs/astro-ph/9810209}{{\tt arXiv:astro-ph/9810209}}.
%Type = Article
\bibitem[{{Aharonian} et~al.(2007){Aharonian}, {Akhperjanian}, {Bazer-Bachi},
  {Behera}, {Beilicke}, {Benbow}, {Berge}, {Bernl{\"o}hr}, {Boisson}, {Bolz},
  {Borrel}, {Boutelier}, {Braun}, {Brion}, {Brown}, {B{\"u}hler},
  {B{\"u}sching}, {Bulik}, {Carrigan}, {Chadwick}, {Clapson}, {Chounet},
  {Coignet}, {Cornils}, {Costamante}, {Degrange}, {Dickinson},
  {Djannati-Ata{\"\i}}, {Domainko}, {Drury}, {Dubus}, {Dyks}, {Egberts},
  {Emmanoulopoulos}, {Espigat}, {Farnier}, {Feinstein}, {Fiasson},
  {F{\"o}rster}, {Fontaine}, {Funk}, {Funk}, {F{\"u}{\ss}ling}, {Gallant},
  {Giebels}, {Glicenstein}, {Gl{\"u}ck}, {Goret}, {Hadjichristidis}, {Hauser},
  {Hauser}, {Heinzelmann}, {Henri}, {Hermann}, {Hinton}, {Hoffmann}, {Hofmann},
  {Holleran}, {Hoppe}, {Horns}, {Jacholkowska}, {de Jager}, {Kendziorra},
  {Kerschhaggl}, {Kh{\'e}lifi}, {Komin}, {Kosack}, {Lamanna}, {Latham}, {Le
  Gallou}, {Lemi{\`e}re}, {Lemoine-Goumard}, {Lenain}, {Lohse}, {Martin},
  {Martineau-Huynh}, {Marcowith}, {Masterson}, {Maurin}, {McComb}, {Moderski},
  {Moulin}, {de Naurois}, {Nedbal}, {Nolan}, {Olive}, {Orford}, {Osborne},
  {Ostrowski}, {Panter}, {Pedaletti}, {Pelletier}, {Petrucci}, {Pita},
  {P{\"u}hlhofer}, {Punch}, {Ranchon}, {Raubenheimer}, {Raue}, {Rayner},
  {Renaud}, {Ripken}, {Rob}, {Rolland}, {Rosier-Lees}, {Rowell}, {Rudak},
  {Ruppel}, {Sahakian}, {Santangelo}, {Saug{\'e}}, {Schlenker}, {Schlickeiser},
  {Schr{\"o}der}, {Schwanke}, {Schwarzburg}, {Schwemmer}, {Shalchi}, {Sol},
  {Spangler}, {Stawarz}, {Steenkamp}, {Stegmann}, {Superina}, {Tam},
  {Tavernet}, {Terrier}, {van Eldik}, {Vasileiadis}, {Venter}, {Vialle},
  {Vincent}, {Vivier}, {V{\"o}lk}, {Volpe}, {Wagner}, {Ward}, and
  {Zdziarski}}]{PKS2155_bigflare}
\bibinfo{author}{F.~{Aharonian}}, \bibinfo{author}{A.~G. {Akhperjanian}},
  \bibinfo{author}{A.~R. {Bazer-Bachi}}, \bibinfo{author}{B.~{Behera}},
  \bibinfo{author}{M.~{Beilicke}}, \bibinfo{author}{W.~{Benbow}},
  \bibinfo{author}{D.~{Berge}}, \bibinfo{author}{K.~{Bernl{\"o}hr}},
  \bibinfo{author}{C.~{Boisson}}, \bibinfo{author}{O.~{Bolz}},
  \bibinfo{author}{V.~{Borrel}}, \bibinfo{author}{T.~{Boutelier}},
  \bibinfo{author}{I.~{Braun}}, \bibinfo{author}{E.~{Brion}},
  \bibinfo{author}{A.~M. {Brown}}, \bibinfo{author}{R.~{B{\"u}hler}},
  \bibinfo{author}{I.~{B{\"u}sching}}, \bibinfo{author}{T.~{Bulik}},
  \bibinfo{author}{S.~{Carrigan}}, \bibinfo{author}{P.~M. {Chadwick}},
  \bibinfo{author}{A.~C. {Clapson}}, \bibinfo{author}{L.~M. {Chounet}},
  \bibinfo{author}{G.~{Coignet}}, \bibinfo{author}{R.~{Cornils}},
  \bibinfo{author}{L.~{Costamante}}, \bibinfo{author}{B.~{Degrange}},
  \bibinfo{author}{H.~J. {Dickinson}},
  \bibinfo{author}{A.~{Djannati-Ata{\"\i}}}, \bibinfo{author}{W.~{Domainko}},
  \bibinfo{author}{L.~O. {Drury}}, \bibinfo{author}{G.~{Dubus}},
  \bibinfo{author}{J.~{Dyks}}, \bibinfo{author}{K.~{Egberts}},
  \bibinfo{author}{D.~{Emmanoulopoulos}}, \bibinfo{author}{P.~{Espigat}},
  \bibinfo{author}{C.~{Farnier}}, \bibinfo{author}{F.~{Feinstein}},
  \bibinfo{author}{A.~{Fiasson}}, \bibinfo{author}{A.~{F{\"o}rster}},
  \bibinfo{author}{G.~{Fontaine}}, \bibinfo{author}{S.~{Funk}},
  \bibinfo{author}{S.~{Funk}}, \bibinfo{author}{M.~{F{\"u}{\ss}ling}},
  \bibinfo{author}{Y.~A. {Gallant}}, \bibinfo{author}{B.~{Giebels}},
  \bibinfo{author}{J.~F. {Glicenstein}}, \bibinfo{author}{B.~{Gl{\"u}ck}},
  \bibinfo{author}{P.~{Goret}}, \bibinfo{author}{C.~{Hadjichristidis}},
  \bibinfo{author}{D.~{Hauser}}, \bibinfo{author}{M.~{Hauser}},
  \bibinfo{author}{G.~{Heinzelmann}}, \bibinfo{author}{G.~{Henri}},
  \bibinfo{author}{G.~{Hermann}}, \bibinfo{author}{J.~A. {Hinton}},
  \bibinfo{author}{A.~{Hoffmann}}, \bibinfo{author}{W.~{Hofmann}},
  \bibinfo{author}{M.~{Holleran}}, \bibinfo{author}{S.~{Hoppe}},
  \bibinfo{author}{D.~{Horns}}, \bibinfo{author}{A.~{Jacholkowska}},
  \bibinfo{author}{O.~C. {de Jager}}, \bibinfo{author}{E.~{Kendziorra}},
  \bibinfo{author}{M.~{Kerschhaggl}}, \bibinfo{author}{B.~{Kh{\'e}lifi}},
  \bibinfo{author}{N.~{Komin}}, \bibinfo{author}{K.~{Kosack}},
  \bibinfo{author}{G.~{Lamanna}}, \bibinfo{author}{I.~J. {Latham}},
  \bibinfo{author}{R.~{Le Gallou}}, \bibinfo{author}{A.~{Lemi{\`e}re}},
  \bibinfo{author}{M.~{Lemoine-Goumard}}, \bibinfo{author}{J.~P. {Lenain}},
  \bibinfo{author}{T.~{Lohse}}, \bibinfo{author}{J.~M. {Martin}},
  \bibinfo{author}{O.~{Martineau-Huynh}}, \bibinfo{author}{A.~{Marcowith}},
  \bibinfo{author}{C.~{Masterson}}, \bibinfo{author}{G.~{Maurin}},
  \bibinfo{author}{T.~J.~L. {McComb}}, \bibinfo{author}{R.~{Moderski}},
  \bibinfo{author}{E.~{Moulin}}, \bibinfo{author}{M.~{de Naurois}},
  \bibinfo{author}{D.~{Nedbal}}, \bibinfo{author}{S.~J. {Nolan}},
  \bibinfo{author}{J.~P. {Olive}}, \bibinfo{author}{K.~J. {Orford}},
  \bibinfo{author}{J.~L. {Osborne}}, \bibinfo{author}{M.~{Ostrowski}},
  \bibinfo{author}{M.~{Panter}}, \bibinfo{author}{G.~{Pedaletti}},
  \bibinfo{author}{G.~{Pelletier}}, \bibinfo{author}{P.~O. {Petrucci}},
  \bibinfo{author}{S.~{Pita}}, \bibinfo{author}{G.~{P{\"u}hlhofer}},
  \bibinfo{author}{M.~{Punch}}, \bibinfo{author}{S.~{Ranchon}},
  \bibinfo{author}{B.~C. {Raubenheimer}}, \bibinfo{author}{M.~{Raue}},
  \bibinfo{author}{S.~M. {Rayner}}, \bibinfo{author}{M.~{Renaud}},
  \bibinfo{author}{J.~{Ripken}}, \bibinfo{author}{L.~{Rob}},
  \bibinfo{author}{L.~{Rolland}}, \bibinfo{author}{S.~{Rosier-Lees}},
  \bibinfo{author}{G.~{Rowell}}, \bibinfo{author}{B.~{Rudak}},
  \bibinfo{author}{J.~{Ruppel}}, \bibinfo{author}{V.~{Sahakian}},
  \bibinfo{author}{A.~{Santangelo}}, \bibinfo{author}{L.~{Saug{\'e}}},
  \bibinfo{author}{S.~{Schlenker}}, \bibinfo{author}{R.~{Schlickeiser}},
  \bibinfo{author}{R.~{Schr{\"o}der}}, \bibinfo{author}{U.~{Schwanke}},
  \bibinfo{author}{S.~{Schwarzburg}}, \bibinfo{author}{S.~{Schwemmer}},
  \bibinfo{author}{A.~{Shalchi}}, \bibinfo{author}{H.~{Sol}},
  \bibinfo{author}{D.~{Spangler}}, \bibinfo{author}{{\L}.~{Stawarz}},
  \bibinfo{author}{R.~{Steenkamp}}, \bibinfo{author}{C.~{Stegmann}},
  \bibinfo{author}{G.~{Superina}}, \bibinfo{author}{P.~H. {Tam}},
  \bibinfo{author}{J.~P. {Tavernet}}, \bibinfo{author}{R.~{Terrier}},
  \bibinfo{author}{C.~{van Eldik}}, \bibinfo{author}{G.~{Vasileiadis}},
  \bibinfo{author}{C.~{Venter}}, \bibinfo{author}{J.~P. {Vialle}},
  \bibinfo{author}{P.~{Vincent}}, \bibinfo{author}{M.~{Vivier}},
  \bibinfo{author}{H.~J. {V{\"o}lk}}, \bibinfo{author}{F.~{Volpe}},
  \bibinfo{author}{S.~J. {Wagner}}, \bibinfo{author}{M.~{Ward}},
  \bibinfo{author}{A.~A. {Zdziarski}},
\newblock \bibinfo{title}{{An Exceptional Very High Energy Gamma-Ray Flare of
  PKS 2155-304}},
\newblock \bibinfo{journal}{\apjl} \bibinfo{volume}{664} (\bibinfo{year}{2007})
  \bibinfo{pages}{L71--L74}. \DOIprefix\doi{10.1086/520635}.
  \href{http://arxiv.org/abs/0706.0797}{{\tt arXiv:0706.0797}}.
%Type = Article
\bibitem[{{Aharonian} et~al.(2009){Aharonian}, {Akhperjanian}, {Anton}, {Barres
  de Almeida}, {Bazer-Bachi}, {Becherini}, {Behera}, {Benbow}, {Bernl{\"o}hr},
  {Boisson}, {Bochow}, {Borrel}, {Brion}, {Brucker}, {Brun}, {B{\"u}hler},
  {Bulik}, {B{\"u}sching}, {Boutelier}, {Chadwick}, {Charbonnier}, {Chaves},
  {Cheesebrough}, {Chounet}, {Clapson}, {Coignet}, {Costamante}, {Dalton},
  {Daniel}, {Davids}, {Degrange}, {Deil}, {Dickinson}, {Djannati-Ata{\"\i}},
  {Domainko}, {O'C. Drury}, {Dubois}, {Dubus}, {Dyks}, {Dyrda}, {Egberts},
  {Emmanoulopoulos}, {Espigat}, {Farnier}, {Feinstein}, {Fiasson},
  {F{\"o}rster}, {Fontaine}, {F{\"u}{\ss}ling}, {Gabici}, {Gallant},
  {G{\'e}rard}, {Giebels}, {Glicenstein}, {Gl{\"u}ck}, {Goret}, {G{\"o}hring},
  {Hauser}, {Hauser}, {Heinz}, {Heinzelmann}, {Henri}, {Hermann}, {Hinton},
  {Hoffmann}, {Hofmann}, {Holleran}, {Hoppe}, {Horns}, {Jacholkowska}, {de
  Jager}, {Jahn}, {Jung}, {Katarzy{\'n}ski}, {Katz}, {Kaufmann}, {Kendziorra},
  {Kerschhaggl}, {Khangulyan}, {Kh{\'e}lifi}, {Keogh}, {Klu{\'z}niak},
  {Kneiske}, {Komin}, {Kosack}, {Lamanna}, {Lenain}, {Lohse}, {Marandon},
  {Martin}, {Martineau-Huynh}, {Marcowith}, {Maurin}, {McComb}, {Medina},
  {Moderski}, {Monard}, {Moulin}, {Naumann-Godo}, {de Naurois}, {Nedbal},
  {Nekrassov}, {Niemiec}, {Nolan}, {Ohm}, {Olive}, {de O{\~n}a Wilhelmi},
  {Orford}, {Ostrowski}, {Panter}, {Paz Arribas}, {Pedaletti}, {Pelletier},
  {Petrucci}, {Pita}, {P{\"u}hlhofer}, {Punch}, {Quirrenbach}, {Raubenheimer},
  {Raue}, {Rayner}, {Renaud}, {Rieger}, {Ripken}, {Rob}, {Rosier-Lees},
  {Rowell}, {Rudak}, {Rulten}, {Ruppel}, {Sahakian}, {Santangelo},
  {Schlickeiser}, {Sch{\"o}ck}, {Schr{\"o}der}, {Schwanke}, {Schwarzburg},
  {Schwemmer}, {Shalchi}, {Sikora}, {Skilton}, {Sol}, {Spangler}, {Stawarz},
  {Steenkamp}, {Stegmann}, {Superina}, {Szostek}, {Tam}, {Tavernet}, {Terrier},
  {Tibolla}, {Tluczykont}, {van Eldik}, {Vasileiadis}, {Venter}, {Venter},
  {Vialle}, {Vincent}, {Vivier}, {V{\"o}lk}, {Volpe}, {Wagner}, {Ward},
  {Zdziarski}, and {Zech}}]{PKS2155_Chandra_night}
\bibinfo{author}{F.~{Aharonian}}, \bibinfo{author}{A.~G. {Akhperjanian}},
  \bibinfo{author}{G.~{Anton}}, \bibinfo{author}{U.~{Barres de Almeida}},
  \bibinfo{author}{A.~R. {Bazer-Bachi}}, \bibinfo{author}{Y.~{Becherini}},
  \bibinfo{author}{B.~{Behera}}, \bibinfo{author}{W.~{Benbow}},
  \bibinfo{author}{K.~{Bernl{\"o}hr}}, \bibinfo{author}{C.~{Boisson}},
  \bibinfo{author}{A.~{Bochow}}, \bibinfo{author}{V.~{Borrel}},
  \bibinfo{author}{E.~{Brion}}, \bibinfo{author}{J.~{Brucker}},
  \bibinfo{author}{P.~{Brun}}, \bibinfo{author}{R.~{B{\"u}hler}},
  \bibinfo{author}{T.~{Bulik}}, \bibinfo{author}{I.~{B{\"u}sching}},
  \bibinfo{author}{T.~{Boutelier}}, \bibinfo{author}{P.~M. {Chadwick}},
  \bibinfo{author}{A.~{Charbonnier}}, \bibinfo{author}{R.~C.~G. {Chaves}},
  \bibinfo{author}{A.~{Cheesebrough}}, \bibinfo{author}{L.~M. {Chounet}},
  \bibinfo{author}{A.~C. {Clapson}}, \bibinfo{author}{G.~{Coignet}},
  \bibinfo{author}{L.~{Costamante}}, \bibinfo{author}{M.~{Dalton}},
  \bibinfo{author}{M.~K. {Daniel}}, \bibinfo{author}{I.~D. {Davids}},
  \bibinfo{author}{B.~{Degrange}}, \bibinfo{author}{C.~{Deil}},
  \bibinfo{author}{H.~J. {Dickinson}},
  \bibinfo{author}{A.~{Djannati-Ata{\"\i}}}, \bibinfo{author}{W.~{Domainko}},
  \bibinfo{author}{L.~{O'C. Drury}}, \bibinfo{author}{F.~{Dubois}},
  \bibinfo{author}{G.~{Dubus}}, \bibinfo{author}{J.~{Dyks}},
  \bibinfo{author}{M.~{Dyrda}}, \bibinfo{author}{K.~{Egberts}},
  \bibinfo{author}{D.~{Emmanoulopoulos}}, \bibinfo{author}{P.~{Espigat}},
  \bibinfo{author}{C.~{Farnier}}, \bibinfo{author}{F.~{Feinstein}},
  \bibinfo{author}{A.~{Fiasson}}, \bibinfo{author}{A.~{F{\"o}rster}},
  \bibinfo{author}{G.~{Fontaine}}, \bibinfo{author}{M.~{F{\"u}{\ss}ling}},
  \bibinfo{author}{S.~{Gabici}}, \bibinfo{author}{Y.~A. {Gallant}},
  \bibinfo{author}{L.~{G{\'e}rard}}, \bibinfo{author}{B.~{Giebels}},
  \bibinfo{author}{J.~F. {Glicenstein}}, \bibinfo{author}{B.~{Gl{\"u}ck}},
  \bibinfo{author}{P.~{Goret}}, \bibinfo{author}{D.~{G{\"o}hring}},
  \bibinfo{author}{D.~{Hauser}}, \bibinfo{author}{M.~{Hauser}},
  \bibinfo{author}{S.~{Heinz}}, \bibinfo{author}{G.~{Heinzelmann}},
  \bibinfo{author}{G.~{Henri}}, \bibinfo{author}{G.~{Hermann}},
  \bibinfo{author}{J.~A. {Hinton}}, \bibinfo{author}{A.~{Hoffmann}},
  \bibinfo{author}{W.~{Hofmann}}, \bibinfo{author}{M.~{Holleran}},
  \bibinfo{author}{S.~{Hoppe}}, \bibinfo{author}{D.~{Horns}},
  \bibinfo{author}{A.~{Jacholkowska}}, \bibinfo{author}{O.~C. {de Jager}},
  \bibinfo{author}{C.~{Jahn}}, \bibinfo{author}{I.~{Jung}},
  \bibinfo{author}{K.~{Katarzy{\'n}ski}}, \bibinfo{author}{U.~{Katz}},
  \bibinfo{author}{S.~{Kaufmann}}, \bibinfo{author}{E.~{Kendziorra}},
  \bibinfo{author}{M.~{Kerschhaggl}}, \bibinfo{author}{D.~{Khangulyan}},
  \bibinfo{author}{B.~{Kh{\'e}lifi}}, \bibinfo{author}{D.~{Keogh}},
  \bibinfo{author}{W.~{Klu{\'z}niak}}, \bibinfo{author}{T.~{Kneiske}},
  \bibinfo{author}{N.~{Komin}}, \bibinfo{author}{K.~{Kosack}},
  \bibinfo{author}{G.~{Lamanna}}, \bibinfo{author}{J.~P. {Lenain}},
  \bibinfo{author}{T.~{Lohse}}, \bibinfo{author}{V.~{Marandon}},
  \bibinfo{author}{J.~M. {Martin}}, \bibinfo{author}{O.~{Martineau-Huynh}},
  \bibinfo{author}{A.~{Marcowith}}, \bibinfo{author}{D.~{Maurin}},
  \bibinfo{author}{T.~J.~L. {McComb}}, \bibinfo{author}{M.~C. {Medina}},
  \bibinfo{author}{R.~{Moderski}}, \bibinfo{author}{L.~A.~G. {Monard}},
  \bibinfo{author}{E.~{Moulin}}, \bibinfo{author}{M.~{Naumann-Godo}},
  \bibinfo{author}{M.~{de Naurois}}, \bibinfo{author}{D.~{Nedbal}},
  \bibinfo{author}{D.~{Nekrassov}}, \bibinfo{author}{J.~{Niemiec}},
  \bibinfo{author}{S.~J. {Nolan}}, \bibinfo{author}{S.~{Ohm}},
  \bibinfo{author}{J.~F. {Olive}}, \bibinfo{author}{E.~{de O{\~n}a Wilhelmi}},
  \bibinfo{author}{K.~J. {Orford}}, \bibinfo{author}{M.~{Ostrowski}},
  \bibinfo{author}{M.~{Panter}}, \bibinfo{author}{M.~{Paz Arribas}},
  \bibinfo{author}{G.~{Pedaletti}}, \bibinfo{author}{G.~{Pelletier}},
  \bibinfo{author}{P.~O. {Petrucci}}, \bibinfo{author}{S.~{Pita}},
  \bibinfo{author}{G.~{P{\"u}hlhofer}}, \bibinfo{author}{M.~{Punch}},
  \bibinfo{author}{A.~{Quirrenbach}}, \bibinfo{author}{B.~C. {Raubenheimer}},
  \bibinfo{author}{M.~{Raue}}, \bibinfo{author}{S.~M. {Rayner}},
  \bibinfo{author}{M.~{Renaud}}, \bibinfo{author}{F.~{Rieger}},
  \bibinfo{author}{J.~{Ripken}}, \bibinfo{author}{L.~{Rob}},
  \bibinfo{author}{S.~{Rosier-Lees}}, \bibinfo{author}{G.~{Rowell}},
  \bibinfo{author}{B.~{Rudak}}, \bibinfo{author}{C.~B. {Rulten}},
  \bibinfo{author}{J.~{Ruppel}}, \bibinfo{author}{V.~{Sahakian}},
  \bibinfo{author}{A.~{Santangelo}}, \bibinfo{author}{R.~{Schlickeiser}},
  \bibinfo{author}{F.~M. {Sch{\"o}ck}}, \bibinfo{author}{R.~{Schr{\"o}der}},
  \bibinfo{author}{U.~{Schwanke}}, \bibinfo{author}{S.~{Schwarzburg}},
  \bibinfo{author}{S.~{Schwemmer}}, \bibinfo{author}{A.~{Shalchi}},
  \bibinfo{author}{M.~{Sikora}}, \bibinfo{author}{J.~L. {Skilton}},
  \bibinfo{author}{H.~{Sol}}, \bibinfo{author}{D.~{Spangler}},
  \bibinfo{author}{{\L}.~{Stawarz}}, \bibinfo{author}{R.~{Steenkamp}},
  \bibinfo{author}{C.~{Stegmann}}, \bibinfo{author}{G.~{Superina}},
  \bibinfo{author}{A.~{Szostek}}, \bibinfo{author}{P.~H. {Tam}},
  \bibinfo{author}{J.~P. {Tavernet}}, \bibinfo{author}{R.~{Terrier}},
  \bibinfo{author}{O.~{Tibolla}}, \bibinfo{author}{M.~{Tluczykont}},
  \bibinfo{author}{C.~{van Eldik}}, \bibinfo{author}{G.~{Vasileiadis}},
  \bibinfo{author}{C.~{Venter}}, \bibinfo{author}{L.~{Venter}},
  \bibinfo{author}{J.~P. {Vialle}}, \bibinfo{author}{P.~{Vincent}},
  \bibinfo{author}{M.~{Vivier}}, \bibinfo{author}{H.~J. {V{\"o}lk}},
  \bibinfo{author}{F.~{Volpe}}, \bibinfo{author}{S.~J. {Wagner}},
  \bibinfo{author}{M.~{Ward}}, \bibinfo{author}{A.~A. {Zdziarski}},
  \bibinfo{author}{A.~{Zech}},
\newblock \bibinfo{title}{{Simultaneous multiwavelength observations of the
  second exceptional {\ensuremath{\gamma}}-ray flare of PKS 2155-304 in July
  2006}},
\newblock \bibinfo{journal}{\aap} \bibinfo{volume}{502} (\bibinfo{year}{2009})
  \bibinfo{pages}{749--770}. \DOIprefix\doi{10.1051/0004-6361/200912128}.
  \href{http://arxiv.org/abs/0906.2002}{{\tt arXiv:0906.2002}}.
%Type = Article
\bibitem[{{H.~E.~S.~S. Collaboration} et~al.(2010){H.~E.~S.~S. Collaboration},
  {Abramowski}, {Acero}, {Aharonian}, {Akhperjanian}, {Anton}, {Barres de
  Almeida}, {Bazer-Bachi}, {Becherini}, {Benbow}, {Bernl{\"o}hr}, {Bochow},
  {Boisson}, {Bolmont}, {Borrel}, {Brucker}, {Brun}, {Brun}, {B{\"u}hler},
  {Bulik}, {B{\"u}sching}, {Boutelier}, {Chadwick}, {Charbonnier}, {Chaves},
  {Cheesebrough}, {Chounet}, {Clapson}, {Coignet}, {Conrad}, {Costamante},
  {Dalton}, {Daniel}, {Davids}, {Degrange}, {Deil}, {Dickinson},
  {Djannati-Ata{\"\i}}, {Domainko}, {O'C. Drury}, {Dubois}, {Dubus}, {Dyks},
  {Dyrda}, {Egberts}, {Eger}, {Espigat}, {Fallon}, {Farnier}, {Fegan},
  {Feinstein}, {Fernandes}, {Fiasson}, {F{\"o}rster}, {Fontaine},
  {F{\"u}{\ss}ling}, {Gabici}, {Gallant}, {G{\'e}rard}, {Gerbig}, {Giebels},
  {Glicenstein}, {Gl{\"u}ck}, {Goret}, {G{\"o}ring}, {Hampf}, {Hauser},
  {Heinz}, {Heinzelmann}, {Henri}, {Hermann}, {Hinton}, {Hoffmann}, {Hofmann},
  {Hofverberg}, {Holleran}, {Hoppe}, {Horns}, {Jacholkowska}, {de Jager},
  {Jahn}, {Jung}, {Katarzy{\'n}ski}, {Katz}, {Kaufmann}, {Kerschhaggl},
  {Khangulyan}, {Kh{\'e}lifi}, {Keogh}, {Klochkov}, {Klu{\'z}niak}, {Kneiske},
  {Komin}, {Kosack}, {Kossakowski}, {Lamanna}, {Lenain}, {Lohse}, {Lu},
  {Marandon}, {Marcowith}, {Masbou}, {Maurin}, {McComb}, {Medina},
  {M{\'e}hault}, {Moderski}, {Moulin}, {Naumann-Godo}, {de Naurois}, {Nedbal},
  {Nekrassov}, {Nguyen}, {Nicholas}, {Niemiec}, {Nolan}, {Ohm}, {Olive}, {de
  O{\~n}a Wilhelmi}, {Opitz}, {Orford}, {Ostrowski}, {Panter}, {Paz Arribas},
  {Pedaletti}, {Pelletier}, {Petrucci}, {Pita}, {P{\"u}hlhofer}, {Punch},
  {Quirrenbach}, {Raubenheimer}, {Raue}, {Rayner}, {Reimer}, {Renaud}, {de los
  Reyes}, {Rieger}, {Ripken}, {Rob}, {Rosier-Lees}, {Rowell}, {Rudak},
  {Rulten}, {Ruppel}, {Ryde}, {Sahakian}, {Santangelo}, {Schlickeiser},
  {Sch{\"o}ck}, {Sch{\"o}nwald}, {Schwanke}, {Schwarzburg}, {Schwemmer},
  {Shalchi}, {Sushch}, {Sikora}, {Skilton}, {Sol}, {Stawarz}, {Steenkamp},
  {Stegmann}, {Stinzing}, {Superina}, {Szostek}, {Tam}, {Tavernet}, {Terrier},
  {Tibolla}, {Tluczykont}, {Valerius}, {van Eldik}, {Vasileiadis}, {Venter},
  {Venter}, {Vialle}, {Viana}, {Vincent}, {Vivier}, {V{\"o}lk}, {Volpe},
  {Vorobiov}, {Wagner}, {Ward}, {Zdziarski}, {Zech}, and
  {Zechlin}}]{PKS2155_quiescent}
\bibinfo{author}{{H.~E.~S.~S. Collaboration}},
  \bibinfo{author}{A.~{Abramowski}}, \bibinfo{author}{F.~{Acero}},
  \bibinfo{author}{F.~{Aharonian}}, \bibinfo{author}{A.~G. {Akhperjanian}},
  \bibinfo{author}{G.~{Anton}}, \bibinfo{author}{U.~{Barres de Almeida}},
  \bibinfo{author}{A.~R. {Bazer-Bachi}}, \bibinfo{author}{B.~{Becherini},
  Y.~Behera}, \bibinfo{author}{W.~{Benbow}},
  \bibinfo{author}{K.~{Bernl{\"o}hr}}, \bibinfo{author}{A.~{Bochow}},
  \bibinfo{author}{C.~{Boisson}}, \bibinfo{author}{J.~{Bolmont}},
  \bibinfo{author}{V.~{Borrel}}, \bibinfo{author}{J.~{Brucker}},
  \bibinfo{author}{F.~{Brun}}, \bibinfo{author}{P.~{Brun}},
  \bibinfo{author}{R.~{B{\"u}hler}}, \bibinfo{author}{T.~{Bulik}},
  \bibinfo{author}{I.~{B{\"u}sching}}, \bibinfo{author}{T.~{Boutelier}},
  \bibinfo{author}{P.~M. {Chadwick}}, \bibinfo{author}{A.~{Charbonnier}},
  \bibinfo{author}{R.~C.~G. {Chaves}}, \bibinfo{author}{A.~{Cheesebrough}},
  \bibinfo{author}{L.~M. {Chounet}}, \bibinfo{author}{A.~C. {Clapson}},
  \bibinfo{author}{G.~{Coignet}}, \bibinfo{author}{J.~{Conrad}},
  \bibinfo{author}{L.~{Costamante}}, \bibinfo{author}{M.~{Dalton}},
  \bibinfo{author}{M.~K. {Daniel}}, \bibinfo{author}{I.~D. {Davids}},
  \bibinfo{author}{B.~{Degrange}}, \bibinfo{author}{C.~{Deil}},
  \bibinfo{author}{H.~J. {Dickinson}},
  \bibinfo{author}{A.~{Djannati-Ata{\"\i}}}, \bibinfo{author}{W.~{Domainko}},
  \bibinfo{author}{L.~{O'C. Drury}}, \bibinfo{author}{F.~{Dubois}},
  \bibinfo{author}{G.~{Dubus}}, \bibinfo{author}{J.~{Dyks}},
  \bibinfo{author}{M.~{Dyrda}}, \bibinfo{author}{K.~{Egberts}},
  \bibinfo{author}{P.~{Eger}}, \bibinfo{author}{P.~{Espigat}},
  \bibinfo{author}{L.~{Fallon}}, \bibinfo{author}{C.~{Farnier}},
  \bibinfo{author}{S.~{Fegan}}, \bibinfo{author}{F.~{Feinstein}},
  \bibinfo{author}{M.~V. {Fernandes}}, \bibinfo{author}{A.~{Fiasson}},
  \bibinfo{author}{A.~{F{\"o}rster}}, \bibinfo{author}{G.~{Fontaine}},
  \bibinfo{author}{M.~{F{\"u}{\ss}ling}}, \bibinfo{author}{S.~{Gabici}},
  \bibinfo{author}{Y.~A. {Gallant}}, \bibinfo{author}{L.~{G{\'e}rard}},
  \bibinfo{author}{D.~{Gerbig}}, \bibinfo{author}{B.~{Giebels}},
  \bibinfo{author}{J.~F. {Glicenstein}}, \bibinfo{author}{B.~{Gl{\"u}ck}},
  \bibinfo{author}{P.~{Goret}}, \bibinfo{author}{D.~{G{\"o}ring}},
  \bibinfo{author}{D.~{Hampf}}, \bibinfo{author}{M.~{Hauser}},
  \bibinfo{author}{S.~{Heinz}}, \bibinfo{author}{G.~{Heinzelmann}},
  \bibinfo{author}{G.~{Henri}}, \bibinfo{author}{G.~{Hermann}},
  \bibinfo{author}{J.~A. {Hinton}}, \bibinfo{author}{A.~{Hoffmann}},
  \bibinfo{author}{W.~{Hofmann}}, \bibinfo{author}{P.~{Hofverberg}},
  \bibinfo{author}{M.~{Holleran}}, \bibinfo{author}{S.~{Hoppe}},
  \bibinfo{author}{D.~{Horns}}, \bibinfo{author}{A.~{Jacholkowska}},
  \bibinfo{author}{O.~C. {de Jager}}, \bibinfo{author}{C.~{Jahn}},
  \bibinfo{author}{I.~{Jung}}, \bibinfo{author}{K.~{Katarzy{\'n}ski}},
  \bibinfo{author}{U.~{Katz}}, \bibinfo{author}{S.~{Kaufmann}},
  \bibinfo{author}{M.~{Kerschhaggl}}, \bibinfo{author}{D.~{Khangulyan}},
  \bibinfo{author}{B.~{Kh{\'e}lifi}}, \bibinfo{author}{D.~{Keogh}},
  \bibinfo{author}{D.~{Klochkov}}, \bibinfo{author}{W.~{Klu{\'z}niak}},
  \bibinfo{author}{T.~{Kneiske}}, \bibinfo{author}{N.~{Komin}},
  \bibinfo{author}{K.~{Kosack}}, \bibinfo{author}{R.~{Kossakowski}},
  \bibinfo{author}{G.~{Lamanna}}, \bibinfo{author}{J.~P. {Lenain}},
  \bibinfo{author}{T.~{Lohse}}, \bibinfo{author}{C.~C. {Lu}},
  \bibinfo{author}{V.~{Marandon}}, \bibinfo{author}{A.~{Marcowith}},
  \bibinfo{author}{J.~{Masbou}}, \bibinfo{author}{D.~{Maurin}},
  \bibinfo{author}{T.~J.~L. {McComb}}, \bibinfo{author}{M.~C. {Medina}},
  \bibinfo{author}{J.~{M{\'e}hault}}, \bibinfo{author}{R.~{Moderski}},
  \bibinfo{author}{E.~{Moulin}}, \bibinfo{author}{M.~{Naumann-Godo}},
  \bibinfo{author}{M.~{de Naurois}}, \bibinfo{author}{D.~{Nedbal}},
  \bibinfo{author}{D.~{Nekrassov}}, \bibinfo{author}{N.~{Nguyen}},
  \bibinfo{author}{B.~{Nicholas}}, \bibinfo{author}{J.~{Niemiec}},
  \bibinfo{author}{S.~J. {Nolan}}, \bibinfo{author}{S.~{Ohm}},
  \bibinfo{author}{J.~F. {Olive}}, \bibinfo{author}{E.~{de O{\~n}a Wilhelmi}},
  \bibinfo{author}{B.~{Opitz}}, \bibinfo{author}{K.~J. {Orford}},
  \bibinfo{author}{M.~{Ostrowski}}, \bibinfo{author}{M.~{Panter}},
  \bibinfo{author}{M.~{Paz Arribas}}, \bibinfo{author}{G.~{Pedaletti}},
  \bibinfo{author}{G.~{Pelletier}}, \bibinfo{author}{P.~O. {Petrucci}},
  \bibinfo{author}{S.~{Pita}}, \bibinfo{author}{G.~{P{\"u}hlhofer}},
  \bibinfo{author}{M.~{Punch}}, \bibinfo{author}{A.~{Quirrenbach}},
  \bibinfo{author}{B.~C. {Raubenheimer}}, \bibinfo{author}{M.~{Raue}},
  \bibinfo{author}{S.~M. {Rayner}}, \bibinfo{author}{O.~{Reimer}},
  \bibinfo{author}{M.~{Renaud}}, \bibinfo{author}{R.~{de los Reyes}},
  \bibinfo{author}{F.~{Rieger}}, \bibinfo{author}{J.~{Ripken}},
  \bibinfo{author}{L.~{Rob}}, \bibinfo{author}{S.~{Rosier-Lees}},
  \bibinfo{author}{G.~{Rowell}}, \bibinfo{author}{B.~{Rudak}},
  \bibinfo{author}{C.~B. {Rulten}}, \bibinfo{author}{J.~{Ruppel}},
  \bibinfo{author}{F.~{Ryde}}, \bibinfo{author}{V.~{Sahakian}},
  \bibinfo{author}{A.~{Santangelo}}, \bibinfo{author}{R.~{Schlickeiser}},
  \bibinfo{author}{F.~M. {Sch{\"o}ck}}, \bibinfo{author}{A.~{Sch{\"o}nwald}},
  \bibinfo{author}{U.~{Schwanke}}, \bibinfo{author}{S.~{Schwarzburg}},
  \bibinfo{author}{S.~{Schwemmer}}, \bibinfo{author}{A.~{Shalchi}},
  \bibinfo{author}{I.~{Sushch}}, \bibinfo{author}{M.~{Sikora}},
  \bibinfo{author}{J.~L. {Skilton}}, \bibinfo{author}{H.~{Sol}},
  \bibinfo{author}{{\L}.~{Stawarz}}, \bibinfo{author}{R.~{Steenkamp}},
  \bibinfo{author}{C.~{Stegmann}}, \bibinfo{author}{F.~{Stinzing}},
  \bibinfo{author}{G.~{Superina}}, \bibinfo{author}{A.~{Szostek}},
  \bibinfo{author}{P.~H. {Tam}}, \bibinfo{author}{J.~P. {Tavernet}},
  \bibinfo{author}{R.~{Terrier}}, \bibinfo{author}{O.~{Tibolla}},
  \bibinfo{author}{M.~{Tluczykont}}, \bibinfo{author}{K.~{Valerius}},
  \bibinfo{author}{C.~{van Eldik}}, \bibinfo{author}{G.~{Vasileiadis}},
  \bibinfo{author}{C.~{Venter}}, \bibinfo{author}{L.~{Venter}},
  \bibinfo{author}{J.~P. {Vialle}}, \bibinfo{author}{A.~{Viana}},
  \bibinfo{author}{P.~{Vincent}}, \bibinfo{author}{M.~{Vivier}},
  \bibinfo{author}{H.~J. {V{\"o}lk}}, \bibinfo{author}{F.~{Volpe}},
  \bibinfo{author}{S.~{Vorobiov}}, \bibinfo{author}{S.~J. {Wagner}},
  \bibinfo{author}{M.~{Ward}}, \bibinfo{author}{A.~A. {Zdziarski}},
  \bibinfo{author}{A.~{Zech}}, \bibinfo{author}{H.~S. {Zechlin}},
\newblock \bibinfo{title}{{VHE {\ensuremath{\gamma}}-ray emission of PKS
  2155-304: spectral and temporal variability}},
\newblock \bibinfo{journal}{\aap} \bibinfo{volume}{520} (\bibinfo{year}{2010})
  \bibinfo{pages}{A83}. \DOIprefix\doi{10.1051/0004-6361/201014484}.
  \href{http://arxiv.org/abs/1005.3702}{{\tt arXiv:1005.3702}}.
%Type = Article
\bibitem[{{Punch} et~al.(1992){Punch}, {Akerlof}, {Cawley}, {Chantell},
  {Fegan}, {Fennell}, {Gaidos}, {Hagan}, {Hillas}, {Jiang}, {Kerrick}, {Lamb},
  {Lawrence}, {Lewis}, {Meyer}, {Mohanty}, {O'Flaherty}, {Reynolds}, {Rovero},
  {Schubnell}, {Sembroski}, {Weekes}, {Whitaker}, and {Wilson}}]{Whipple:421}
\bibinfo{author}{M.~{Punch}}, \bibinfo{author}{C.~W. {Akerlof}},
  \bibinfo{author}{M.~F. {Cawley}}, \bibinfo{author}{M.~{Chantell}},
  \bibinfo{author}{D.~J. {Fegan}}, \bibinfo{author}{S.~{Fennell}},
  \bibinfo{author}{J.~A. {Gaidos}}, \bibinfo{author}{J.~{Hagan}},
  \bibinfo{author}{A.~M. {Hillas}}, \bibinfo{author}{Y.~{Jiang}},
  \bibinfo{author}{A.~D. {Kerrick}}, \bibinfo{author}{R.~C. {Lamb}},
  \bibinfo{author}{M.~A. {Lawrence}}, \bibinfo{author}{D.~A. {Lewis}},
  \bibinfo{author}{D.~I. {Meyer}}, \bibinfo{author}{G.~{Mohanty}},
  \bibinfo{author}{K.~S. {O'Flaherty}}, \bibinfo{author}{P.~T. {Reynolds}},
  \bibinfo{author}{A.~C. {Rovero}}, \bibinfo{author}{M.~S. {Schubnell}},
  \bibinfo{author}{G.~{Sembroski}}, \bibinfo{author}{T.~C. {Weekes}},
  \bibinfo{author}{T.~{Whitaker}}, \bibinfo{author}{C.~{Wilson}},
\newblock \bibinfo{title}{{Detection of TeV photons from the active galaxy
  Markarian 421}},
\newblock \bibinfo{journal}{\nat} \bibinfo{volume}{358} (\bibinfo{year}{1992})
  \bibinfo{pages}{477--478}. \DOIprefix\doi{10.1038/358477a0}.
%Type = Article
\bibitem[{{Albert} et~al.(2022){Albert}, {Alfaro}, {Alvarez}, {Camacho},
  {Arteaga-Vel{\'a}zquez}, {Arunbabu}, {Rojas}, {Solares}, {Baghmanyan},
  {Belmont-Moreno}, {Caballero-Mora}, {Capistr{\'a}n}, {Carrami{\~n}ana},
  {Casanova}, {Cotti}, {Cotzomi}, {de Le{\'o}n}, {de La Fuente}, {Hernandez},
  {Duvernois}, {Durocher}, {D{\'\i}az-V{\'e}lez}, {Engel}, {Espinoza}, {Fan},
  {Alonso}, {Fraija}, {Garcia}, {Garc{\'\i}a-Gonz{\'a}lez}, {Garfias},
  {Gonz{\'a}lez}, {Goodman}, {Harding}, {Hona}, {Huang}, {Hueyotl-Zahuantitla},
  {H{\"u}ntemeyer}, {Iriarte}, {Joshi}, {Lara}, {Lee}, {Lee}, {Vargas},
  {Linneman}, {Longinotti}, {Luis-Raya}, {Malone}, {Martinez},
  {Mart{\'\i}nez-Castro}, {Matthews}, {Miranda-Romagnoli}, {Moreno},
  {Mostaf{\'a}}, {Nayerhoda}, {Nellen}, {Newbold}, {Noriega-Papaqui},
  {Peisker}, {Araujo}, {P{\'e}rez-P{\'e}rez}, {Rho}, {Rosa-Gonz{\'a}lez},
  {Salazar}, {Greus}, {Sandoval}, {Schneider}, {Serna-Franco}, {Smith},
  {Springer}, {Tollefson}, {Torres}, {Torres-Escobedo}, {Ure{\~n}a-Mena},
  {Villase{\~n}or}, {Wang}, {Weisgarber}, {Willox}, {Zhou}, {de Le{\'o}n}, and
  {The Hawc Collaboration}}]{HAWC:421}
\bibinfo{author}{A.~{Albert}}, \bibinfo{author}{R.~{Alfaro}},
  \bibinfo{author}{C.~{Alvarez}}, \bibinfo{author}{J.~R.~A. {Camacho}},
  \bibinfo{author}{J.~C. {Arteaga-Vel{\'a}zquez}}, \bibinfo{author}{K.~P.
  {Arunbabu}}, \bibinfo{author}{D.~A. {Rojas}}, \bibinfo{author}{H.~A.~A.
  {Solares}}, \bibinfo{author}{V.~{Baghmanyan}},
  \bibinfo{author}{E.~{Belmont-Moreno}}, \bibinfo{author}{K.~S.
  {Caballero-Mora}}, \bibinfo{author}{T.~{Capistr{\'a}n}},
  \bibinfo{author}{A.~{Carrami{\~n}ana}}, \bibinfo{author}{S.~{Casanova}},
  \bibinfo{author}{U.~{Cotti}}, \bibinfo{author}{J.~{Cotzomi}},
  \bibinfo{author}{S.~C. {de Le{\'o}n}}, \bibinfo{author}{E.~{de La Fuente}},
  \bibinfo{author}{R.~D. {Hernandez}}, \bibinfo{author}{M.~A. {Duvernois}},
  \bibinfo{author}{M.~{Durocher}}, \bibinfo{author}{J.~C.
  {D{\'\i}az-V{\'e}lez}}, \bibinfo{author}{K.~{Engel}},
  \bibinfo{author}{C.~{Espinoza}}, \bibinfo{author}{K.~L. {Fan}},
  \bibinfo{author}{M.~F. {Alonso}}, \bibinfo{author}{N.~{Fraija}},
  \bibinfo{author}{D.~{Garcia}}, \bibinfo{author}{J.~A.
  {Garc{\'\i}a-Gonz{\'a}lez}}, \bibinfo{author}{F.~{Garfias}},
  \bibinfo{author}{M.~M. {Gonz{\'a}lez}}, \bibinfo{author}{J.~A. {Goodman}},
  \bibinfo{author}{J.~P. {Harding}}, \bibinfo{author}{B.~{Hona}},
  \bibinfo{author}{D.~{Huang}}, \bibinfo{author}{F.~{Hueyotl-Zahuantitla}},
  \bibinfo{author}{P.~{H{\"u}ntemeyer}}, \bibinfo{author}{A.~{Iriarte}},
  \bibinfo{author}{V.~{Joshi}}, \bibinfo{author}{A.~{Lara}},
  \bibinfo{author}{W.~H. {Lee}}, \bibinfo{author}{J.~{Lee}},
  \bibinfo{author}{H.~L. {Vargas}}, \bibinfo{author}{J.~T. {Linneman}},
  \bibinfo{author}{A.~L. {Longinotti}}, \bibinfo{author}{G.~{Luis-Raya}},
  \bibinfo{author}{K.~{Malone}}, \bibinfo{author}{O.~{Martinez}},
  \bibinfo{author}{J.~{Mart{\'\i}nez-Castro}}, \bibinfo{author}{J.~A.
  {Matthews}}, \bibinfo{author}{P.~{Miranda-Romagnoli}},
  \bibinfo{author}{E.~{Moreno}}, \bibinfo{author}{M.~{Mostaf{\'a}}},
  \bibinfo{author}{A.~{Nayerhoda}}, \bibinfo{author}{L.~{Nellen}},
  \bibinfo{author}{M.~{Newbold}}, \bibinfo{author}{R.~{Noriega-Papaqui}},
  \bibinfo{author}{A.~{Peisker}}, \bibinfo{author}{Y.~P. {Araujo}},
  \bibinfo{author}{E.~G. {P{\'e}rez-P{\'e}rez}}, \bibinfo{author}{C.~D. {Rho}},
  \bibinfo{author}{D.~{Rosa-Gonz{\'a}lez}}, \bibinfo{author}{H.~{Salazar}},
  \bibinfo{author}{F.~S. {Greus}}, \bibinfo{author}{A.~{Sandoval}},
  \bibinfo{author}{M.~{Schneider}}, \bibinfo{author}{J.~{Serna-Franco}},
  \bibinfo{author}{A.~J. {Smith}}, \bibinfo{author}{R.~W. {Springer}},
  \bibinfo{author}{K.~{Tollefson}}, \bibinfo{author}{I.~{Torres}},
  \bibinfo{author}{R.~{Torres-Escobedo}},
  \bibinfo{author}{F.~{Ure{\~n}a-Mena}}, \bibinfo{author}{L.~{Villase{\~n}or}},
  \bibinfo{author}{X.~{Wang}}, \bibinfo{author}{T.~{Weisgarber}},
  \bibinfo{author}{E.~{Willox}}, \bibinfo{author}{H.~{Zhou}},
  \bibinfo{author}{C.~{de Le{\'o}n}}, \bibinfo{author}{{The Hawc
  Collaboration}},
\newblock \bibinfo{title}{{Long-term Spectra of the Blazars Mrk 421 and Mrk 501
  at TeV Energies Seen by HAWC}},
\newblock \bibinfo{journal}{\apj} \bibinfo{volume}{929} (\bibinfo{year}{2022})
  \bibinfo{pages}{125}. \DOIprefix\doi{10.3847/1538-4357/ac58f6}.
  \href{http://arxiv.org/abs/2106.03946}{{\tt arXiv:2106.03946}}.
%Type = Article
\bibitem[{{Aleksi{\'c}} et~al.(2012){Aleksi{\'c}}, {Alvarez}, {Antonelli},
  {Antoranz}, {Ansoldi}, {Asensio}, {Backes}, {Barres de Almeida}, {Barrio},
  {Bastieri}, {Becerra Gonz{\'a}lez}, {Bednarek}, {Berger}, {Bernardini},
  {Biland}, {Blanch}, {Bock}, {Boller}, {Bonnoli}, {Borla Tridon}, {Bretz},
  {Ca{\~n}ellas}, {Carmona}, {Carosi}, {Colin}, {Colombo}, {Contreras},
  {Cortina}, {Cossio}, {Covino}, {Da Vela}, {Dazzi}, {De Angelis}, {De Caneva},
  {De Cea del Pozo}, {De Lotto}, {Delgado Mendez}, {Diago Ortega}, {Doert},
  {Dom{\'\i}nguez}, {Dominis Prester}, {Dorner}, {Doro}, {Eisenacher},
  {Elsaesser}, {Ferenc}, {Fonseca}, {Font}, {Fruck}, {Garc{\'\i}a L{\'o}pez},
  {Garczarczyk}, {Garrido Terrats}, {Gaug}, {Giavitto}, {Godinovi{\'c}},
  {Gonz{\'a}lez Mu{\~n}oz}, {Gozzini}, {Hadasch}, {H{\"a}fner}, {Herrero},
  {Hildebrand}, {Hose}, {Hrupec}, {Huber}, {Jankowski}, {Jogler}, {Kadenius},
  {Kellermann}, {Klepser}, {Kr{\"a}henb{\"u}hl}, {Krause}, {La Barbera},
  {Lelas}, {Leonardo}, {Lewandowska}, {Lindfors}, {Lombardi}, {L{\'o}pez},
  {L{\'o}pez-Coto}, {L{\'o}pez-Oramas}, {Lorenz}, {Makariev}, {Maneva},
  {Mankuzhiyil}, {Mannheim}, {Maraschi}, {Mariotti}, {Mart{\'\i}nez}, {Mazin},
  {Meucci}, {Miranda}, {Mirzoyan}, {Mold{\'o}n}, {Moralejo}, {Munar-Adrover},
  {Niedzwiecki}, {Nieto}, {Nilsson}, {Nowak}, {Orito}, {Paiano}, {Paneque},
  {Paoletti}, {Pardo}, {Paredes}, {Partini}, {Perez-Torres}, {Persic}, {Pilia},
  {Pochon}, {Prada}, {Prada Moroni}, {Prandini}, {Puerto Gimenez}, {Puljak},
  {Reichardt}, {Reinthal}, {Rhode}, {Rib{\'o}}, {Rico}, {R{\"u}gamer},
  {Saggion}, {Saito}, {Saito}, {Salvati}, {Satalecka}, {Scalzotto}, {Scapin},
  {Schultz}, {Schweizer}, {Shore}, {Sillanp{\"a}{\"a}}, {Sitarek}, {Snidaric},
  {Sobczynska}, {Spanier}, {Spiro}, {Stamatescu}, {Stamerra}, {Steinke},
  {Storz}, {Strah}, {Sun}, {Suri{\'c}}, {Takalo}, {Takami}, {Tavecchio},
  {Temnikov}, {Terzi{\'c}}, {Tescaro}, {Teshima}, {Tibolla}, {Torres},
  {Treves}, {Uellenbeck}, {Vogler}, {Wagner}, {Weitzel}, {Zabalza}, {Zandanel},
  {Zanin}, {Berdyugin}, {Buson}, {J{\"a}rvel{\"a}}, {Larsson},
  {L{\"a}hteenm{\"a}ki}, and {Tammi}}]{MAGIC:1215}
\bibinfo{author}{J.~{Aleksi{\'c}}}, \bibinfo{author}{E.~A. {Alvarez}},
  \bibinfo{author}{L.~A. {Antonelli}}, \bibinfo{author}{P.~{Antoranz}},
  \bibinfo{author}{S.~{Ansoldi}}, \bibinfo{author}{M.~{Asensio}},
  \bibinfo{author}{M.~{Backes}}, \bibinfo{author}{U.~{Barres de Almeida}},
  \bibinfo{author}{J.~A. {Barrio}}, \bibinfo{author}{D.~{Bastieri}},
  \bibinfo{author}{J.~{Becerra Gonz{\'a}lez}}, \bibinfo{author}{W.~{Bednarek}},
  \bibinfo{author}{K.~{Berger}}, \bibinfo{author}{E.~{Bernardini}},
  \bibinfo{author}{A.~{Biland}}, \bibinfo{author}{O.~{Blanch}},
  \bibinfo{author}{R.~K. {Bock}}, \bibinfo{author}{A.~{Boller}},
  \bibinfo{author}{G.~{Bonnoli}}, \bibinfo{author}{D.~{Borla Tridon}},
  \bibinfo{author}{T.~{Bretz}}, \bibinfo{author}{A.~{Ca{\~n}ellas}},
  \bibinfo{author}{E.~{Carmona}}, \bibinfo{author}{A.~{Carosi}},
  \bibinfo{author}{P.~{Colin}}, \bibinfo{author}{E.~{Colombo}},
  \bibinfo{author}{J.~L. {Contreras}}, \bibinfo{author}{J.~{Cortina}},
  \bibinfo{author}{L.~{Cossio}}, \bibinfo{author}{S.~{Covino}},
  \bibinfo{author}{P.~{Da Vela}}, \bibinfo{author}{F.~{Dazzi}},
  \bibinfo{author}{A.~{De Angelis}}, \bibinfo{author}{G.~{De Caneva}},
  \bibinfo{author}{E.~{De Cea del Pozo}}, \bibinfo{author}{B.~{De Lotto}},
  \bibinfo{author}{C.~{Delgado Mendez}}, \bibinfo{author}{A.~{Diago Ortega}},
  \bibinfo{author}{M.~{Doert}}, \bibinfo{author}{A.~{Dom{\'\i}nguez}},
  \bibinfo{author}{D.~{Dominis Prester}}, \bibinfo{author}{D.~{Dorner}},
  \bibinfo{author}{M.~{Doro}}, \bibinfo{author}{D.~{Eisenacher}},
  \bibinfo{author}{D.~{Elsaesser}}, \bibinfo{author}{D.~{Ferenc}},
  \bibinfo{author}{M.~V. {Fonseca}}, \bibinfo{author}{L.~{Font}},
  \bibinfo{author}{C.~{Fruck}}, \bibinfo{author}{R.~J. {Garc{\'\i}a
  L{\'o}pez}}, \bibinfo{author}{M.~{Garczarczyk}}, \bibinfo{author}{D.~{Garrido
  Terrats}}, \bibinfo{author}{M.~{Gaug}}, \bibinfo{author}{G.~{Giavitto}},
  \bibinfo{author}{N.~{Godinovi{\'c}}}, \bibinfo{author}{A.~{Gonz{\'a}lez
  Mu{\~n}oz}}, \bibinfo{author}{S.~R. {Gozzini}},
  \bibinfo{author}{D.~{Hadasch}}, \bibinfo{author}{D.~{H{\"a}fner}},
  \bibinfo{author}{A.~{Herrero}}, \bibinfo{author}{D.~{Hildebrand}},
  \bibinfo{author}{J.~{Hose}}, \bibinfo{author}{D.~{Hrupec}},
  \bibinfo{author}{B.~{Huber}}, \bibinfo{author}{F.~{Jankowski}},
  \bibinfo{author}{T.~{Jogler}}, \bibinfo{author}{V.~{Kadenius}},
  \bibinfo{author}{H.~{Kellermann}}, \bibinfo{author}{S.~{Klepser}},
  \bibinfo{author}{T.~{Kr{\"a}henb{\"u}hl}}, \bibinfo{author}{J.~{Krause}},
  \bibinfo{author}{A.~{La Barbera}}, \bibinfo{author}{D.~{Lelas}},
  \bibinfo{author}{E.~{Leonardo}}, \bibinfo{author}{N.~{Lewandowska}},
  \bibinfo{author}{E.~{Lindfors}}, \bibinfo{author}{S.~{Lombardi}},
  \bibinfo{author}{M.~{L{\'o}pez}}, \bibinfo{author}{R.~{L{\'o}pez-Coto}},
  \bibinfo{author}{A.~{L{\'o}pez-Oramas}}, \bibinfo{author}{E.~{Lorenz}},
  \bibinfo{author}{M.~{Makariev}}, \bibinfo{author}{G.~{Maneva}},
  \bibinfo{author}{N.~{Mankuzhiyil}}, \bibinfo{author}{K.~{Mannheim}},
  \bibinfo{author}{L.~{Maraschi}}, \bibinfo{author}{M.~{Mariotti}},
  \bibinfo{author}{M.~{Mart{\'\i}nez}}, \bibinfo{author}{D.~{Mazin}},
  \bibinfo{author}{M.~{Meucci}}, \bibinfo{author}{J.~M. {Miranda}},
  \bibinfo{author}{R.~{Mirzoyan}}, \bibinfo{author}{J.~{Mold{\'o}n}},
  \bibinfo{author}{A.~{Moralejo}}, \bibinfo{author}{P.~{Munar-Adrover}},
  \bibinfo{author}{A.~{Niedzwiecki}}, \bibinfo{author}{D.~{Nieto}},
  \bibinfo{author}{K.~{Nilsson}}, \bibinfo{author}{N.~{Nowak}},
  \bibinfo{author}{R.~{Orito}}, \bibinfo{author}{S.~{Paiano}},
  \bibinfo{author}{D.~{Paneque}}, \bibinfo{author}{R.~{Paoletti}},
  \bibinfo{author}{S.~{Pardo}}, \bibinfo{author}{J.~M. {Paredes}},
  \bibinfo{author}{S.~{Partini}}, \bibinfo{author}{M.~A. {Perez-Torres}},
  \bibinfo{author}{M.~{Persic}}, \bibinfo{author}{M.~{Pilia}},
  \bibinfo{author}{J.~{Pochon}}, \bibinfo{author}{F.~{Prada}},
  \bibinfo{author}{P.~G. {Prada Moroni}}, \bibinfo{author}{E.~{Prandini}},
  \bibinfo{author}{I.~{Puerto Gimenez}}, \bibinfo{author}{I.~{Puljak}},
  \bibinfo{author}{I.~{Reichardt}}, \bibinfo{author}{R.~{Reinthal}},
  \bibinfo{author}{W.~{Rhode}}, \bibinfo{author}{M.~{Rib{\'o}}},
  \bibinfo{author}{J.~{Rico}}, \bibinfo{author}{S.~{R{\"u}gamer}},
  \bibinfo{author}{A.~{Saggion}}, \bibinfo{author}{K.~{Saito}},
  \bibinfo{author}{T.~Y. {Saito}}, \bibinfo{author}{M.~{Salvati}},
  \bibinfo{author}{K.~{Satalecka}}, \bibinfo{author}{V.~{Scalzotto}},
  \bibinfo{author}{V.~{Scapin}}, \bibinfo{author}{C.~{Schultz}},
  \bibinfo{author}{T.~{Schweizer}}, \bibinfo{author}{S.~N. {Shore}},
  \bibinfo{author}{A.~{Sillanp{\"a}{\"a}}}, \bibinfo{author}{J.~{Sitarek}},
  \bibinfo{author}{I.~{Snidaric}}, \bibinfo{author}{D.~{Sobczynska}},
  \bibinfo{author}{F.~{Spanier}}, \bibinfo{author}{S.~{Spiro}},
  \bibinfo{author}{V.~{Stamatescu}}, \bibinfo{author}{A.~{Stamerra}},
  \bibinfo{author}{B.~{Steinke}}, \bibinfo{author}{J.~{Storz}},
  \bibinfo{author}{N.~{Strah}}, \bibinfo{author}{S.~{Sun}},
  \bibinfo{author}{T.~{Suri{\'c}}}, \bibinfo{author}{L.~{Takalo}},
  \bibinfo{author}{H.~{Takami}}, \bibinfo{author}{F.~{Tavecchio}},
  \bibinfo{author}{P.~{Temnikov}}, \bibinfo{author}{T.~{Terzi{\'c}}},
  \bibinfo{author}{D.~{Tescaro}}, \bibinfo{author}{M.~{Teshima}},
  \bibinfo{author}{O.~{Tibolla}}, \bibinfo{author}{D.~F. {Torres}},
  \bibinfo{author}{A.~{Treves}}, \bibinfo{author}{M.~{Uellenbeck}},
  \bibinfo{author}{P.~{Vogler}}, \bibinfo{author}{R.~M. {Wagner}},
  \bibinfo{author}{Q.~{Weitzel}}, \bibinfo{author}{V.~{Zabalza}},
  \bibinfo{author}{F.~{Zandanel}}, \bibinfo{author}{R.~{Zanin}},
  \bibinfo{author}{A.~{Berdyugin}}, \bibinfo{author}{S.~{Buson}},
  \bibinfo{author}{E.~{J{\"a}rvel{\"a}}}, \bibinfo{author}{S.~{Larsson}},
  \bibinfo{author}{A.~{L{\"a}hteenm{\"a}ki}}, \bibinfo{author}{J.~{Tammi}},
\newblock \bibinfo{title}{{Discovery of VHE {\ensuremath{\gamma}}-rays from the
  blazar 1ES 1215+303 with the MAGIC telescopes and simultaneous
  multi-wavelength observations}},
\newblock \bibinfo{journal}{\aap} \bibinfo{volume}{544} (\bibinfo{year}{2012})
  \bibinfo{pages}{A142}. \DOIprefix\doi{10.1051/0004-6361/201219133}.
  \href{http://arxiv.org/abs/1203.0490}{{\tt arXiv:1203.0490}}.
%Type = Article
\bibitem[{{Abeysekara} et~al.(2017){Abeysekara}, {Archambault}, {Archer},
  {Benbow}, {Bird}, {Buchovecky}, {Buckley}, {Bugaev}, {Byrum}, {Cerruti},
  {Chen}, {Ciupik}, {Cui}, {Dickinson}, {Eisch}, {Errando}, {Falcone}, {Feng},
  {Finley}, {Fleischhack}, {Fortson}, {Furniss}, {Gillanders}, {Griffin},
  {Grube}, {H{\"u}tten}, {H{\r{a}}kansson}, {Hanna}, {Holder}, {Humensky},
  {Johnson}, {Kaaret}, {Kar}, {Kertzman}, {Kieda}, {Krause}, {Krennrich},
  {Kumar}, {Lang}, {Maier}, {McArthur}, {McCann}, {Meagher}, {Moriarty},
  {Mukherjee}, {Nguyen}, {Nieto}, {Ong}, {Otte}, {Park}, {Pelassa}, {Pohl},
  {Popkow}, {Pueschel}, {Quinn}, {Ragan}, {Reynolds}, {Richards}, {Roache},
  {Rulten}, {Santander}, {Sembroski}, {Shahinyan}, {Staszak}, {Telezhinsky},
  {Tucci}, {Tyler}, {Wakely}, {Weiner}, {Weinstein}, {Wilhelm}, {Williams},
  {VERITAS Collaboration}, {Fegan}, {Giebels}, {Horan}, {Fermi-LAT
  Collaboration}, {Berdyugin}, {Kuan}, {Lindfors}, {Nilsson}, {Oksanen},
  {Prokoph}, {Reinthal}, {Takalo}, and {Zefi}}]{VERITAS:1215_flare}
\bibinfo{author}{A.~U. {Abeysekara}}, \bibinfo{author}{S.~{Archambault}},
  \bibinfo{author}{A.~{Archer}}, \bibinfo{author}{W.~{Benbow}},
  \bibinfo{author}{R.~{Bird}}, \bibinfo{author}{M.~{Buchovecky}},
  \bibinfo{author}{J.~H. {Buckley}}, \bibinfo{author}{V.~{Bugaev}},
  \bibinfo{author}{K.~{Byrum}}, \bibinfo{author}{M.~{Cerruti}},
  \bibinfo{author}{X.~{Chen}}, \bibinfo{author}{L.~{Ciupik}},
  \bibinfo{author}{W.~{Cui}}, \bibinfo{author}{H.~J. {Dickinson}},
  \bibinfo{author}{J.~D. {Eisch}}, \bibinfo{author}{M.~{Errando}},
  \bibinfo{author}{A.~{Falcone}}, \bibinfo{author}{Q.~{Feng}},
  \bibinfo{author}{J.~P. {Finley}}, \bibinfo{author}{H.~{Fleischhack}},
  \bibinfo{author}{L.~{Fortson}}, \bibinfo{author}{A.~{Furniss}},
  \bibinfo{author}{G.~H. {Gillanders}}, \bibinfo{author}{S.~{Griffin}},
  \bibinfo{author}{J.~{Grube}}, \bibinfo{author}{M.~{H{\"u}tten}},
  \bibinfo{author}{N.~{H{\r{a}}kansson}}, \bibinfo{author}{D.~{Hanna}},
  \bibinfo{author}{J.~{Holder}}, \bibinfo{author}{T.~B. {Humensky}},
  \bibinfo{author}{C.~A. {Johnson}}, \bibinfo{author}{P.~{Kaaret}},
  \bibinfo{author}{P.~{Kar}}, \bibinfo{author}{M.~{Kertzman}},
  \bibinfo{author}{D.~{Kieda}}, \bibinfo{author}{M.~{Krause}},
  \bibinfo{author}{F.~{Krennrich}}, \bibinfo{author}{S.~{Kumar}},
  \bibinfo{author}{M.~J. {Lang}}, \bibinfo{author}{G.~{Maier}},
  \bibinfo{author}{S.~{McArthur}}, \bibinfo{author}{A.~{McCann}},
  \bibinfo{author}{K.~{Meagher}}, \bibinfo{author}{P.~{Moriarty}},
  \bibinfo{author}{R.~{Mukherjee}}, \bibinfo{author}{T.~{Nguyen}},
  \bibinfo{author}{D.~{Nieto}}, \bibinfo{author}{R.~A. {Ong}},
  \bibinfo{author}{A.~N. {Otte}}, \bibinfo{author}{N.~{Park}},
  \bibinfo{author}{V.~{Pelassa}}, \bibinfo{author}{M.~{Pohl}},
  \bibinfo{author}{A.~{Popkow}}, \bibinfo{author}{E.~{Pueschel}},
  \bibinfo{author}{J.~{Quinn}}, \bibinfo{author}{K.~{Ragan}},
  \bibinfo{author}{P.~T. {Reynolds}}, \bibinfo{author}{G.~T. {Richards}},
  \bibinfo{author}{E.~{Roache}}, \bibinfo{author}{C.~{Rulten}},
  \bibinfo{author}{M.~{Santander}}, \bibinfo{author}{G.~H. {Sembroski}},
  \bibinfo{author}{K.~{Shahinyan}}, \bibinfo{author}{D.~{Staszak}},
  \bibinfo{author}{I.~{Telezhinsky}}, \bibinfo{author}{J.~V. {Tucci}},
  \bibinfo{author}{J.~{Tyler}}, \bibinfo{author}{S.~P. {Wakely}},
  \bibinfo{author}{O.~M. {Weiner}}, \bibinfo{author}{A.~{Weinstein}},
  \bibinfo{author}{A.~{Wilhelm}}, \bibinfo{author}{D.~A. {Williams}},
  \bibinfo{author}{{VERITAS Collaboration}}, \bibinfo{author}{S.~{Fegan}},
  \bibinfo{author}{B.~{Giebels}}, \bibinfo{author}{D.~{Horan}},
  \bibinfo{author}{{Fermi-LAT Collaboration}},
  \bibinfo{author}{A.~{Berdyugin}}, \bibinfo{author}{J.~{Kuan}},
  \bibinfo{author}{E.~{Lindfors}}, \bibinfo{author}{K.~{Nilsson}},
  \bibinfo{author}{A.~{Oksanen}}, \bibinfo{author}{H.~{Prokoph}},
  \bibinfo{author}{R.~{Reinthal}}, \bibinfo{author}{L.~{Takalo}},
  \bibinfo{author}{F.~{Zefi}},
\newblock \bibinfo{title}{{A Luminous and Isolated Gamma-ray Flare from the
  Blazar B2 1215+30}},
\newblock \bibinfo{journal}{\apj} \bibinfo{volume}{836} (\bibinfo{year}{2017})
  \bibinfo{pages}{205}. \DOIprefix\doi{10.3847/1538-4357/836/2/205}.
  \href{http://arxiv.org/abs/1701.01067}{{\tt arXiv:1701.01067}}.
%Type = Article
\bibitem[{{Albert} et~al.(2021){Albert}, {Alvarez}, {Angeles Camacho},
  {Arteaga-Vel{\'a}zquez}, {Arunbabu}, {Avila Rojas}, {Ayala Solares},
  {Baghmanyan}, {Belmont-Moreno}, {BenZvi}, {Brisbois}, {Caballero-Mora},
  {Capistr{\'a}n}, {Carrami{\~n}ana}, {Casanova}, {Cotti}, {Cotzomi},
  {Couti{\~n}o de Le{\'o}n}, {De la Fuente}, {Dingus}, {DuVernois}, {Durocher},
  {D{\'\i}az-V{\'e}lez}, {Engel}, {Espinoza}, {Fan}, {Fern{\'a}ndez Alonso},
  {Fleischhack}, {Fraija}, {Galv{\'a}n-G{\'a}mez}, {Garc{\'\i}a},
  {Garc{\'\i}a-Gonz{\'a}lez}, {Garfias}, {Gonz{\'a}lez}, {Goodman}, {Harding},
  {Hern{\'a}ndez}, {Hona}, {Huang}, {Hueyotl-Zahuantitla}, {H{\"u}ntemeyer},
  {Iriarte}, {Jardin-Blicq}, {Joshi}, {Kieda}, {Kunde}, {Lara}, {Lee},
  {Le{\'o}n Vargas}, {Linnemann}, {Longinotti}, {Luis-Raya}, {Lundeen},
  {Malone}, {Mart{\'\i}nez}, {Martinez-Castellanos}, {Mart{\'\i}nez-Castro},
  {Matthews}, {Miranda-Romagnoli}, {Morales-Soto}, {Moreno}, {Mostaf{\'a}},
  {Nayerhoda}, {Nellen}, {Newbold}, {Nisa}, {Noriega-Papaqui}, {Olivera-Nieto},
  {Peisker}, {P{\'e}rez-P{\'e}rez}, {Rho}, {Rosa-Gonz{\'a}lez}, {Ruiz-Velasco},
  {Salazar}, {Greus}, {Sandoval}, {Schneider}, {Schoorlemmer}, {Smith},
  {Springer}, {Tollefson}, {Torres}, {Torres-Escobedo}, {Ure{\~n}a-Mena},
  {Villase{\~n}or}, {Weisgarber}, {Willox}, {Zepeda}, {Zhou}, {de Le{\'o}n},
  and {HAWC Collaboration}}]{HAWC:AGN_survey}
\bibinfo{author}{A.~{Albert}}, \bibinfo{author}{C.~{Alvarez}},
  \bibinfo{author}{J.~R. {Angeles Camacho}}, \bibinfo{author}{J.~C.
  {Arteaga-Vel{\'a}zquez}}, \bibinfo{author}{K.~P. {Arunbabu}},
  \bibinfo{author}{D.~{Avila Rojas}}, \bibinfo{author}{H.~A. {Ayala Solares}},
  \bibinfo{author}{V.~{Baghmanyan}}, \bibinfo{author}{E.~{Belmont-Moreno}},
  \bibinfo{author}{S.~Y. {BenZvi}}, \bibinfo{author}{C.~{Brisbois}},
  \bibinfo{author}{K.~S. {Caballero-Mora}},
  \bibinfo{author}{T.~{Capistr{\'a}n}}, \bibinfo{author}{A.~{Carrami{\~n}ana}},
  \bibinfo{author}{S.~{Casanova}}, \bibinfo{author}{U.~{Cotti}},
  \bibinfo{author}{J.~{Cotzomi}}, \bibinfo{author}{S.~{Couti{\~n}o de
  Le{\'o}n}}, \bibinfo{author}{E.~{De la Fuente}}, \bibinfo{author}{B.~L.
  {Dingus}}, \bibinfo{author}{M.~A. {DuVernois}},
  \bibinfo{author}{M.~{Durocher}}, \bibinfo{author}{J.~C.
  {D{\'\i}az-V{\'e}lez}}, \bibinfo{author}{K.~{Engel}},
  \bibinfo{author}{C.~{Espinoza}}, \bibinfo{author}{K.~L. {Fan}},
  \bibinfo{author}{M.~{Fern{\'a}ndez Alonso}},
  \bibinfo{author}{H.~{Fleischhack}}, \bibinfo{author}{N.~{Fraija}},
  \bibinfo{author}{A.~{Galv{\'a}n-G{\'a}mez}},
  \bibinfo{author}{D.~{Garc{\'\i}a}}, \bibinfo{author}{J.~A.
  {Garc{\'\i}a-Gonz{\'a}lez}}, \bibinfo{author}{F.~{Garfias}},
  \bibinfo{author}{M.~M. {Gonz{\'a}lez}}, \bibinfo{author}{J.~A. {Goodman}},
  \bibinfo{author}{J.~P. {Harding}}, \bibinfo{author}{S.~{Hern{\'a}ndez}},
  \bibinfo{author}{B.~{Hona}}, \bibinfo{author}{D.~{Huang}},
  \bibinfo{author}{F.~{Hueyotl-Zahuantitla}},
  \bibinfo{author}{P.~{H{\"u}ntemeyer}}, \bibinfo{author}{A.~{Iriarte}},
  \bibinfo{author}{A.~{Jardin-Blicq}}, \bibinfo{author}{V.~{Joshi}},
  \bibinfo{author}{D.~{Kieda}}, \bibinfo{author}{G.~J. {Kunde}},
  \bibinfo{author}{A.~{Lara}}, \bibinfo{author}{W.~H. {Lee}},
  \bibinfo{author}{H.~{Le{\'o}n Vargas}}, \bibinfo{author}{J.~T. {Linnemann}},
  \bibinfo{author}{A.~L. {Longinotti}}, \bibinfo{author}{G.~{Luis-Raya}},
  \bibinfo{author}{J.~{Lundeen}}, \bibinfo{author}{K.~{Malone}},
  \bibinfo{author}{O.~{Mart{\'\i}nez}},
  \bibinfo{author}{I.~{Martinez-Castellanos}},
  \bibinfo{author}{J.~{Mart{\'\i}nez-Castro}}, \bibinfo{author}{J.~A.
  {Matthews}}, \bibinfo{author}{P.~{Miranda-Romagnoli}}, \bibinfo{author}{J.~A.
  {Morales-Soto}}, \bibinfo{author}{E.~{Moreno}},
  \bibinfo{author}{M.~{Mostaf{\'a}}}, \bibinfo{author}{A.~{Nayerhoda}},
  \bibinfo{author}{L.~{Nellen}}, \bibinfo{author}{M.~{Newbold}},
  \bibinfo{author}{M.~U. {Nisa}}, \bibinfo{author}{R.~{Noriega-Papaqui}},
  \bibinfo{author}{L.~{Olivera-Nieto}}, \bibinfo{author}{A.~{Peisker}},
  \bibinfo{author}{E.~G. {P{\'e}rez-P{\'e}rez}}, \bibinfo{author}{C.~D. {Rho}},
  \bibinfo{author}{D.~{Rosa-Gonz{\'a}lez}},
  \bibinfo{author}{E.~{Ruiz-Velasco}}, \bibinfo{author}{H.~{Salazar}},
  \bibinfo{author}{F.~S. {Greus}}, \bibinfo{author}{A.~{Sandoval}},
  \bibinfo{author}{M.~{Schneider}}, \bibinfo{author}{H.~{Schoorlemmer}},
  \bibinfo{author}{A.~J. {Smith}}, \bibinfo{author}{R.~W. {Springer}},
  \bibinfo{author}{K.~{Tollefson}}, \bibinfo{author}{I.~{Torres}},
  \bibinfo{author}{R.~{Torres-Escobedo}},
  \bibinfo{author}{F.~{Ure{\~n}a-Mena}}, \bibinfo{author}{L.~{Villase{\~n}or}},
  \bibinfo{author}{T.~{Weisgarber}}, \bibinfo{author}{E.~{Willox}},
  \bibinfo{author}{A.~{Zepeda}}, \bibinfo{author}{H.~{Zhou}},
  \bibinfo{author}{C.~{de Le{\'o}n}}, \bibinfo{author}{{HAWC Collaboration}},
\newblock \bibinfo{title}{{A Survey of Active Galaxies at TeV Photon Energies
  with the HAWC Gamma-Ray Observatory}},
\newblock \bibinfo{journal}{\apj} \bibinfo{volume}{907} (\bibinfo{year}{2021})
  \bibinfo{pages}{67}. \DOIprefix\doi{10.3847/1538-4357/abca9a}.
  \href{http://arxiv.org/abs/2009.09039}{{\tt arXiv:2009.09039}}.
%Type = Article
\bibitem[{{MAGIC Collaboration} et~al.(2019){MAGIC Collaboration}, {Acciari},
  {Ansoldi}, {Antonelli}, {Arbet Engels}, {Baack}, {Babi{\'c}}, {Banerjee},
  {Barres de Almeida}, {Barrio}, {Becerra Gonz{\'a}lez}, {Bednarek},
  {Bellizzi}, {Bernardini}, {Berti}, {Besenrieder}, {Bhattacharyya},
  {Bigongiari}, {Biland}, {Blanch}, {Bonnoli}, {Bo{\v{s}}njak}, {Busetto},
  {Carosi}, {Carosi}, {Ceribella}, {Chai}, {Chilingaryan}, {Cikota}, {Colak},
  {Colin}, {Colombo}, {Contreras}, {Cortina}, {Covino}, {D'Amico}, {D'Elia},
  {da Vela}, {Dazzi}, {de Angelis}, {de Lotto}, {Delfino}, {Delgado},
  {Depaoli}, {di Pierro}, {di Venere}, {Do Souto Espi{\~n}eira}, {Dominis
  Prester}, {Donini}, {Dorner}, {Doro}, {Elsaesser}, {Fallah Ramazani},
  {Fattorini}, {Fern{\'a}ndez-Barral}, {Ferrara}, {Fidalgo}, {Foffano},
  {Fonseca}, {Font}, {Fruck}, {Fukami}, {Gallozzi}, {Garc{\'\i}a L{\'o}pez},
  {Garczarczyk}, {Gasparyan}, {Gaug}, {Giglietto}, {Giordano}, {Godinovi{\'c}},
  {Green}, {Guberman}, {Hadasch}, {Hahn}, {Herrera}, {Hoang}, {Hrupec},
  {H{\"u}tten}, {Inada}, {Inoue}, {Ishio}, {Iwamura}, {Jouvin}, {Kerszberg},
  {Kubo}, {Kushida}, {Lamastra}, {Lelas}, {Leone}, {Lindfors}, {Lombardi},
  {Longo}, {L{\'o}pez}, {L{\'o}pez-Coto}, {L{\'o}pez-Oramas}, {Loporchio},
  {Machado de Oliveira Fraga}, {Maggio}, {Majumdar}, {Makariev}, {Mallamaci},
  {Maneva}, {Manganaro}, {Mannheim}, {Maraschi}, {Mariotti}, {Mart{\'\i}nez},
  {Masuda}, {Mazin}, {Mi{\'c}anovi{\'c}}, {Miceli}, {Minev}, {Miranda},
  {Mirzoyan}, {Molina}, {Moralejo}, {Morcuende}, {Moreno}, {Moretti},
  {Munar-Adrover}, {Neustroev}, {Nigro}, {Nilsson}, {Ninci}, {Nishijima},
  {Noda}, {Nogu{\'e}s}, {N{\"o}the}, {Nozaki}, {Paiano}, {Palacio},
  {Palatiello}, {Paneque}, {Paoletti}, {Paredes}, {Pe{\~n}il}, {Peresano},
  {Persic}, {Prada Moroni}, {Prandini}, {Puljak}, {Rhode}, {Rib{\'o}}, {Rico},
  {Righi}, {Rugliancich}, {Saha}, {Sahakyan}, {Saito}, {Sakurai}, {Satalecka},
  {Schmidt}, {Schweizer}, {Sitarek}, {{\v{S}}nidari{\'c}}, {Sobczynska},
  {Somero}, {Stamerra}, {Strom}, {Strzys}, {Suda}, {Suri{\'c}}, {Takahashi},
  {Tavecchio}, {Temnikov}, {Terzi{\'c}}, {Teshima}, {Torres-Alb{\`a}}, {Tosti},
  {Tsujimoto}, {Vagelli}, {van Scherpenberg}, {Vanzo}, {Vazquez Acosta},
  {Vigorito}, {Vitale}, {Vovk}, {Will}, {Zari{\'c}}, and
  {Nava}}]{MAGIC:GRB190114C}
\bibinfo{author}{{MAGIC Collaboration}}, \bibinfo{author}{V.~A. {Acciari}},
  \bibinfo{author}{S.~{Ansoldi}}, \bibinfo{author}{L.~A. {Antonelli}},
  \bibinfo{author}{A.~{Arbet Engels}}, \bibinfo{author}{D.~{Baack}},
  \bibinfo{author}{A.~{Babi{\'c}}}, \bibinfo{author}{B.~{Banerjee}},
  \bibinfo{author}{U.~{Barres de Almeida}}, \bibinfo{author}{J.~A. {Barrio}},
  \bibinfo{author}{J.~{Becerra Gonz{\'a}lez}}, \bibinfo{author}{W.~{Bednarek}},
  \bibinfo{author}{L.~{Bellizzi}}, \bibinfo{author}{E.~{Bernardini}},
  \bibinfo{author}{A.~{Berti}}, \bibinfo{author}{J.~{Besenrieder}},
  \bibinfo{author}{W.~{Bhattacharyya}}, \bibinfo{author}{C.~{Bigongiari}},
  \bibinfo{author}{A.~{Biland}}, \bibinfo{author}{O.~{Blanch}},
  \bibinfo{author}{G.~{Bonnoli}}, \bibinfo{author}{{\v{Z}}.~{Bo{\v{s}}njak}},
  \bibinfo{author}{G.~{Busetto}}, \bibinfo{author}{A.~{Carosi}},
  \bibinfo{author}{R.~{Carosi}}, \bibinfo{author}{G.~{Ceribella}},
  \bibinfo{author}{Y.~{Chai}}, \bibinfo{author}{A.~{Chilingaryan}},
  \bibinfo{author}{S.~{Cikota}}, \bibinfo{author}{S.~M. {Colak}},
  \bibinfo{author}{U.~{Colin}}, \bibinfo{author}{E.~{Colombo}},
  \bibinfo{author}{J.~L. {Contreras}}, \bibinfo{author}{J.~{Cortina}},
  \bibinfo{author}{S.~{Covino}}, \bibinfo{author}{G.~{D'Amico}},
  \bibinfo{author}{V.~{D'Elia}}, \bibinfo{author}{P.~{da Vela}},
  \bibinfo{author}{F.~{Dazzi}}, \bibinfo{author}{A.~{de Angelis}},
  \bibinfo{author}{B.~{de Lotto}}, \bibinfo{author}{M.~{Delfino}},
  \bibinfo{author}{J.~{Delgado}}, \bibinfo{author}{D.~{Depaoli}},
  \bibinfo{author}{F.~{di Pierro}}, \bibinfo{author}{L.~{di Venere}},
  \bibinfo{author}{E.~{Do Souto Espi{\~n}eira}}, \bibinfo{author}{D.~{Dominis
  Prester}}, \bibinfo{author}{A.~{Donini}}, \bibinfo{author}{D.~{Dorner}},
  \bibinfo{author}{M.~{Doro}}, \bibinfo{author}{D.~{Elsaesser}},
  \bibinfo{author}{V.~{Fallah Ramazani}}, \bibinfo{author}{A.~{Fattorini}},
  \bibinfo{author}{A.~{Fern{\'a}ndez-Barral}}, \bibinfo{author}{G.~{Ferrara}},
  \bibinfo{author}{D.~{Fidalgo}}, \bibinfo{author}{L.~{Foffano}},
  \bibinfo{author}{M.~V. {Fonseca}}, \bibinfo{author}{L.~{Font}},
  \bibinfo{author}{C.~{Fruck}}, \bibinfo{author}{S.~{Fukami}},
  \bibinfo{author}{S.~{Gallozzi}}, \bibinfo{author}{R.~J. {Garc{\'\i}a
  L{\'o}pez}}, \bibinfo{author}{M.~{Garczarczyk}},
  \bibinfo{author}{S.~{Gasparyan}}, \bibinfo{author}{M.~{Gaug}},
  \bibinfo{author}{N.~{Giglietto}}, \bibinfo{author}{F.~{Giordano}},
  \bibinfo{author}{N.~{Godinovi{\'c}}}, \bibinfo{author}{D.~{Green}},
  \bibinfo{author}{D.~{Guberman}}, \bibinfo{author}{D.~{Hadasch}},
  \bibinfo{author}{A.~{Hahn}}, \bibinfo{author}{J.~{Herrera}},
  \bibinfo{author}{J.~{Hoang}}, \bibinfo{author}{D.~{Hrupec}},
  \bibinfo{author}{M.~{H{\"u}tten}}, \bibinfo{author}{T.~{Inada}},
  \bibinfo{author}{S.~{Inoue}}, \bibinfo{author}{K.~{Ishio}},
  \bibinfo{author}{Y.~{Iwamura}}, \bibinfo{author}{L.~{Jouvin}},
  \bibinfo{author}{D.~{Kerszberg}}, \bibinfo{author}{H.~{Kubo}},
  \bibinfo{author}{J.~{Kushida}}, \bibinfo{author}{A.~{Lamastra}},
  \bibinfo{author}{D.~{Lelas}}, \bibinfo{author}{F.~{Leone}},
  \bibinfo{author}{E.~{Lindfors}}, \bibinfo{author}{S.~{Lombardi}},
  \bibinfo{author}{F.~{Longo}}, \bibinfo{author}{M.~{L{\'o}pez}},
  \bibinfo{author}{R.~{L{\'o}pez-Coto}},
  \bibinfo{author}{A.~{L{\'o}pez-Oramas}}, \bibinfo{author}{S.~{Loporchio}},
  \bibinfo{author}{B.~{Machado de Oliveira Fraga}},
  \bibinfo{author}{C.~{Maggio}}, \bibinfo{author}{P.~{Majumdar}},
  \bibinfo{author}{M.~{Makariev}}, \bibinfo{author}{M.~{Mallamaci}},
  \bibinfo{author}{G.~{Maneva}}, \bibinfo{author}{M.~{Manganaro}},
  \bibinfo{author}{K.~{Mannheim}}, \bibinfo{author}{L.~{Maraschi}},
  \bibinfo{author}{M.~{Mariotti}}, \bibinfo{author}{M.~{Mart{\'\i}nez}},
  \bibinfo{author}{S.~{Masuda}}, \bibinfo{author}{D.~{Mazin}},
  \bibinfo{author}{S.~{Mi{\'c}anovi{\'c}}}, \bibinfo{author}{D.~{Miceli}},
  \bibinfo{author}{M.~{Minev}}, \bibinfo{author}{J.~M. {Miranda}},
  \bibinfo{author}{R.~{Mirzoyan}}, \bibinfo{author}{E.~{Molina}},
  \bibinfo{author}{A.~{Moralejo}}, \bibinfo{author}{D.~{Morcuende}},
  \bibinfo{author}{V.~{Moreno}}, \bibinfo{author}{E.~{Moretti}},
  \bibinfo{author}{P.~{Munar-Adrover}}, \bibinfo{author}{V.~{Neustroev}},
  \bibinfo{author}{C.~{Nigro}}, \bibinfo{author}{K.~{Nilsson}},
  \bibinfo{author}{D.~{Ninci}}, \bibinfo{author}{K.~{Nishijima}},
  \bibinfo{author}{K.~{Noda}}, \bibinfo{author}{L.~{Nogu{\'e}s}},
  \bibinfo{author}{M.~{N{\"o}the}}, \bibinfo{author}{S.~{Nozaki}},
  \bibinfo{author}{S.~{Paiano}}, \bibinfo{author}{J.~{Palacio}},
  \bibinfo{author}{M.~{Palatiello}}, \bibinfo{author}{D.~{Paneque}},
  \bibinfo{author}{R.~{Paoletti}}, \bibinfo{author}{J.~M. {Paredes}},
  \bibinfo{author}{P.~{Pe{\~n}il}}, \bibinfo{author}{M.~{Peresano}},
  \bibinfo{author}{M.~{Persic}}, \bibinfo{author}{P.~G. {Prada Moroni}},
  \bibinfo{author}{E.~{Prandini}}, \bibinfo{author}{I.~{Puljak}},
  \bibinfo{author}{W.~{Rhode}}, \bibinfo{author}{M.~{Rib{\'o}}},
  \bibinfo{author}{J.~{Rico}}, \bibinfo{author}{C.~{Righi}},
  \bibinfo{author}{A.~{Rugliancich}}, \bibinfo{author}{L.~{Saha}},
  \bibinfo{author}{N.~{Sahakyan}}, \bibinfo{author}{T.~{Saito}},
  \bibinfo{author}{S.~{Sakurai}}, \bibinfo{author}{K.~{Satalecka}},
  \bibinfo{author}{K.~{Schmidt}}, \bibinfo{author}{T.~{Schweizer}},
  \bibinfo{author}{J.~{Sitarek}}, \bibinfo{author}{I.~{{\v{S}}nidari{\'c}}},
  \bibinfo{author}{D.~{Sobczynska}}, \bibinfo{author}{A.~{Somero}},
  \bibinfo{author}{A.~{Stamerra}}, \bibinfo{author}{D.~{Strom}},
  \bibinfo{author}{M.~{Strzys}}, \bibinfo{author}{Y.~{Suda}},
  \bibinfo{author}{T.~{Suri{\'c}}}, \bibinfo{author}{M.~{Takahashi}},
  \bibinfo{author}{F.~{Tavecchio}}, \bibinfo{author}{P.~{Temnikov}},
  \bibinfo{author}{T.~{Terzi{\'c}}}, \bibinfo{author}{M.~{Teshima}},
  \bibinfo{author}{N.~{Torres-Alb{\`a}}}, \bibinfo{author}{L.~{Tosti}},
  \bibinfo{author}{S.~{Tsujimoto}}, \bibinfo{author}{V.~{Vagelli}},
  \bibinfo{author}{J.~{van Scherpenberg}}, \bibinfo{author}{G.~{Vanzo}},
  \bibinfo{author}{M.~{Vazquez Acosta}}, \bibinfo{author}{C.~F. {Vigorito}},
  \bibinfo{author}{V.~{Vitale}}, \bibinfo{author}{I.~{Vovk}},
  \bibinfo{author}{M.~{Will}}, \bibinfo{author}{D.~{Zari{\'c}}},
  \bibinfo{author}{L.~{Nava}},
\newblock \bibinfo{title}{{Teraelectronvolt emission from the
  {\ensuremath{\gamma}}-ray burst GRB 190114C}},
\newblock \bibinfo{journal}{\nat} \bibinfo{volume}{575} (\bibinfo{year}{2019})
  \bibinfo{pages}{455--458}. \DOIprefix\doi{10.1038/s41586-019-1750-x}.
  \href{http://arxiv.org/abs/2006.07249}{{\tt arXiv:2006.07249}}.
%Type = Article
\bibitem[{{Huang} et~al.(2022){Huang}, {Hu}, {Chen}, {Zha}, {Liu}, {Yao},
  {Cao}, and {Experiment}}]{LHAASO:GRB221009A}
\bibinfo{author}{Y.~{Huang}}, \bibinfo{author}{S.~{Hu}},
  \bibinfo{author}{S.~{Chen}}, \bibinfo{author}{M.~{Zha}},
  \bibinfo{author}{C.~{Liu}}, \bibinfo{author}{Z.~{Yao}},
  \bibinfo{author}{Z.~{Cao}}, \bibinfo{author}{T.~L. {Experiment}},
\newblock \bibinfo{title}{{LHAASO observed GRB 221009A with more than 5000 VHE
  photons up to around 18 TeV}},
\newblock \bibinfo{journal}{GRB Coordinates Network} \bibinfo{volume}{32677}
  (\bibinfo{year}{2022}) \bibinfo{pages}{1}. \URLprefix
  \url{https://gcn.gsfc.nasa.gov/gcn3/32677.gcn3}.

\end{thebibliography}

\end{document}